\documentstyle[psfig,rotating]{mn}

\newcommand{\beq}{\begin{equation}}
\newcommand{\eeq}{\end{equation}}
\newcommand{\DS}{\displaystyle}
\newcommand{\SS}{\scriptstyle}
\newcommand{\SSS}{\scriptscriptstyle}
\newcommand{\th}{\thinspace}
\newcommand{\vect}[1]{{\mathbf{ #1}}}
\newcommand{\dvect}[1]{{\dot{\mathbf{ #1}}}}
\newcommand{\bldr}[1] {{\textbf{\textit{ #1}}}}
\newcommand{\Menv}{\mbox{$M_{\rm env}$}}
\newcommand{\Renv}{\mbox{$R_{\rm env}$}}
\newcommand{\Mc}{\mbox{$M_{\rm c}$}}
\newcommand{\Rc}{\mbox{$R_{\rm c}$}}
\newcommand{\tMS}{\mbox{$t_{\rm MS}$}}
\newcommand{\tBGB}{\mbox{$t_{\rm BGB}$}}
\newcommand{\RZAMS}{\mbox{$R_{\rm ZAMS}$}}
\newcommand{\MHeF}{\mbox{$M_{\rm HeF}$}}
\newcommand{\MFGB}{\mbox{$M_{\rm FGB}$}}
\newcommand{\McHeI}{\mbox{$M_{\rm c,HeI}$}}
\newcommand{\McDU}{\mbox{$M_{\rm c,{\SSS DU}}$}}
\newcommand{\Mche}{\mbox{$M_{\rm c,He}$}}
\newcommand{\McBGB}{\mbox{$M_{\rm c,BGB}$}}
\newcommand{\McBAGB}{\mbox{$M_{\rm c,BAGB}$}}
\newcommand{\McSN}{\mbox{$M_{\rm c,SN}$}}
\newcommand{\MCh}{\mbox{$M_{\rm Ch}$}}
\newcommand{\Mcon}{\mbox{$M_{{\rm c} 1}$}}
\newcommand{\Mctw}{\mbox{$M_{{\rm c} 2}$}}
\newcommand{\Rcon}{\mbox{$R_{{\rm c} 1}$}}
\newcommand{\Rctw}{\mbox{$R_{{\rm c} 2}$}}
\newcommand{\Menvon}{\mbox{$M_{{\rm env} 1}$}}
\newcommand{\Menvtwd}{\mbox{$M'_{{\rm env} 2}$}}
\newcommand{\Mctwd}{\mbox{$M'_{{\rm c} 2}$}}
\newcommand{\Rctwd}{\mbox{$R'_{{\rm c} 2}$}}
\newcommand{\Msun}{\hbox{$\th {\rm M}_{\odot}$}}
\newcommand{\Lsun}{\hbox{$\th {\rm L}_{\odot}$}}
\newcommand{\Rsun}{\hbox{$\th {\rm R}_{\odot}$}}

\title{
Evolution of Binary Stars and the 
Effect of Tides on Binary Populations
}

\author[J. R. Hurley, C. A. Tout and O. R. Pols]
  {Jarrod R. Hurley$^{1,3}$,
  Christopher A. Tout$^1$ 
  and Onno R. Pols$^{2,4}$\\ 
  $^1$Institute of Astronomy, Madingley Road, Cambridge CB3 0HA, UK \\
  $^2$ Department of Mathematics, P.O. Box 28M, Monash University, Victoria, 
       3800, Australia \\ 
  $^3$ Current address: Department of Astrophysics, American Museum of
                        Natural History, Central Park West at 79th Street,
                        New York, NY 10024 \\
  $^4$ Current address: Astronomical Institute, Utrecht University, Postbus 80000, 
                        3508 TA Utrecht, The Netherlands \\ 
  E-mail: {\rm jhurley@amnh.org, cat@ast.cam.ac.uk, 
               O.R.Pols@astro.uu.nl} }

\pagerange{\pageref{firstpage}--\pageref{lastpage}}
\pubyear{2000}

\begin{document}
\label{firstpage}

\maketitle

\begin{abstract}

We present a rapid binary evolution algorithm that enables modelling
of even the most complex binary systems.
In addition to all aspects of single star evolution, features such as mass
transfer, mass accretion, common-envelope evolution, collisions,
supernova kicks and angular momentum loss mechanisms are included.
In particular, circularization and synchronization of the orbit by tidal 
interactions are calculated for convective, radiative and degenerate 
damping mechanisms.
We use this algorithm to study the formation and evolution of various
binary systems.
We also investigate the effect that tidal friction has on the outcome of 
binary evolution. 
Using the rapid binary code, we generate a series of large binary populations 
and evaluate the formation rate of interesting individual species
and events.
By comparing the results for populations with and without tidal friction
we quantify the hitherto ignored systematic effect of tides
and show that modelling of tidal evolution in binary systems is necessary in
order to draw accurate conclusions from population synthesis work.
Tidal synchronism is important but because orbits generally circularize 
before Roche-lobe overflow the outcome of the interactions of systems with 
the same semi-latus rectum is almost independent of eccentricity. 
It is not necessary to include a distribution of eccentricities in 
population synthesis of interacting binaries, however, the initial separations 
should be distributed according to the observed distribution of semi-latera   
recta rather than periods or semi-major axes.  

\end{abstract}


\section{Introduction}
\label{s:intro}

The evolution of binary stars does not differ from that of single stars 
unless they get in each other's way. 
If the binary orbit is wide enough the individual stars are not affected 
by the presence of a companion so that standard stellar evolution theory is all 
that is required to describe their evolution. 
However if the stars become close they can interact with consequences 
for the evolution and appearance of the stars as well as the nature of 
the orbit. 

The effective gravitational potential in a frame rotating with a circular 
binary system forms equipotential surfaces called Roche surfaces. 
A sphere of the volume enclosed by the critical Roche surface defines the 
Roche-lobe radius of each star. 
If either star fills its Roche-lobe then gas flows from the outer 
layers of that star through the inner Lagrangian point that connects the 
two Roche-lobes. 
Some or all of this gas may be captured by the companion star so that mass 
transfer occurs and, as a result, the subsequent evolution of both 
stars takes a different course from that of isolated stars. 
When the Roche-lobe filling star is a giant, with a convective envelope, or 
is significantly more massive than its companion then, as described by 
Paczy\'{n}ski (1976), the transferred mass may not be captured by the 
companion but instead accumulates in a common envelope surrounding both stars. 
The outcome of common-envelope evolution is still not fully understood but  
possible scenarios include loss of the envelope as the two cores spiral-in 
to form a closer binary or coalescence of the two stars. 

Even in a detached system it is still possible for the stars to interact 
tidally.  
Tides can synchronize the spin of the stars with the orbit and circularize 
an eccentric orbit as the binary tends towards an 
equilibrium state of minimum energy. 
Further, if one star is losing mass in a stellar wind the 
companion may accrete some of the material with consequences for the orbit.
For a discussion and review of the processes involved in close binary 
evolution and the various kinds of binaries or exotic stars that can result 
see Pringle \& Wade (1985) and Wijers, Davies \& Tout~(1996). 

The effects of close binary evolution are observed in many systems, 
such as cataclysmic variables, X-ray binaries and Algols, and in the 
presence of stars such as blue stragglers which cannot be explained by 
single star evolution. 
While many of the processes involved are not understood in detail we do 
have a qualitative picture of how binaries evolve and can hope to construct 
a model that correctly follows them through the various phases of evolution.  
Initial conditions are the mass and composition of the stars, the period 
(or separation) and eccentricity of the orbit. 
In order to conduct statistical studies of complete binary populations, 
i.e. population synthesis, such a model must be able to produce 
any type of binary that is observed in enough detail but at the same time   
be computationally efficient. 
By comparing results from the model with observed populations we can 
enhance our understanding of both binary evolution and the initial 
distributions (e.g. Eggleton, Fitchett \& Tout 1989; 
Tutukov \& Yungelson 1996; Terman, Taam \& Savage 1998; 
Nelson \& Eggleton 2001). 

Models for binary evolution have been presented in the past (e.g. Whyte \& 
Eggleton 1985; Pols \& Marinus 1994; Portegies Zwart \& Verbunt 1996). 
The model we present here supersedes the work of Tout et al. (1997) 
primarily by including eccentric orbits and stellar spins, 
which are subject to tidal circularization and synchronization. 
Amongst other improvements the possibility of mass accretion from a wind is 
included.  
Our model incorporates the detailed single star evolution (SSE) formulae of 
Hurley, Pols \& Tout (2000, hereinafter PapI) 
which allow for a wider range of stellar types than the description of stellar 
evolution used by Tout et al.\ (1997). 
This requires an updating of the treatment of processes such as Roche-lobe 
overflow, common-envelope evolution and coalescence by collision. 

Throughout this paper we refer to one star as the primary, 
mass $M_1$, stellar type $k_1$ etc., and the other as the 
secondary (or companion), mass $M_2$ and stellar type $k_2$.
At any time the primary is the star filling, or closest to filling, 
its Roche-lobe. 
Numerical values of mass, luminosity and radius are in solar units unless 
indicated otherwise. 
The stellar types correspond to the evolutionary phases desiginated by 
the rapid SSE algorithm of PapI, which are: 

\begin{eqnarray*}
0 & = & \mbox{ MS star } M \la 0.7 \mbox{ deeply or fully convective} \\
1 & = & \mbox{ MS star } M \ga 0.7 \\
2 & = & \mbox{ Hertzsprung Gap (HG)} \\
3 & = & \mbox{ First Giant Branch (GB)} \\
4 & = & \mbox{ Core Helium Burning (CHeB)} \\
5 & = & \mbox{ Early Asymptotic Giant Branch (EAGB)} \\
6 & = & \mbox{ Thermally Pulsing AGB (TPAGB)} \\
7 & = & \mbox{ Naked Helium Star MS (HeMS)} \\
8 & = & \mbox{ Naked Helium Star Hertzsprung Gap (HeHG)} \\
9 & = & \mbox{ Naked Helium Star Giant Branch (HeGB)} \\
10 & = & \mbox{ Helium White Dwarf (HeWD)} \\
11 & = & \mbox{ Carbon/Oxygen White Dwarf (COWD)} \\
12 & = & \mbox{ Oxygen/Neon White Dwarf (ONeWD)} \\
13 & = & \mbox{ Neutron Star (NS)} \\
14 & = & \mbox{ Black Hole (BH)} \\
15 & = & \mbox{ massless remnant.} 
\end{eqnarray*}

The SSE algorithm provides the stellar luminosity $L$, radius $R$, 
core mass $M_{\rm c}$, core radius $R_{\rm c}$, and spin frequency 
$\Omega_{\rm spin}$, for each of the component stars as they evolve. 
A prescription for mass loss from stellar winds is included in 
the SSE algorithm. 
The algorithm covers all the evolution phases from the zero-age main-sequence 
(ZAMS), up to and including the remnant stages, and is valid for all masses 
in the range 0.1 to $100 \Msun$ and metallicities from $Z = 10^{-4}$ to 
$Z = 0.03$. 
This rapid binary evolution algorithm is a natural 
extension of the SSE algorithm. 
Many of the formulae and much of the terminology contained in PapI are 
utilised in this current paper and therefore the interested reader is 
encouraged to review PapI. 

In Section~2 we describe the binary evolution algorithm in detail. 
Section~3 contains illustrative examples of Algol and cataclysmic variable 
evolution and also compares our model with the results of certain binary 
cases highlighted by Tout et al. (1997) and other authors. 
We present the results of population synthesis to examine the effects of tides 
on binary evolution in Section~4 and conclusions are given in Section~5.

\section{Binary Evolution}
\label{s:bsealg}

Before describing our treatment of Roche-lobe overflow (RLOF) in detail we 
describe interaction in detached systems, via wind accretion and tides. 
We also consider how mass and angular momentum loss from an individual 
component affects the system. 

Two stars, bound through their mutual gravity, move in 
elliptical orbits about their common centre-of-mass. 
In plane polar coordinates, $r$, the separation of the stars, and $\theta$,  
the phase angle, the equations of motion for an elliptical orbit are 
\beq
r = \frac{a \left( 1 - e^2 \right)}{1 + e \cos \theta} 
\eeq
and 
\beq 
h = r^2 \dot{\theta} 
\eeq
where $a$ is the semi-major axis of the ellipse and $e$ the eccentricity. 
The specific angular momentum of the system, $h = | \vect{h} |$, is given by  
\beq 
\vect{h} = \vect{r} \times \vect{v} 
\eeq
where both $\vect{r}$ and $\vect{v} = \dvect{r}$ lie in the orbital plane.  
Note that the semi-latus rectum 
\beq\label{e:ltoh}
l = a \left( 1 - e^2 \right) = \frac{h^2}{G M_{\rm b}} \, , 
\eeq
where 
\beq 
M_{\rm b} = M_1 + M_2 
\eeq 
is the total mass of the system, is constant if orbital angular momentum 
is conserved. 
These equations consider the stars as point masses interacting by gravity  
alone. 
Perturbing effects, such as tidal forces, are not taken into account.

\subsection{Wind Accretion}
\label{s:windac}

If the primary star loses mass in a wind at a rate ${\dot{M}}_{\rm 1W}$, 
which is determined according to a prescription for single stars such as 
that given in PapI, the secondary can accrete some of the material  
as it orbits through it. 
Wind accretion is important for the evolution of $\zeta$-Aurigae and VV 
Cephei systems (Che-Bohnenstengel \& Reimers 1986; Hack et al. 1992), 
symbiotic stars (Kenyon 1986; Iben \& Tutukov 1996) and 
massive X-ray binaries (Lamers, van den Heuvel \& Petterson 1976). 

The mean accretion rate, on to the secondary, can be estimated according 
to a Bondi \& Hoyle (1944) mechanism to be 
\beq\label{e:macc}  
\left\langle {\dot{M}}_{\rm 2A} \right\rangle = \frac{-1}{\sqrt{1 - e^2}} 
\left[ \frac{G M_2}{v^2_{\rm\SSS W}} \right]^2 \frac{{\alpha}_{\rm\SSS W}}
{2 a^2} \frac{1}{\left( 1 + v^2 \right)^{3/2}} {\dot{M}}_{\rm 1W} 
\eeq
where 
\begin{eqnarray} 
v^2 & = & \frac{v^2_{\rm orb}}{v^2_{\rm\SSS W}} \\  
v^2_{\rm orb} & = & \frac{G M_{\rm b}}{a} \, .  
\end{eqnarray} 
The wind velocity is difficult to determine accurately from observations.  
We set it proportional to the escape velocity from the surface, 
\beq 
v^2_{\rm\SSS W} = 2 {\beta}_{\rm\SSS W} \frac{G M_1}{R_1} \, , 
\eeq 
where the value of ${\beta}_{\rm\SSS W}$ must depend on the spectral type. 
Observations of luminous stars indicate that ${\beta}_{\rm\SSS W}$ 
decreases from about 7 for O stars to about 0.5 for A and F stars 
(Lamers, Snow \& Lindholm 1995). 
Cool super-giant wind velocities are observed to be $5 - 35 \, {\rm km} \,
{\rm s}^{-1}$ (Ku\v{c}inskas 1999) where the lower limit should roughly
correspond to the wind from the largest stars, about $900 \Rsun$. 
This suggests ${\beta}_{\rm\SSS W} \simeq 1/8$. 

Averages over an orbital period are justified because the duration of 
mass loss is typically much greater than $P$.  
Eq.~(\ref{e:macc}) is strictly only valid under the fast wind assumption  
($v_{\rm\SSS W} \gg v_{\rm orb}$). 
If the orbital and wind velocities are comparable in size then the binary 
motion disturbs the shape of the wind and the assumption of 
spherical symmetry is violated. 
The parameter ${\alpha}_{\rm\SSS W}$ is taken to be 3/2 appropriate for 
Bondi-Hoyle accretion and agrees with the lower limit found by 
Boffin \& Jorissen (1988) when modelling wind accretion from a supergiant 
to explain the Barium star $\zeta\,$Capricorni. 

It is necessary that the secondary does not accrete more mass than 
is lost by the primary, as may follow from eq.~(\ref{e:macc}) for highly 
eccentric orbits, so we enforce the condition 
\beq 
| {\dot{M}}_{\rm 2A} | \leq 0.8 \, | {\dot{M}}_{\rm 1W} | 
\eeq 
to ensure this, and for orbital angular momentum considerations 
(see Section~\ref{s:delorb}). 
Use of this condition is rare because we expect any system with 
appreciable mass accretion to also be experiencing strong tidal evolution 
(see Section~\ref{s:tidal}). 
We note that there is no special reason for the choice of 0.8 as the upper 
limit. 
If the mass-ratio of the binary system is close to unity it is possible 
for both stars to lose mass at the same time and 
the primary may accrete some mass from the secondary. 
In this case any interaction between the wind material from the two stars 
is ignored and eq.~(\ref{e:macc}) is simply repeated with the roles of the 
stars reversed. 

If a star loses mass in a stellar wind then angular momentum is lost too. 
When the companion accretes some of the material a fraction of the
angular momentum lost from the the intrinsic spin of the primary goes into
the spin angular momentum of the companion. 
Therefore 
\beq
{\dot{J}}_{{\rm spin}1} = \frac{2}{3} {\dot{M}}_{\rm 1W} h_1 + \frac{2}{3} 
{\mu}_{\rm\SSS W} {\dot{M}}_{\rm 1A} h_2 
\eeq
where ${\dot{M}}_{\rm 1W} < 0$ and ${\dot{M}}_{\rm 1A} > 0$ 
(see also Section~7.2 of PapI). 
We assume that the specific angular momentum in the wind is transferred 
with perfect efficiency, i.e. ${\mu}_{\rm\SSS W} = 1$, whereas Ruffert (1999), 
who performed a numerical 3-D study of wind accretion using a high
resolution hydrodynamic code, found that the efficiency is roughly
between 0-70\%, depending on model parameters. 

Motivated by observations of RS$\,$CVn binaries in which the more evolved star 
becomes the less massive before Roche-lobe overflow has begun, Tout \& 
Eggleton (1988) suggested that mass loss may be tidally enhanced by the 
presence of a moderately close companion. 
They give a simple descriptive formula 
\beq\label{e:menhan} 
\dot{M} = {\dot{M}}_{\rm R} \left\{ 1 + B_{\rm\SSS W} \max \left( \frac{1}{2} , 
\frac{R}{R_{\rm L}} \right)^6 \right\} 
\eeq
for the enhanced mass-loss rate where ${\dot{M}}_{\rm R}$ is the Reimers rate  
(cf. PapI), $R_{\rm L}$ is the Roche-lobe radius and 
$B_{\rm\SSS W} \simeq 10^4$. 
It is uncertain whether such enhanced mass loss rates are realistic for a 
wide range of binary systems, so we include eq.~(\ref{e:menhan}) in the model 
with $B_{\rm\SSS W}$ as a variable parameter, but typically 
$B_{\rm\SSS W} = 0$ is used.

\subsection{Orbital Changes due to Mass Variations}
\label{s:delorb}

The orbital angular momentum can be expressed as 
\beq 
J_{\rm orb} = M_1 a_1^2 {\Omega}_{\rm orb} + M_2 a_2^2 {\Omega}_{\rm orb}  
\eeq 
which naively gives 
\beq\label{e:jorbcon}
{\dot{J}}_{\rm orb} = \left[ \left( {\dot{M}}_{\rm 1W} + 
{\dot{M}}_{\rm 2A} \right) 
a_1^2 + \left( {\dot{M}}_{\rm 2W} + {\dot{M}}_{\rm 1A} \right) a_2^2 \right] 
{\Omega}_{\rm orb} 
\eeq
so that if the primary loses mass in a wind the orbit loses an amount 
$\Delta M_1 a_1^2 {\Omega}_{\rm orb}$ of angular momentum but if the 
secondary accretes some of this mass then an amount $\Delta M_2 a_1^2 
{\Omega}_{\rm orb}$ of angular momentum is returned to the orbit. 
This is exactly the case of conservative mass exchange where the momentum 
gained by the secondary must be taken from the primary. 
However if the fast wind approximation is considered more carefully 
(Gair \& Snellgrove private communication, for details see Hurley 2000) 
then, averaged over an orbit 
\beq\label{e:edote}
\frac{\left\langle \delta e  \right\rangle}{e} =
- \left\langle \delta M_2 \right\rangle \left[ \frac{1}{M_{\rm b}} +
\frac{1}{2 M_2}
\right] \, , 
\eeq
and 
\beq\label{e:adota}
\frac{\left\langle \delta a  \right\rangle}{a} =
- \frac{\delta M_1}{M_{\rm b}} - \left[ \frac{2 - e^2}{M_2} +
\frac{1 + e^2}{M_{\rm b}}
\right] \frac{\left\langle \delta M_2 \right\rangle}{1 - e^2} \, .
\eeq
The change to the orbital angular momentum is then 
\beq
\frac{\left\langle \delta J  \right\rangle}{J} =
\frac{\delta M_1 M_2 - \left\langle \delta M_2 \right\rangle M_1}
{M_1 M_{\rm b}}
\eeq
and
\beq
\left\langle \delta J  \right\rangle = \left[ \left( \frac{M_2}{M_{\rm b}}
\right)^2 \delta M_1 - \left( \frac{M_1}{M_{\rm b}} \right)^2 \frac{M_2}{M_1}
\left\langle \delta M_2 \right\rangle \right] a^2 {\Omega}_{\rm orb}
\eeq
where
\beq
a_1^2 = \left( \frac{M_2}{M_{\rm b}} \right)^2 a^2 \quad , \quad
a_2^2 = \left( \frac{M_1}{M_{\rm b}} \right)^2 a^2
\eeq
so that
\beq
{\dot{J}}_{\rm orb} = \left( {\dot{M}}_{\rm 1W} a_1^2 - \frac{M_2}{M_1}
{\dot{M}}_{\rm 2A} a_2^2 \right) {\Omega}_{\rm orb} \, .
\eeq
This breaks down if $| \Delta M_2 | \rightarrow | \Delta M_1 |$ which should 
only occur in extreme cases, e.g. when $R \sim R_{\rm L}$ so that 
$v_{\rm orb} \sim v_{\rm\SSS W}$. 
The restriction to eq.~(\ref{e:macc}) avoids this problem. 
If mass loss from both stars is considered then 
\beq\label{e:jdot} 
{\dot{J}}_{\rm orb} = \left[ \left( {\dot{M}}_{\rm 1W} - \frac{M_1}{M_2} 
{\dot{M}}_{\rm 1A} \right) a_1^2 + \left( {\dot{M}}_{\rm 2W} - \frac{M_2}{M_1} 
{\dot{M}}_{\rm 2A} \right) a_2^2 \right] {\Omega}_{\rm orb} \, .
\eeq
In an eccentric orbit this model takes into account the fact that more 
matter is accreted at periastron than apastron so that mass accretion 
results in circularization of the orbit. 
The secondary loses momentum owing to the drag it experiences in moving 
through the primary wind and it is 
assumed that these losses are taken from the system in the wind. 
The model gives $\dot{e}/e < 0$ in all cases. 
Boffin \& Jorissen (1988) predicted a modest increase in eccentricity,  
based on Huang (1956), due to neglecting changes to the orbital velocity.  
This was subsequently corrected by Theuns, Boffin \& Jorissen~(1996) 
although their treatment still allows eccentricity growth in certain cases. 
Another difference stems from the time averages where Huang seems 
to assume, incorrectly, that the time average operator is multiplicative. 
Eq.~(\ref{e:edote}) gives a circularization timescale  
which, for small eccentricity, is 
\beq 
\frac{1}{{\tau}_{\rm circ}} = \frac{| \dot{e} |}{e} \approx 
q_2^2 \left( \frac{R_1}{a} \right)^2 
\frac{| {\dot{M}}_{\rm 1W}|}{M_{\rm b}} 
\eeq 
where $q_2 = M_2/M_1$ is the mass-ratio of the secondary star. 
For a $2.0 \Msun$ star on the giant branch the timescale is 
\beq 
{\tau}_{\rm circ} \sim 1.263 \times 10^{10} \frac{1}{q_2^2} \left( 
\frac{a}{R_1} \right)^2 \: {\rm yr} 
\eeq 
and we show in Section~\ref{s:tidal} that the equivalent timescale 
due to damping of tides raised on the surface of the $2.0 \Msun$ star is 
\beq 
{\tau}_{\rm circ} \sim 0.43 \frac{1}{q_2^2} \left( 
\frac{a}{R_1} \right)^2 \: {\rm yr} \, . 
\eeq 
Therefore the changes in eccentricity owing to mass variations are very  
small compared to the orbital changes brought about tidal friction. 
Even on the asymptotic giant branch, where the mass loss rates increase 
significantly, this is still true because the deep convective envelope 
even more rapidly damps the tides.

\subsection{Tidal Evolution}
\label{s:tidal}

Observations of close binary stars reveal that stellar rotation tends to 
synchronize with the orbital motion and can be slower (Levato 1974) 
or faster (Strassmeier 1996) than for single stars of the same type. 
The stars do not need to be in contact for this co-rotation to be achieved 
so a mechanism such as tidal friction must be the cause. 
Tidal interaction is important in changing the orbit of a close 
detached binary. 
The degree of interaction is critically dependent on the ratio of the 
stellar radius to the separation of the stars (Zahn 1977; Hut 1981). 
The presence of a companion introduces a 
tidal, or differential, force which acts to elongate 
the star along the line between the centres of mass, producing tidal bulges. 
If the rotational period of the star is shorter than the orbital period 
then frictional forces on the surface of the star will drag the bulge axis 
ahead of the line of centres. 

The tidal field can be decomposed into two parts: 
the {\it equilibrium tide} which, if all forms of dissipation are 
neglected, is described by assuming that the star is always in 
hydrostatic equilibrium 
and the {\it dynamical tide} which arises from stellar oscillations. 
If dissipative processes are at work within the star the equilibrium tide 
will lag or precede the line of centres. 
The resulting torque transfers angular momentum between the stellar spin 
and the orbit. 
Energy is dissipated in the tides which diminishes the total energy. 
Thus the orbital parameters change and either asymptotically approach an 
equilibrium state or lead to an accelerated spiralling in of the two stars 
(Hut 1980). 
The equilibrium state is characterized by co-rotation and a circular orbit, 
corresponding to a minimum energy for a given total angular momentum and 
alignment of the spin-orbit axes (which we do not consider in detail). 

The main difficulty in a treatment of tidal evolution arises in identifying 
the physical processes which are responsible for the tidal torque, that is, 
the dissipation mechanisms which cause the tides to deviate from an 
instantaneous equilibrium shape and thus to be misaligned with the line of 
centres (see Zahn 1992 and Tassoul \& Tassoul 1996 for recent reviews of 
tidal evolution theory). 

\subsubsection{The Equilibrium Tide with Convective Damping}
\label{s:beqtid}

The most efficient form of dissipation which may operate on the 
equilibrium tide is turbulent viscosity in the convective regions of 
a star. 
All other forms of dissipation produce timescales that normally 
exceed stellar nuclear lifetimes. 
This mechanism provides a satisfactory interpretation of the behaviour of 
stars possessing relatively deep convective envelopes. 
However, it should be noted that this is only an interpretation because no 
completely satisfactory description of stellar convection is available. 
Nevertheless, estimates based on the eddy-viscosity treatment of convection, 
i.e. the mixing-length theory, seem to adequately represent the tidal 
timescales for such stars (Zahn 1989). 
 
To investigate the change of binary parameters due to tidal friction Hut (1981) 
considers a simple model in which only equilibrium tides are described, with 
very small deviations in position and amplitude with respect to the 
equipotential surfaces. 
The resulting tidal evolution equations for the tide raised on a star of 
mass $M$ due to the presence of its companion with mass $m$ are 
\begin{eqnarray}\label{e:dedt}
\frac{de}{dt} & = & - 27 \left( \frac{k}{T} \right)_{\rm c} 
q \left( 1 + q \right) 
\left( \frac{R}{a} \right)^8 \frac{e}{\left( 1 - e^2 \right)^{13/2}} 
 \nonumber \\  & & \times 
\left[ f_3 \left( e^2 \right) - \frac{11}{18} \left( 1 - e^2 \right)^{3/2} 
f_4 \left( e^2 \right) \frac{{\Omega}_{\rm spin}}{{\Omega}_{\rm orb}} \right] 
\end{eqnarray}
\begin{eqnarray}\label{e:dodt}
\frac{d {\Omega}_{\rm spin}}{dt} & = & 3 \left( \frac{k}{T} \right)_{\rm c}  
\frac{q^2}{r_g^2} \left( \frac{R}{a} \right)^6  
\frac{{\Omega}_{\rm orb}}{\left( 1 - e^2 \right)^{6}} 
 \nonumber \\ 
 & & \times \left[ f_2 \left( e^2 \right) - \left( 1 - e^2 \right)^{3/2} 
f_5 \left( e^2 \right) \frac{{\Omega}_{\rm spin}}{{\Omega}_{\rm orb}} \right] 
\end{eqnarray}
where the $f_n$'s are polynomial expressions in $e^2$ given by Hut~(1981). 
The apsidal motion constant of the primary $k$ takes into account the 
structure of the star 
and $T$ is the timescale on which significant changes in the orbit take 
place through tidal evolution, i.e. the damping timescale 
defined in terms of the time it takes for the tidal bulge to catch-up with 
the current position of the line of centres. 
Furthermore  $q = m/M$ ($= q_2$ as defined in this work) 
is the mass-ratio of the stars and $r_{\rm g}$ is  
the radius of gyration where $r_{\rm g}^2 = I/(M R^2)$.
This set of equations are valid for any value of the eccentricity and 
further discussion of the weak friction model can be found in Alexander (1973) 
and Kopal (1978). 

From eq.~(\ref{e:dodt}) the synchronization timescale can be defined as 
\beq\label{e:ctsyn}
\frac{1}{{\tau}_{\rm sync}} = \frac{{\dot{\Omega}}_{\rm spin}}
{\left( {\Omega}_{\rm spin} - {\Omega}_{\rm orb} \right)} = 
3 \left( \frac{k}{T} \right)_{\rm c} q_2^2 \frac{M R^2}{I} 
\left( \frac{R}{a} \right)^6
\eeq
and from eq.~(\ref{e:dedt}) the circularization timescale is 
\beq\label{e:ctcir}
\frac{1}{{\tau}_{\rm circ}} =  
\frac{21}{2} \left( \frac{k}{T} \right)_{\rm c} q_2 \left( 1 + q_2 \right) 
\left( \frac{R}{a} \right)^8 
\eeq
when $e \approx 0$ and, in the case of ${\tau}_{\rm circ}$, when 
${\Omega}_{\rm spin} = {\Omega}_{\rm orb}$. 
Because $a > R$, ${\tau}_{\rm sync} < {\tau}_{\rm circ}$ and co-rotation is  
achieved before the orbit circularizes. 
When the orbit is nearly circular, $e \simeq 0$, it is stable 
($\dot{e} < 0$) if ${\Omega}_{\rm spin}/{\Omega}_{\rm orb} < 18/11$,  
in agreement with Zahn (1977). 
If ${\Omega}_{\rm spin} > {\Omega}_{\rm orb}$ then $\dot{\Omega}_{\rm spin} < 
0$ and $\dot{a} > 0$ which is as expected, i.e. transfer of angular momentum 
from the star to the orbit. 

In this work it is more convenient to use the tidal circularization 
timescale in the form given by Rasio et al.\ (1996), 
\beq\label{e:c2tcir}
\frac{1}{{\tau}_{\rm circ}} =  
\frac{f_{\rm conv}}{{\tau}_{\rm conv}} \frac{\Menv}{M} q_2 \left( 1 + q_2 \right) 
\left( \frac{R}{a} \right)^8 \, , 
\eeq
so we use 
\beq
\left( \frac{k}{T} \right)_{\rm c} = \frac{2}{21} \frac{f_{\rm conv}}
{{\tau}_{\rm conv}} \frac{\Menv}{M} \: {\rm yr}^{-1} 
\eeq
in Hut's equations. 
Here 
\beq\label{e:tconv} 
{\tau}_{\rm conv} = 0.4311 \left( \frac{\Menv \Renv \left( R - \frac{1}{2} 
\Renv \right)}{3 L} \right)^{1/3} \: {\rm yr} 
\eeq 
is the eddy turnover timescale (timescale on which the largest convective 
cells turnover) and \Renv\ is the depth of the convective envelope.  
Eq.~(\ref{e:tconv}) is essentially the same as eq.~(4) of 
Rasio et al.\ (1996) except that we use the radius in the middle of 
the convective zone, i.e. $r \simeq ( R - \frac{1}{2} R_{\rm env} )$,
rather than at the base, as the typical position of a convective element.  
The numerical factor $f_{\rm conv}$ is 
\beq
f_{\rm conv} = \min \left( 1 , \left( \frac{P_{\rm tid}}{2 {\tau}_{\rm conv}} 
\right)^2 \right) 
\eeq
where $P_{\rm tid}$ is the tidal pumping timescale given by 
\beq
\frac{1}{P_{\rm tid}} = \left| \frac{1}{P_{\rm orb}} - \frac{1}{P_{\rm spin}} 
\right| \, .
\eeq
As noted by Rasio et al.\ (1996), theoretically and observationally, 
$f_{\rm conv} = 1$ as long as ${\tau}_{\rm conv} \ll P_{\rm orb}$. 
However, if $P_{\rm tid} < {\tau}_{\rm conv}$ then the largest convective 
cells can no longer contribute to viscosity, because the velocity field they 
are damping changes direction before they can transfer momentum. 
Only the eddies that turn-over in a time less than the pumping timescale 
contribute so that the average length and the velocity of these cells 
are both smaller by the same factor $2 {\tau}_{\rm conv} / P_{\rm tid}$. 
The factor of two arises because tides come around twice in each period. 

We implement tidal evolution during a time-step $\Delta t$ by calculating 
$\dot{e}$ and ${\dot{\Omega}}_{\rm spin}$ from eqs.~(\ref{e:dedt}) and 
(\ref{e:dodt}) 
and ensuring that the star is not spun down (or up) past the equilibrium 
spin at which no angular momentum can be transferred, 
\beq
{\Omega}_{\rm spin,eq} = f_2 \left( e^2 \right) {\Omega}_{\rm orb} 
\left[ \frac{1}{f_5 \left( e^2 \right) \left( 1 - e^2 \right)^{3/2}} \right] 
\, .
\eeq 
The rate of change of the rotational angular momentum is then given by 
\beq
{\dot{J}}_{\rm spin} = \left[ k'_2 \left( M - \Mc \right) R^2 + 
k'_3 \Mc {\Rc}^2 \right]  
{\dot{\Omega}}_{\rm spin} \, , 
\eeq
where $k'_2 = 0.1$ and $k'_3 = 0.21$ (see Section~7.2 of PapI which contains  
a description of how $J_{\rm spin}$ and $\Omega_{\rm spin}$ are calculated), 
so that 
\begin{eqnarray*}
J_{\rm orb} & = & J_{\rm orb} - {\dot{J}}_{\rm spin} \Delta t \\ 
J_{\rm spin} & = & J_{\rm spin} + {\dot{J}}_{\rm spin} \Delta t \\ 
e & = & e + \dot{e} \Delta t 
\end{eqnarray*}
and the change in semi-major axis is taken care of by the change in 
orbital angular momentum. 
A description of this type has been used successfully for red giants 
by Karakas, Tout \& Lattanzio~(2000) to model the eccentricities of 
Barium stars. 

The procedure for tidal circularization and synchronization given above is 
valid for all stars with convective envelopes. 
These have stellar types 
\begin{eqnarray*}
k & = & 1 \quad \& \quad M < 1.25 \\ 
k & \in & \left\{ 0, 2, 3, 5, 6, 8, 9 \right\} \, .
\end{eqnarray*}
If the companion has an appreciable convective envelope too then the 
roles of the two stars can be reversed and the process repeated. 
Details of how to obtain the mass of the convective envelope can be found 
in Section~7.2 of PapI. 
The radial extent of the convective envelope is calculated in a similar 
fashion. 
For all giant-like stars ($k \in \left\{  3,5,6,8,9 \right\}$) 
$\Renv = R - \Rc$. 
For main-sequence stars 
\beq 
R_{\rm env,0} = \left\{ \begin{array} {l@{\:}l} 
R & M \leq 0.35 \\ 
R' \left( \frac{1.25 - M}{0.9} \right)^{1/2} 
& 0.35 < M < 1.25 \\ 
0.0 & M \geq 1.25 
\end{array} \right. 
\eeq 
and then 
\beq 
\Renv = R_{\rm env,0} \left( 1 - \tau \right)^{1/4} 
\eeq 
where 
\beq 
\tau = \frac{t}{\tMS} \, 
\eeq 
and $R'$ is the radius of a MS star with $M = 0.35 \Msun$ at $\tau$ 
(see PapI). 
For Hertzsprung gap stars the convective envelope grows as the stars evolve 
so that 
\beq 
\Renv = {\tau}^{1/2} \left( R - \Rc \right) 
\eeq 
with  
\beq 
\tau = \frac{t - \tMS}{\tBGB - \tMS} \, . 
\eeq 
The main-sequence lifetime, $\tMS$, and the time taken for a star to reach 
the base of the GB, $\tBGB$, are given by eqns.~(4) and (5) of PapI.

\subsubsection{The Dynamical Tide with Radiative Damping}
\label{s:bdytid}

For stars with radiative envelopes, another mechanism is required to explain 
the observed synchronization of close binaries. 
A star can experience a range of oscillations that arise from, 
and are driven by, the tidal field: the dynamical tide (Zahn 1975). 

Zahn (1977) derived a circularization timescale 
\beq\label{e:rtcir}
\frac{1}{{\tau}_{\rm circ}} = 
\frac{21}{2} \left( \frac{G M}{R^3} \right)^{1/2} 
q_2 \left( 1 + q_2 \right)^{11/6} E_2 \left( \frac{R}{a} \right)^{21/2} 
\eeq
for radiative damping of the dynamical tide. 
Comparison with eq.~(\ref{e:ctcir}) gives 
\beq
\left( \frac{k}{T} \right)_{\rm r} = 1.9782 \times 10^4 \frac{M R^2}{a^5} 
\left( 1 + q_2 \right)^{5/6} E_2 \: {\rm yr}^{-1} 
\eeq
where $E_2$ is a second-order tidal coefficient which can be fitted to 
values given by Zahn (1975),  
\beq 
E_2 = 1.592 \times 10^{-9} M^{2.84} \, . 
\eeq 
We use $\left( k/T \right)_{\rm r}$, along with $r_g^2 = k'_2$, 
in Hut's equations to model tides raised on stars with a radiative envelope. 
These have types 
\begin{eqnarray*}
k & = & 1 \quad M \geq 1.25 \\ 
k & \in & \left\{ 4,7 \right\} \, .
\end{eqnarray*}
The corresponding synchronization timescale is given by 
\beq\label{e:rtsyn}
\frac{1}{{\tau}_{\rm synch}} = 
52^{5/3} \left( \frac{G M}{R^3} \right)^{1/2} \frac{M R^2}{I} 
\, q_2^2 \left( 1 + q_2 \right)^{5/6} E_2 \left( \frac{R}{a} \right)^{17/2} 
\eeq
so that 
\beq 
\frac{1}{{\tau}_{\rm circ}} = K \frac{1}{{\tau}_{\rm synch}} \quad , \quad  
K \approx 0.075 \, \frac{1 + q_2}{q_2} \left( \frac{R}{a} \right)^2  
\eeq 
and once again ${\tau}_{\rm circ} > {\tau}_{\rm synch}$ for realistic  
mass-ratios.

\subsubsection{Tides on Degenerate Stars}
\label{s:bdgtid}

Campbell (1984) examined the tidal effects in double-degenerate binaries and 
derived a synchronization timescale  
\beq\label{e:dtsyn}
\frac{1}{{\tau}_{\rm synch}} = 
\frac{1}{1.3 \times 10^7} \, q_2^2 \left( \frac{L}{M} \right)^{5/7}  
\left( \frac{R}{a} \right)^{6} \: {\rm yr}^{-1} 
\eeq
for a tide raised on a WD of mass $M$.  
This assumes that the WD on which the tides act is initially 
non-rotating in inertial space and ${\tau}_{\rm synch}$ provides an 
upper limit if this is not true. 
The synchronization timescale would also be shorter if tidally excited 
non-radial oscillations in the star were considered. 
For WD-WD binaries the orbit should already be circular and a 
circularization timescale is not relevant. 
However a synchronization timescale is applicable because the companion may be 
spun-up by mass transfer. 
The formulation also applies to WD-NS binaries which are in most cases  
initially eccentric so that a circularization mechanism is required. 

Comparing the synchronization timescale with that from eq.~(\ref{e:ctsyn}) 
for convective damping gives 
\beq
\left( \frac{k}{T} \right)_{\rm d} = 2.564 \times 10^{-8} r_g^2  
\left( \frac{L}{M} \right)^{5/7} \: {\rm yr}^{-1} 
\eeq
which can be used along with $r_{\rm g}^2 = k'_3$ in Hut's equations to give an 
estimate of the effect of tidal forces on the orbital parameters owing to 
degenerate damping.

In most cases a WD primary is expected to be effectively non-rotating because 
the progenitor star would have spun-down considerably on the AGB through  
mass loss and a large radius increase. 
However, it should be noted that if the primary is initially rotating it may 
be better to take eq.~(\ref{e:dtsyn}) as representative of the 
circularization time. 
Assuming similar results to those derived for convective and radiative damping, 
this would mean an underestimate of ${\tau}_{\rm circ}$. 
Therefore as eq.~(\ref{e:dtsyn}) is an overestimate of ${\tau}_{\rm synch}$,  
and in the absence of a better understanding, we can hope that one 
approximation corrects for the other. 

\subsubsection{Relative Strengths of the Damping Mechanisms}
\label{s:brltid}

Timescales for the synchronization and circularization of binary stars with 
radiative envelopes are generally orders of magnitude larger than those 
characterizing the equilibrium tide in convective-envelope stars. 
As an illustration consider a binary consisting of two MS stars, with masses  
$1.0 \Msun$ and $2.0 \Msun$ respectively, separated by a distance of 
$10.0 \Rsun$. 
The circularization timescale owing to tides raised on the $1.0 \Msun$ star is 
${\tau}_{\rm circ,c} \simeq 2.5 \times 10^8\,$yr as given by 
eq.~(\ref{e:ctcir}) for convective damping while that due to radiative 
damping of the tides on the $2.0 \Msun$ star is, according to 
eq.~(\ref{e:rtcir}), ${\tau}_{\rm circ,r} \simeq 1.8 \times 10^{10}\,$yr,  
an order of magnitude greater than the nuclear lifetime of the star. 
Tidal friction on the $1.0 \Msun$ star dominates changes to the 
orbital parameters because the convective timescale, ${\tau}_{\rm circ,c}$,  
is an order of magnitude less than the MS lifetime of the $2.0 \Msun$ star. 
For comparison a WD-NS binary with a separation 10 times the WD radius 
would have ${\tau}_{\rm circ,d} \ga 10^{14}\,$yr. 
 
\begin{table}
\begin{center}
\begin{tabular}{|rcccc|}
\hline
 $M$ & $\RZAMS$ & $\tMS$/yr & $(a/R)_{\rm sync}$ & $(a/R)_{\rm circ}$ \\
\hline
 0.5 & 0.46 & $1.29 \times 10^{11}$ & 125.46 & 32.76 \\
 0.8 & 0.73 & $2.65 \times 10^{10}$ & 109.81 & 24.64 \\
 1.0 & 0.89 & $1.10 \times 10^{10}$ & 107.61 & 20.36 \\
 1.2 & 1.14 & $5.62 \times 10^{9}$ & 124.58 & 14.71 \\
 1.6 & 1.48 & $2.25 \times 10^{9}$ & 6.80 & 3.94 \\
 2.0 & 1.61 & $1.16 \times 10^{9}$ & 6.75 & 3.92 \\
 3.2 & 2.05 & $3.18 \times 10^{8}$ & 6.69 & 3.89 \\
 5.0 & 2.64 & $1.04 \times 10^{8}$ & 6.68 & 3.88 \\
 7.0 & 3.20 & $4.89 \times 10^{7}$ & 6.74 & 3.91 \\
10.0 & 3.94 & $2.43 \times 10^{7}$ & 6.89 & 3.98 \\
\hline
\end{tabular}
\caption{
The fractional separations at which the synchronization and 
circularization timescales are equal to 1/4 of the main-sequence lifetime 
for equal-mass binaries starting on the ZAMS. Only the contribution from one 
star has been taken into account.
Column~2 gives the ZAMS radius of the star. 
}
\label{t:syncir} 
\end{center}
\end{table}

Comparing the circularization timescales for a binary consisting of a 
$0.5 \Msun$ MS star and a $0.8 \Msun$ WD gives 
\begin{eqnarray*}
{\tau}_{\rm circ,c} & \approx & 60 \left( a / \Rsun \right)^8 \: {\rm yr} \\ 
{\tau}_{\rm circ,d} & \approx & 10^{19} \left( a / \Rsun \right)^6 \: {\rm yr}  
\end{eqnarray*}
and thus the degenerate damping is only dominant for WD-WD and WD-NS 
systems in which the separation can become very small. 

Following Zahn (1977) the limiting separations for 
synchronization and circularization can be defined as the initial separations 
$(a/R)_{\rm sync}$ and $(a/R)_{\rm circ}$ for which the characteristic 
timescales are equal. 
As an example consider equal-mass MS binaries with 
both ${\tau}_{\rm sync}$ and ${\tau}_{\rm circ}$ set equal to one-quarter 
of the MS lifetime. The results are given in Table~\ref{t:syncir}. 
The tidal damping is clearly more efficient for stars with convective 
envelopes and only becomes more so if the binary includes a 
giant-like star. 

We note that Zahn (1977) obtains larger values of $(a/R)_{\rm circ}$ for 
radiative stars than those given in Table~\ref{t:syncir}. 
This seems to arise  from using MS lifetimes up to six times greater  
than those found here.  
However, using his $(a/R)_{\rm circ}$ with 
${\tau}_{\rm sync} = {\tau}_{\rm circ}$ gives larger $(a/R)_{\rm sync}$ 
values than he records with the end result that, while the values 
of $(a/R)_{\rm circ}$ calculated here are lower than his, the values of 
$(a/R)_{\rm sync}$ are much the same.

\subsection{Gravitational Radiation and Magnetic Braking}
\label{s:grvbr}

Coalescing neutron stars are a source of gravitational waves and may 
produce $\gamma$-ray bursts (Hartmann 1996). 
Such merging requires angular momentum loss from the system. 
This can be achieved by gravitational radiation from 
close binary systems driving the system to 
a mass transfer state, possibly followed by coalescence. 
It is also necessary for the formation of AM~CVn systems 
(Gonz\'{a}lez P\'{e}rez 1999) and adequate for cataclysmic variables (CVs) 
below the period gap (Faulkner 1971). 
Eggleton (private communication) uses the weak-field 
approximation of general relativity to derive the orbital changes due to 
gravitational radiation from two point masses as 
\beq\label{e:jdotgr}
\frac{{\dot{J}}_{\rm gr}}{J_{\rm orb}} = - 8.315 \times 10^{-10} 
\frac{M_1 M_2 M_{\rm b}}{a^4} \frac{1 + \frac{7}{8} e^2}
{\left( 1 - e^2 \right)^{5/2}} 
 \: {\rm yr}^{-1} 
\eeq
\beq\label{e:edotgr}
\frac{\dot{e}}{e} = - 8.315 \times 10^{-10} 
\frac{M_1 M_2 M_{\rm b}}{a^4} \frac{\frac{19}{6} + \frac{121}{96} e^2}
{\left( 1 - e^2 \right)^{5/2}} \: {\rm yr}^{-1} \, .
\eeq
These formulae do not take into account any contribution from distortions 
in spherical symmetry of the stars. 
Kuznetsov et al.\ (1998) show that for a binary consisting of two neutron 
stars this contribution only becomes noticeable when the separation 
decreases to less than seven times the size of the neutron star. 
Neglecting it then causes less than a 10\% error in the small time remaining 
to coalescence.  

Gravitational radiation is only efficient for CVs  
with orbital periods less than 3 hours and so cannot explain the mass  
transfer rates of such objects with periods up to 10 hours 
(Faulkner 1971; Zangrilli, Tout \& Bianchini 1997). 
However, it is possible for orbital angular momentum to be lost from the 
system via magnetic braking of the tidally coupled primary by its own 
magnetic wind. 
Rappaport, Verbunt \& Joss (1983) used mass-transfer rates deduced from 
observations of CVs, and the Skumanich (1972) braking law,
to parameterize the rate of angular momentum loss by magnetic braking.
We use  
\beq\label{e:mgbrke} 
{\dot{J}}_{\rm spin} = 
{\dot{J}}_{\rm mb} = - 5.83 \times 10^{-16} \frac{\Menv}{M} \left( 
R \Omega_{\rm spin} \right)^{3} \: \Msun {\rm R}^2_{\odot} {\rm yr}^{-2} \, , 
\eeq 
where all masses and the radius are in solar units and ${\Omega}_{\rm spin}$ 
in units of years, 
to alter the spin angular momentum of the component star and allow 
tides to alter $J_{\rm orb}$. 
This is effective for any star with an appreciable convective envelope. 
We note that the non-parametric formulation for magnetic braking 
given by St\c{e}pie\'{n} (1995), 
derived and calibrated from observations of spin-down of single stars, 
should be investigated in future versions of our binary model. 

The main-sequence primary in a CV has $R_1 \simeq 
R_{\rm L1} \simeq M_1$ so it is possible to derive the relation 
$P / {\rm hr} \approx 10.0 M_1 / \Msun$ for the period (Verbunt 1984). 
This with eqs.~(\ref{e:jdotgr}) and (\ref{e:mgbrke}) gives
\beq 
\frac{\left| {\dot{J}}_{\rm mb} \right|}{\left| {\dot{J}}_{\rm gr} \right|} 
= 9.815 \times 10^9 \frac{\left( M_{\rm b} M_1^2 \right)^{2/3}}{M_2^2} 
\eeq 
so that magnetic braking dominates even as $M_1 \rightarrow 0.0$. 
Observations of CVs reveal a gap between 2 and 
$3\,$hr in the otherwise smooth period-mass distribution. 
This period gap can be explained if magnetic braking becomes less 
effective when the orbital period falls below $3\,$hr, 
corresponding roughly to a main-sequence primary which has 
just become fully convective (Rappaport, Verbunt \& Joss 1983; 
Zangrilli, Tout \& Bianchini 1997). 
We therefore do not apply magnetic braking when the primary is a fully 
convective main-sequence star, $M_1 < 0.35$. 
Braking proceeds only by gravitational radiation for such systems.

\subsection{Supernovae Kicks}
\label{s:supkck}

When a star explodes as a supernova and becomes a neutron star or 
black hole it receives a velocity kick, due to any asymmetry in the 
explosion (Lyne \& Lorimer 1994). 
This may disrupt the binary. 
Evidence for such a kick 
includes observations of pulsars with velocities in excess of those of 
ordinary stars (Hansen \& Phinney 1997), double-NS binaries 
(Fryer \& Kalogera 1997) and X-ray binaries 
with large eccentricities (Kaspi et al.~1996). 
The state of the binary post-supernova depends on the orbital parameters  
at the moment of explosion and the kick velocity. 
A complete derivation is given in Appendix~\ref{s:app3b}. 

If the binary survives the supernova explosion then it is quite likely that 
the orbital parameters, $a$ and $e$,  
change markedly from those of the initial orbit. 
In addition, the mass lost by the primary star carries away an amount 
$\Delta J_{\rm orb} = \Delta M_1 a_1^2 {\Omega}_{\rm orb}$ of orbital 
angular momentum from the system. 
Before the primary core collapses it has spin angular momentum 
\beq 
J_{\rm spin} = k'_3 \Mc \Rc^2 {\Omega}_{\rm spin} 
\eeq 
which we conserve in the neutron star, spinning it up as it shrinks.

\subsection{Roche-Lobe Overflow}
\label{s:rlof}

Mass transfer occurs in close binary systems following the onset of Roche-lobe 
overflow.
This can be triggered either by a star expanding to fill its Roche-lobe as a 
result of stellar evolution or by angular momentum losses causing 
contraction of the orbit. 
The Roche-lobe radius of a star is fitted by Eggleton (1983) with  
\beq\label{e:rla}
\frac{R_{\rm L1}}{a} = \frac{0.49 q_1^{2/3}}{0.6 q_1^{2/3} + \ln 
\left( 1 + q_1^{1/3} \right)} \, , 
\eeq
in terms of the semi-major axis of the orbit and the mass-ratio 
$q_1 = M_1/M_2$ of the primary star, 
accurate to within 2\% for $0 < q_1 < \infty $. 
Note that eq.~(\ref{e:rla}) may also be used to obtain the Roche-lobe radius 
of the secondary by using the appropriate mass-ratio, i.e. $q_2$. 
The theory of RLOF is based on two stars in a circular orbit in which complete 
co-rotation has been achieved. 
In most cases this is adequate because tidal friction generally acts to 
remove any eccentricity on a timescale shorter than the evolution timescale 
of the binary. 
However it may be possble for RLOF to occur in an eccentric orbit if the 
binary is formed by tidal capture so that the primary, the more evolved star, 
has not spent all its life as part of the present system. 
It is also possible that stars may form in a close eccentic binary but it is 
generally expected that some initial circularization occurs as part of  
the formation process. 
Eccentric RLOF could also be envisaged if a star in an eccentric 
orbit is rapidly expanding, such as on the AGB, so that the nuclear and 
circularization timescales are similar. 
If any of these rare cases arise then, for want of a more detailed treatment, 
we subject them to instant synchronization at the onset of RLOF. 

When $R_1 > R_{\rm L1}$ mass is transferred to the companion star with the 
primary losing an amount of mass $\Delta M_{1 {\rm\SSS R}}$ and the secondary 
gaining some fraction $\Delta M_{2 {\rm\SSS R}}$ of this mass. 
All normal processes of binary evolution, as described in the preceding 
sections of this paper, are also treated during RLOF. 
This means that mass variations due to stellar winds and the related changes 
to $J_{\rm orb}$ and $J_{\rm spin}$ are still accounted for as are changes 
resulting from tidal friction. 
In addition, if a degenerate secondary star is not altered by the tidal 
evolution treatment then its spin is changed according to 
\beq 
\Delta J_{{\rm spin} 2} = k'_3 \Delta M_{2 {\rm\SSS R}} R_2^2 \, 
{\Omega}_{\rm K2} 
\eeq  
where we assume that the accreted material comes from the inner edge of 
an accretion disk which has 
\beq 
{\Omega}_{{\rm K} 2} = \sqrt{\frac{G M_2}{R_2^3}} \, . 
\eeq 
Care is taken to conserve the total angular momentum of the system. 
Independent of the transfer the tides should maintain co-rotation of the 
primary and the orbit during the Roche phase. 
We assume that mass changes due to RLOF do not affect the orbital 
angular momentum if the mass transfer is conservative. 
Any material not accepted by the secondary is lost from the system 
taking with it specific angular momentum from the primary so that 
$a \left( M_1 + M_2 \right) =$ constant. 
During RLOF $R_{\rm L1}$ is used as the effective radius of the primary when 
calculating ${\dot{M}}_{\rm 1W}$ and any associated angular momentum 
changes, as well as in the tidal expressions. 

The treatment of RLOF presented here is a substantially revised version of that 
described by Tout et al.\ (1997) including essential changes required 
to make it compatible with the updated stellar evolution treatment and the 
explicit treatment of angular momentum. 
Following Tout et al.\ (1997) we describe the stability of mass transfer 
using the radius-mass exponents $\zeta$ defined by Webbink (1985).  

\subsubsection{Dynamical Mass Transfer}
\label{s:dynmt} 

If ${\zeta}_{\rm ad} < {\zeta}_{\rm L}$ the radius of the primary increases 
faster than the Roche-lobe on conservative mass transfer. 
The mass loss rate from the primary is limited only by the sonic expansion 
rate of its envelope as mass is transferred through the inner Lagrangian point 
connecting the Roche-lobes of the two stars.  
Stars with deep surface convection zones, and degenerate stars, are 
unstable to such dynamical timescale mass-loss unless the primary is rather 
less massive than the secondary so that ${\zeta}_{\rm L}$ is more negative 
than ${\zeta}_{\rm ad}$. 
Thus dynamical mass transfer occurs for giants ($k_1 \in \left\{3,5,6,8,9
\right\}$), low-mass MS stars ($k_1 = 0$) and WDs 
($k_1 \in \left\{10,11,12\right\}$) when $q_1 > q_{\rm crit}$. 
This critical mass-ratio is defined by ${\zeta}_{\rm ad} = {\zeta}_{\rm L}$ 
where ${\zeta}_{\rm L} \approx 2.13 q_1 - 1.67$ (Tout et al. 1997) for 
conservative mass transfer. 
For normal giants $R \propto M^{-x}$, where $x \simeq 0.3$ varies with 
composition (as given by eq.~(47) of PapI), so that ${\zeta}_{\rm ad} = - x$ 
if ${\zeta}_{\rm ad} \approx {\zeta}_{\rm eq}$. 
However the behaviour of the radius deviates from this relation when the 
mass of the giant envelope is small. 
To detailed stellar models we fit 
\beq 
{\zeta}_{\rm ad} \approx {\zeta}_{\rm eq} \approx 
- x + 2 \left( \frac{\Mc}{M} \right)^5 
\eeq 
and use 
\beq\label{e:qcrit1} 
q_{\rm crit} = \left( 1.67 - x + 2 \left( \frac{\Mcon}{M_1} \right)^5 \right) 
/ 2.13 \,  
\eeq 
for the critical mass-ratio above which mass transfer proceeds on a 
dynamical timescale. 
We note that a modified criterion is required in cases of non-conservative 
mass transfer and that the assumption of 
${\zeta}_{\rm ad} \approx {\zeta}_{\rm eq}$ is not always true 
(Hjellming 1989). 
Naked helium giants have ${\zeta}_{\rm ad} \approx 0$ so that 
$q_{\rm crit} = 0.784$. 
We treat all cases with $q_1 > q_{\rm crit}$ in which the primary star is a 
giant as common-envelope evolution, described in Section~\ref{s:comenv}. 
For low-mass main-sequence stars $q_{\rm crit} = 0.695$ and for white dwarfs 
$q_{\rm crit} = 0.628$ (see Tout et al. 1997). 
We describe treatment of these cases in Sections~\ref{s:dynmtlm} 
and \ref{s:thmmtd}. 

Based on models of condensed polytropes (Hjellming \& Webbink 1987) an
alternative condition for dynamical mass transfer from a giant primary,
\begin{eqnarray*}
q_{\rm crit} = 0.362 + \left[ 3 \left( 1 - \Mcon/M_1 \right) \right]^{-1} \, ,
\end{eqnarray*}
valid for $\Mcon/M_1 \ga 0.2$, is given by Webbink (1988).
This relation is similar to eq.~(\ref{e:qcrit1}) for $\Mcon/M_1 = 0$
but quickly diverges as $\Mcon/M_1$ increases: 
it is a factor of two larger at $\Mcon/M_1 = 0.6$.
For values of $q_1$ intermediate between the two conditions Tout et al.~(1997)
allow mass transfer to proceed on a thermal timescale so that common-envelope
evolution is avoided.
In the presence of mass loss, and particularly any enhanced mass loss,
this would lead to a significant increase in the number of stable mass
transfer systems.
The formation paths of many binary populations, such as CVs (which require a
common-envelope phase) and symbiotic stars 
(which need to avoid common-envelope), would therefore be affected if 
we were to follow the same scheme. 

\subsubsection{Nuclear Mass Transfer}
\label{s:nucmt} 

If ${\zeta}_{\rm L} < \left( {\zeta}_{\rm ad} , {\zeta}_{\rm eq} \right)$ 
then mass transfer is stable until nuclear evolution causes further 
expansion of the star, that is to say that the mass transfer is not 
self-stimulating. 
The radius of the primary remains constrained to that of its Roche-lobe 
and the star remains in thermal equilibrium. 
Mass transfer proceeds at a rate such that 
$R_1 \approx R_{{\rm L} 1}$ so that the primary just overfills its Roche-lobe. 
We achieve this by transferring mass at a rate that steeply increases with 
the amount by which the Roche-lobe is overfilled, 
\beq\label{e:mtnuc} 
{\dot{M}}_{1 {\rm R}} = F \left( M_1 \right) \left[ \ln \left( R_1 / 
R_{{\rm L} 1} \right) \right]^3 \: \Msun \: {\rm yr}^{-1} \, ,  
\eeq
where 
\beq\label{e:fmstdy}
F \left( M_1 \right) = 3 \times 10^{-6} \left[ \min \left( M_1 , 5.0 
\right) \right]^2 
\eeq
is chosen by experiment to ensure that the mass transfer is steady. 
If the primary is a degenerate star ($k_1 \geq 10$) then it is necessary to 
increase eq.~(\ref{e:fmstdy}) by the factor $10^3/ \max \left( R_1 , 
10^{-4} \right)$. 
In general, nuclear mass transfer occurs on a much longer timescale than  
either dynamical or thermal mass transfer so that systems such as Algols 
and CVs can be observed in this state.

\subsubsection{Thermal Mass Transfer}
\label{s:thmmt} 

If ${\zeta}_{\rm eq} < {\zeta}_{\rm L} < {\zeta}_{\rm ad}$ mass transfer 
is unstable on a thermal timescale. 
The primary does not remain in thermal equilibrium as it loses mass:  
it contracts and remains just filling its Roche-lobe. 
This is an awkward case to treat because the thermal timescale is long 
compared with the orbital period but short compared with the timescale for 
nuclear evolution. 
First we calculate the mass transfer rate at the full equilibrium rate using 
eq.~(\ref{e:mtnuc}). 
This gives an upper limit. 
For giants and giant-like stars ($k_1 \in \left\{ 2,3,4,5,6,8,9 \right\}$) 
we limit this to the thermal rate which we approximate by 
\beq
{\dot{M}}_{\rm\SSS KH} = \frac{M_1}{{\tau}_{\rm\SSS KH1}} 
\eeq
with 
\beq
\frac{{\tau}_{\rm\SSS KH}}{\rm yr} = \frac{10^7 M}{R L} 
\left\{ \begin{array} {l@{\quad}l} 
M & \: k \in \left\{ 0,1,7,10,11,12,13,14 \right\} \\ 
\left( M - M_{\rm c} \right) & \: k \in 
\left\{ 2,3,4,5,6,8,9 \right\}  \end{array} \right. 
\eeq 
where ${\tau}_{\rm\SSS KH}$ is the Kelvin-Helmholtz timescale. 
This still allows the possibility of $R_1 > R_{{\rm L} 1}$ by quite a large 
fraction. 
The most common case of mass transfer at the thermal rate has a 
Hertzsprung gap primary star ($k_1 = 2$) because ${\zeta}_{\rm ad}$ is 
quite large while ${\zeta}_{\rm eq}$ is close to its giant branch value. 
For these stars ${\zeta}_{\rm L}$ can still exceed ${\zeta}_{\rm ad}$ when 
$q_1$ is large. 
We assume ${\zeta}_{\rm ad} \approx 6.85$ (Tout et al.~1997) so that if  
$q_1$ exceeds $q_{\rm crit} = 4$ common-envelope evolution ensues 
as a result of dynamical mass transfer. 
We note that this assumption is rather approximate and can be improved in 
future versions of the algorithm by calibrating to detailed binary 
calculations. 
For all non-giant-like stars we allow mass transfer to proceed at the rate 
given by eq.~(\ref{e:mtnuc}) but limited by the dynamical rate 
\beq
{\dot{M}}_{\rm\SSS DYN} = \frac{M_1}{{\tau}_{\rm\SSS DYN1}}
\eeq
where
\beq\label{e:tdynam}
{\tau}_{\rm\SSS DYN} = 5.05 \times 10^{-5} \sqrt{\frac{R^3}{M}} 
\: {\rm yr} .
\eeq
In this dynamically limited case, if $R_1 > 10 R_{{\rm L} 1}$ as a result of 
significant orbital contraction, the stars are allowed to merge according 
to the appropriate collision prescription (see Section~\ref{s:merger}).

\subsubsection{Dynamical Mass Transfer from Low-Mass Main-Sequence Stars}
\label{s:dynmtlm} 

Low-mass MS stars ($k_1 = 0$) are deeply convective so that mass 
transfer to a companion proceeds dynamically if $q_1 > 0.695$. 
If this is the case we assume that the entire star overflows its 
Roche-lobe on a timescale ${\tau}_{\dot{\SSS M}} = {\tau}_{\rm\SSS DYN1}$ 
and only a single star remains. 

If the secondary star is fairly unevolved ($k_2 \in \left\{ 0,1,2 \right\}$) 
its thermal timescale will be relatively long and it can only 
accrete a fraction 
\beq\label{e:thlim}
\Delta M_{\rm\SSS 2R} = \min \left( 10 \frac{{\tau}_{\dot{\SSS M}}}
{{\tau}_{\rm\SSS KH2}} , 1 \right) \Delta M_{\rm\SSS 1R} \, ,  
\eeq
of the mass, where $\Delta M_{\rm\SSS 1R} = M_1$. 
If the secondary is still on the MS it must be rejuvenated 
(see Section~\ref{s:secresp}). 
When the secondary is a giant or a CHeB star 
($k_2 \in \left\{ 3,4,5,6 \right\}$) its envelope can easily absorb all of 
the primary mass on a dynamical timescale. 
If the secondary is a naked He star or a WD ($k_2 \in \left\{ 
7,8,9,10,11,12 \right\}$) then all the material is accreted dynamically and 
swells up to form a giant envelope around the degenerate star which 
becomes the core of a new giant star. 
This is effectively a reverse of the process which formed the degenerate 
remnant star in the first place when a giant or CHeB star lost its envelope.  
The new star requires an age appropriate to its core, the calculation 
of which is described in Section~\ref{s:comcol} along with a detailed 
description of the outcome.

\subsubsection{Mass Transfer from Degenerate Objects}
\label{s:thmmtd} 

If a WD ($k_1 \in \left\{ 10,11,12 \right\}$) evolves to fill its Roche-lobe 
then the secondary must be a more massive WD or possibly a NS or BH  
($k_2 \in \left\{ 10,11,12,13,14 \right\}$). 
Mass transfer proceeds at a steady rate according to eq.~(\ref{e:mtnuc}), 
unless $q_1 > 0.628$ when the mass transfer becomes dynamical. 
In this case the entire mass is transferred on a timescale 
${\tau}_{\rm\SSS DYN1}$, as given by eq.~(\ref{e:tdynam}), leading to 
coalescence of the two stars. 

When dynamical mass transfer leads to the coalescence of two HeWDs we  
assume that the temperature produced is enough to ignite the triple-$\alpha$ 
reaction. 
If all the material in the new core were rapidly converted to CO then the
nuclear energy released would be greater than the binding energy of the core
so that no remnant is left ($k_2 = 15$).
Here we follow this procedure but it is possible that a naked helium star is 
produced (Webbink 1984).
This alternative could easily be included. 
Dynamical mass transfer of helium on to a CO or ONeWD causes the accreted 
material to swell up and form an envelope around the CO or ONe core 
(Iben, Tutukov \& Yungelson 1996) so that a HeGB star is formed ($k_2 = 9$).
If an ONeWD accretes CO or ONe material this is simply added to the WD 
($k_2 = 12$), unless the new mass exceeds the Chandrasekhar mass, $\MCh$, 
in which case electron capture on $^{24}$Mg nuclei leads to an 
accretion-induced collapse (AIC; Nomoto \& Kondo 1991; van Paradijs et al. 
1997) and the formation of a NS ($k_2 = 13$). 

Dynamical mass transfer from a COWD to another COWD quickly leads to 
the formation of a thick accretion disk (as in all the above cases of 
dynamical transfer) around the more massive WD followed by coalescence 
over a viscous timescale. 
When two COWDs coalesce we form a larger COWD 
($k_2 = 11$) of mass $M_2 + \Delta M_{2 {\rm\SSS R}}$. 
If the mass exceeds $\MCh$ then it explodes as a possible type Ia SN 
(Branch 1998) leaving no remnant ($k_2 = 15$).
However, whether a thermonuclear explosion actually occurs is not clear. 
The temperature produced at the core-disk boundary depends on the 
accretion rate which in turn depends on the viscosity of the disk, all of 
which is uncertain. 
If the temperature is hot enough to ignite carbon and oxygen,  
whether the WD is converted to an ONeWD depends on 
competition between the rate of propagation of the flame inwards, which 
depends on the opacity, and the cooling rate of the WD. 
Saio \& Nomoto (1998) used spherically symmetric evolution models of 
accreting WDs 
to show that carbon burning is ignited at the core-disk boundary and that 
the flame propagates all the way to the centre, by heat conduction, 
converting the material to ONeMg without causing an explosion. 
If this is indeed the case then CO-CO WD merger products with a total mass 
greater than $\MCh$ would become AIC neutron stars rather than 
type Ia SNe. 

When a NS or BH ($k_1 \in \left\{ 13,14 \right\}$) fills its 
Roche-lobe the secondary must be a NS or BH and we allow the two stars  
to merge and form a single remnant of the combined mass. 
Alternatively coalescing NSs may destroy themselves in a $\gamma$-ray burst 
(Hartmann 1996).

\subsubsection{The Response of the Secondary}
\label{s:secresp} 

If the secondary is a MS, HG or CHeB star ($k_2 \in \left\{ 0,1,2,4 \right\}$)  
we transfer an amount of mass given by eq.~(\ref{e:thlim}),  
limited by the thermal timescale of the secondary  
\beq\label{e:taumth}
{\tau}_{\dot{\SSS M}} = \frac{M_2}{{\dot{M}}_{\SSS 1R}} \, < \, 
{\tau}_{\rm\SSS KH2} \, . 
\eeq
A giant secondary can respond to mass transfer by 
dynamically shrinking and no limitation need be applied. 
When a HeMS star accretes mass from a primary of the same type we transfer  
an amount of mass given by eq.~(\ref{e:thlim}). 
The HeMS star must then be rejuvenated. 
However, if mass is transferred to a naked helium star from a primary of any 
hydrogen-rich type, we put all the mass in an envelope around 
the helium core and form a new CHeB or AGB star. 
Details of how the new star is made are given in Section~\ref{s:comcol}. 

At any time the secondary may respond to mass transfer by 
filling its own Roche-lobe so that a contact binary is formed. 
If this occurs we allow the two stars to coalesce 
(see Section~\ref{s:merger}). 
Generally common-envelope evolution arises before contact unless 
the binary formed as a very close system. 

\vspace{0.2cm}
\noindent{\it Rejuvenation} 
\vspace{0.2cm}

We extend the rejuvanation described by Tout et al.~(1997) for MS stars 
to include HG and HeMS stars. 
When such a primary transfers mass it must be aged so that the 
fractional age of the star, in terms of its current evolution phase, 
remains unchanged. 
The process by which this is done is the same as that described in 
Section~7.1 of PapI. 
If any of these stars are accepting mass then they rejuvenate.  
This we do simply by reversing the ageing process for MS stars 
with radiative cores ($0.35 \leq M \leq 1.25$) and for HG secondaries 
(see Section~7.1 of PapI). 
For MS stars with convective cores ($M > 1.25$), and fully 
convective stars ($M < 0.35$), the core grows and mixes in unburnt fuel as 
the star gains mass, so that the star appears even younger. 
The rejuvenation of MS stars is detailed in Tout et al.~(1997). 
The core of a HeMS star is convective with
$\Mc \propto M$ so these secondaries are rejuvenated in the same way.

\vspace{0.2cm}
\noindent{\it Steady Accretion on to Degenerate Objects} 
\vspace{0.2cm}

For transfer of mass to degenerate objects we follow Tout et al.~(1997) 
but the greater variety of stars now makes the procedure more complex. 
Accretion of H-rich material leads to novae if 
${\dot{M}} < 1.03 \times 10^{-7} \Msun \: {\rm yr}^{-1}$, 
and supersoft X-ray sources if $1.03 \times 10^{-7} \leq  
{\dot{M}} < 2.71 \times 10^{-7} \Msun \: {\rm yr}^{-1}$,  
in the same way and for accretion rates 
${\dot{M}} \geq 2.71 \times 10^{-7} \Msun \: {\rm yr}^{-1}$ 
HeWDs become GB stars while CO and ONeWDs become TPAGB stars 
(see Section~\ref{s:comcol} for details).  
For nova systems we assume that all but a fraction $\epsilon$ of the accreted 
material is ejected in the nova explosion, i.e. 
\beq
\Delta M_{\rm\SSS 2R} = \epsilon \Delta M_{\rm\SSS 1R} \, ,  
\eeq
where $\epsilon$ is an input parameter to the model and may be negative. 
Typically we take $\epsilon = 0.001$. 

How the WD secondary responds to steady accretion of He-rich material,  
either directly ($ 7 \leq k_1 \leq 10 $) or as a consequence of steady 
hydrogen burning on the surface of the WD, depends on its composition.  
Woosley, Taam \& Weaver (1986) calculated thermonuclear explosions 
induced by the accretion of helium on to a HeWD. 
Their models showed that accretion at $2 \times 10^{-8} \Msun \: 
{\rm yr}^{-1}$ on to a $0.4 \Msun$ HeWD leads to a violent, centrally 
ignited detonation when the mass of the star reaches $0.66 \Msun$.  
Nomoto \& Sugimoto (1977) had previously found detonations at $0.78 \Msun$.  
Here a HeWD is allowed to accrete material until its mass reaches 
$0.7 \Msun$ when it is destroyed ($k_2 = 15$). 

A COWD accreting helium can accumulate about $0.15 \Msun$ of 
helium-rich material, provided the total mass remains below $\MCh$, before 
helium ignites at the base of the accreted layer. 
This is an edge-lit detonation (ELD). 
The detonation front propagates outward through the helium layer while an 
inward propagating pressure wave compresses the CO core which ignites 
off-centre, followed by an outward detonation which destroys the WD  
(Woosley \& Weaver 1994; Livne \& Arnett 1995). 
Kawai, Saio \& Nomoto (1987), by considering steady state spherically 
accreting models, argue that if the accretion rate exceeds 
$3 \times 10^{-8} \Msun \: {\rm yr}^{-1}$ helium can burn steadily and 
non-degenerately. 
It is not clear that such steady burning is possible if accretion is 
from a disk leaving most of the WD photosphere free to radiate. 
Therefore, once a COWD has accumulated $0.15 \Msun$ of accreted helium, 
we set off an ELD which destroys the WD in a type Ia SN. 
Livne \& Arnett (1995) argue that the properties of type Ia SNe are diverse 
enough to suggest that the progenitors have a range of parameters, such as 
mass. 
The more luminous events might then correspond to more massive progenitors. 
They find that detonations in sub-Chandrasekhar mass WDs are a 
promising explanation for most, if not all, type Ia SNe. 
Such SNe are not standard candles (cf. Section~\ref{s:thmmtd}): 
they exhibit considerable diversity in behaviour. 

In the case of steady transfer of He-rich material on to an ONeWD, or steady 
transfer of C-rich material ($k_1 = 11,12$) on to a CO or ONeWD we add the 
material to the degenerate core of the secondary. 
If the new mass of a white dwarf secondary exceeds the Chandrasekhar 
limit, $\MCh$, then we annihilate it in a type Ia SN, unless it 
is an ONeWD ($k_2 = 12$) which undergoes AIC leaving a NS remnant.   
When such a SN explodes as a result of 
steady mass transfer the primary survives, having 
transferred only just enough material for the secondary mass to reach $\MCh$.  

The amount of mass that a WD, NS or BH can accrete may be 
limited by the Eddington limit (Cameron \& Mock 1967), 
\beq
{\dot{M}}_{\rm Edd} = 2.08 \times 10^{-3} \left( 1 + X \right)^{-1} R_2 
\quad \Msun \: {\rm yr}^{-1} \, , 
\eeq 
where $X$ is the hydrogen mass fraction, so that 
\beq\label{e:meddton} 
\Delta M_{\rm\SSS 2R} = \min \left( {\dot{M}}_{\rm Edd} 
{\tau}_{\dot{\SSS M}} , \Delta M_{\rm\SSS 1R} \right) \, .
\eeq
There is some uncertainty as to whether the Eddington limit should actually 
be applied because the energy generated in excess of the 
limit might be removed from the system in a strong wind or 
asymmetrically through a disk. 
Super-Eddington accretion rates may be important in the formation of 
low-mass X-ray binaries and millisecond pulsars (Webbink \& Kalogera 1997) 
and in models of X-ray emission from quasars due to accretion from a disk 
on to a rapidly rotating BH (Beloborodov 1998). 
Imposing the Eddington limit significantly reduces the formation rate of 
Type Ia SNe (Livio 2000). 
For these reasons the Eddington limit, eq.~(\ref{e:meddton}),  
is only optionally included. 
A NS may gain enough mass to exceed the maximum NS mass of $1.8 \Msun$  
(see Section~6.2.2 of PapI) in which case it becomes a BH.

\subsection{Common-envelopes, Coalescence and Collisions}
\label{s:comcol}

As a result of Roche-lobe overflow it is possible for the binary components 
to come into contact and coalesce or for the binary to reach a 
common-envelope (CE) state. 
The most frequent case of common-envelope evolution involves a giant 
transferring mass to a main-sequence star on a dynamical timescale. 
Although the process is difficult to model, and therefore uncertain, it is 
envisaged that the secondary is not able to accept the overflowing 
material owing to its relatively long thermal timescale. 
The giant envelope overfills the Roche-lobes of both stars so that the 
giant core and the MS star are contained within a common-envelope. 
Owing to its expansion the envelope rotates slower than 
the orbit of the core and the MS star so that friction causes 
them to spiral together and transfer energy to the envelope. 
This process may release sufficient energy to drive off the entire envelope,  
leaving a close binary containing a WD and a MS star, 
or it may lead to coalescence of the giant core and the MS star. 
Evidence for a mechanism such as common-envelope evolution is provided by CVs 
and close double-degenerate binaries whose characteristics can only be  
understood if a significant amount of angular momentum and mass have been 
removed from the precursor system. 
Planetary nebulae containing a close binary at their centre are probably 
the result of a recent common-envelope event (Bond, Liller \& Mannery 1978).  

In contrast to the relatively gentle nature of coalescence it is possible 
that two stars in a close eccentric orbit could collide at 
periastron before either star has a chance to fill its Roche-lobe. 
Furthermore, in a dense environment such as the core of a star cluster, 
two individual stars may be involved in a hyperbolic collision.
Generally this involves formation of a binary via tidal capture 
and then common-envelope evolution or coalescence, owing to the relatively low 
velocity dispersion in star clusters. 
We assume that any collision involving a giant or giant-like star that has a 
dense core and an appreciable envelope leads to common-envelope evolution. 
All other collisions simply result in a merger of the two stars. 
The outcome of a collision depends on the impact parameter,  
which in turn depends on the relative velocity and the relative 
sizes of the stars, as well as their structure. 
Detailed modelling of specific collision cases has been undertaken 
(Rasio \& Shapiro 1991; Benz \& Hills 1992; Bailey \& Davies 1999), 
but in the absence of results covering the full parameter space we employ 
the simple treatment given in the following sections. 

\subsubsection{Common-Envelope Evolution}
\label{s:comenv}

Common-envelope evolution occurs either as a result of a collision between a 
star with a dense core ($k_1 \in \left\{ 2,3,4,5,6,8,9 \right\}$) and any  
other star, or at the onset of Roche-lobe overflow where mass is transferred 
from a giant ($k_1 \in \left\{ 2,3,5,6,8,9 \right\}$) 
on a dynamical timescale and $q_1 > q_{\rm crit}$.  
The primary is therefore a giant or giant-like star with core mass $\Mcon$ 
and core radius $\Rcon$, and an envelope mass $\Menvon = M_1 - \Mcon$. 
To make the calculations simpler we define effective values, marked with a  
prime ($'$), for the secondary. 
If the secondary is a MS star ($k_2 \in \left\{ 0,1,7 \right\}$) 
then it has an effective core mass $\Mctwd = M_2$, and an effective core radius 
$\Rctwd = R_2$, but $\Mctw = 0.0$. 
For a giant or giant-like secondary ($k_2 \in \left\{ 2,3,4,5,6,8,9 \right\}$) 
the effective values are the actual values, i.e. $\Mctwd = \Mctw$ and 
$\Rctwd = \Rctw$, and for degenerate secondaries 
($k_2 \geq 10$) where $\Mctw = M_2$ and $\Rctw = R_2$. 
The effective envelope mass is given by $\Menvtwd = M_2 - \Mctwd$. 

Our treatment of CE evolution follows closely the description 
of Tout et al.\ (1997) where the outcome depends on the initial 
binding energy of the envelope and the initial orbital energy of the two 
cores. 
First we calculate the total binding energy of the envelope,   
\beq\label{e:ebindi}
E_{{\rm bind},i} = - \frac{G}{\lambda} \left[ \frac{M_1 \Menvon}{R_1} + 
\frac{M_2 \Menvtwd}{R_2} \right] \, , 
\eeq
with $\lambda = 0.5$. 
Note that we use the indices $i$ and $f$ to represent initial and 
final quantities respectively, in relation to the CE event. 
The initial orbital energy of the cores is 
\beq
E_{{\rm orb},i} = - \frac{1}{2} \frac{G \Mcon \Mctwd}{a_i} \, , 
\eeq
where $a_i$ is the semi-major axis at the onset of the CE phase. 
We assume that the cores spiral-in, transferring orbital 
energy to the envelope with an efficency ${\alpha}_{\rm\SSS CE}$, 
which is necessarily a free parameter due to uncertainty in its value. 
It is probably not a constant (Reg\H{o}s \& Tout 1995) 
but generally ${\alpha}_{\rm\SSS CE} \approx 1$ is used. 
The final separation $a_f$, if sufficient energy were released to drive 
away the entire envelope, can be calculated via 
\beq\label{e:ebnd2i}
E_{{\rm bind},i} = {\alpha}_{\rm\SSS CE} \left( 
 E_{{\rm orb},f} - E_{{\rm orb},i} \right)\, ,  
\eeq
where 
\beq
E_{{\rm orb},f} = - \frac{1}{2} \frac{G \Mcon \Mctwd}{a_f} \, . 
\eeq
Eq.~(\ref{e:rla}), in conjunction with $a_f$, gives the Roche-lobe of each 
core: $R_{{\rm L}1,f}$ uses $q_1 = \Mcon/\Mctwd$ and $R_{{\rm L}2,f}$ uses   
$q_2 = \Mctwd/\Mcon$. 
If neither core fills its Roche-lobe then a binary composed of the cores with 
separation $a_f$ survives and the entire envelope escapes the 
system. 
The envelope has been removed from one, or both, of the stars  
leaving behind the appropriate remnant (see Section~6 of PapI). 
We assume the envelope ejection is isotropic and both cores emerge 
from the CE in co-rotation with the orbit. 
We note that HG stars ($k \in \left\{ 2,8 \right\}$) have an evolving 
density profile that is not yet as steep as in giants. 
Our assumption that HG stars have a core mass growing linearly in time  
(see PapI) mimics this to some extent, although the actual remnant mass in a 
detailed computation may still be somewhat different. 

If Roche-lobe overflow would have occurred for either core then they could not 
have spiralled in so far, instead coalescing at an earlier time. 
We assume this takes place when $\Rctwd = R_{{\rm L}2}$, or 
$\Rcon = R_{{\rm L}1}$ depending on which core fills its Roche-lobe first, 
so that the corresponding $a = a_{\SSS L}$ can be found from eq.~(\ref{e:rla}) 
and then the binding energy left in the remaining envelope can be calculated 
according to 
\beq
E_{{\rm bind},f} - E_{{\rm bind},i} = {\alpha}_{\rm\SSS CE} \left( \frac{1}{2} 
\frac{G \Mcon \Mctwd}{a_{\SSS L}} + E_{{\rm orb},i} \right) \, .
\eeq
This results in formation of a new giant or giant-like star with 
\begin{eqnarray*}
k_3 & \in & \left\{ 2,3,4,6,8,9 \right\} \\ 
M_3 & = & M_f \\ 
M_{{\rm c} 3} & = & \Mcon + \Mctw \, , 
\end{eqnarray*}
unless $k_2 = 7$ and $k_1 \in \left\{ 2,3,4,5,6 \right\}$, in which case 
$M_{{\rm c} 3} = \Mcon + M_2$ (see later).  
The resultant mass $M_f$ must be determined. 
The star has an envelope binding energy 
\beq
E_{{\rm bind},f} = - \frac{G M_f M_{{\rm env},f}}{\lambda R_f} 
= - \frac{G M_f \left[ M_f - M_{{\rm c} 3} \right]}
{\lambda R_f} 
\eeq
and $R_f$ can be estimated by noting that on a dynamical timescale the 
radius of a giant responds to changes in mass according to 
$R \propto M^{-x}$ (where $x$ is given by eq.~(47) of PapI). 
The radius $R_i$ of the star, if the system were to coalesce immediately,  
satisfies 
\beq
E_{{\rm bind},i} = - \frac{G \left( M_1 + M_2 \right) \left( M_1 + M_2 - 
M_{{\rm c} 3} \right)}{\lambda R_i} 
\eeq
and as CE evolution is over on a dynamical timescale, 
\beq
\frac{R_f}{R_i} = \left[ \frac{M_1 + M_2}{M_f} \right]^x \, .
\eeq
Therefore 
\beq
\frac{E_{{\rm bind},f}}{E_{{\rm bind},i}} = \left( \frac{M_f}{M_1 + M_2} 
\right)^{1 + x} \frac{M_f - M_{{\rm c} 3}}{M_1 + M_2 - M_{{\rm c} 3}} 
\eeq
which we solve for $M_f$ by a Newton-Raphson iteration. 
We assume the new star rotates with the same period as the orbit 
immediately prior to coalesence. 

A different treatment is described by Iben \& Livio (1993).  
Their initial envelope binding energy is not that of the giant (or giants) 
but that of the initial configuration of the CE itself,  
which surrounds the cores with a diameter of about $2 a_i$. 
This amounts to replacing eq.~(\ref{e:ebindi}) with 
\beq
E_{{\rm bind},i} = - G \frac{\left( M_1 + M_2 \right) \left( \Menvon + 
\Menvtwd \right)}{a_i} 
\eeq
(see Yungelson et al. 1994) 
giving a reduction in the envelope binding energy and therefore less 
likelihood that the process ends in coalescence of the cores. 
A similar effect can be mimicked in our treatment by 
increasing ${\alpha}_{\rm\SSS CE}$ to values greater than unity.  
At first sight this may seem unphysical given the definition, but as 
discussed by Iben \& Livio (1993), an increase in ${\alpha}_{\rm\SSS CE}$ 
can be envisaged 
if additional energy sources other than the orbital energy are involved. 
Processes with potential to supply such energy include enhanced nuclear burning 
in shell burning zones of giants, nuclear burning on the surface of a 
degenerate secondary, dynamo generation of magnetic fields and recombination 
of the hydrogen and helium ionization zones in giants. 
Possibly the CE absorbs ordinary nuclear energy in the process 
of swelling up to a diameter of $2 a_i$, but this should occur on a thermal 
timescale. 
Unfortunately, the theoretical determination of reliable values for 
${\alpha}_{\rm\SSS CE}$ 
have proven difficult due to a lack of understanding of the 
processes involved, and our ability to model them.

\subsubsection{Coalescence of CE Cores}
\label{s:comcor}

When two cores spiral into each other the outcome depends on their  
relative density.  
If they are of similar {\it compactness}, we  
assume that they coalesce and mix completely. 
If one core is considerably more compact than the other, that core 
sinks to the centre without mixing while the less dense core mixes with the 
envelope. 
All cores, including remnant stars, are considered to be compact unless the 
core is a MS star. 
The case of an EAGB star ($k = 5$) is particularly interesting because it 
has a carbon core growing within a stationary helium core. 
Detailed models show that the mass inside the hydrogen shell is dense 
enough for the helium core to be used as the core for collision purposes. 

In our model the stellar type $k_3$ of the new star produced as a result 
of coalescence 
is given by the intersection of the $k_1$th column and the $k_2$th row of 
the collision matrix in Table~\ref{t:colmtx}. 
The mass of the new star is $M_3 = M_f$ and the core mass is $M_{{\rm c}3} = 
\Mcon + \Mctw$. 
If, by chance, there is no remaining envelope the coalescence product 
is the remnant remaining if a star of type $k_3$ loses its envelope. 
Note that $k_3 \in \left\{ 13 , 14 \right\}$ signifies the creation of an   
unstable Thorne-\.{Z}ytkow object (Thorne \& \.{Z}ytkow 1977)  
when the merger involves a NS or BH.  
For simplicity our Thorne-\.{Z}ytkow objects rapidly eject their envelopes  
leaving only the NS or BH (Fryer, Benz \& Herant 1996) so that $M_3 = M_2$.  

\begin{table*}
\begin{center}
\begin{tabular}{|cr|rrrrrrrrrrrrrrr|}
\hline
 & & \multicolumn{15}{|c|}{\rule [-3mm]{0mm}{8mm} Primary Star \, $k_1$} \\
 & & \, 0 & \, 1 & \, 2 & \, 3 & \, 4 & \, 5 & \, 6 & \, 7 &
\, 8 & \, 9 & 10 & 11 & 12 & 13 & 14 \\ [0.5ex]
\hline
 & 0 & 1 & 1 & \bldr{2} & \bldr{3} & \bldr{4} & \bldr{5} & \bldr{6} & 4 &
\bldr{6} & \bldr{6} & 3 & 6 & 6 & 13 & 14 \\ [0.5ex]
 & 1 & 1 & 1 & \bldr{2} & \bldr{3} & \bldr{4} & \bldr{5} & \bldr{6} & 4 &
\bldr{6} & \bldr{6} & 3 & 6 & 6 & 13 & 14 \\ [0.5ex]
 & 2 & \bldr{2} & \bldr{2} & \bldr{3} & \bldr{3} & \bldr{4} & \bldr{4} & 
\bldr{5} & \bldr{4} & \bldr{4} & \bldr{4} & \bldr{3} & 
\bldr{5} & \bldr{5} & \bldr{13} & \bldr{14}
\\ [0.5ex]
 & 3 & \bldr{3} & \bldr{3} & \bldr{3} & \bldr{3} & \bldr{4} & \bldr{4} & 
\bldr{5} & \bldr{4} & \bldr{4} & \bldr{4} & \bldr{3} & 
\bldr{5} & \bldr{5} & \bldr{13} & \bldr{14}
\\ [0.5ex]
 & 4 & \bldr{4} & \bldr{4} & \bldr{4} & \bldr{4} & \bldr{4} & \bldr{4} & 
\bldr{4} & \bldr{4} & \bldr{4} & \bldr{4} & \bldr{4} & 
\bldr{4} & \bldr{4} & \bldr{13} & \bldr{14}
\\ [0.5ex]
 & 5 & \bldr{5} & \bldr{5} & \bldr{4} & \bldr{4} & \bldr{4} & \bldr{4} & 
\bldr{4} & \bldr{4} & \bldr{4} & \bldr{4} & \bldr{4} & 
\bldr{4} & \bldr{4} & \bldr{13} & \bldr{14}
\\ [0.5ex]
 & 6 & \bldr{6} & \bldr{6} & \bldr{5} & \bldr{5} & \bldr{4} & \bldr{4} & 
\bldr{6} & \bldr{4} & \bldr{6} & \bldr{6} & \bldr{5} & 
\bldr{6} & \bldr{6} & \bldr{13} & \bldr{14}
\\ [0.5ex]
 & 7 & 4 & 4 & \bldr{4} & \bldr{4} & \bldr{4} & \bldr{4} & \bldr{4} & 7 &
\bldr{8} & \bldr{9} & 7 & 9 & 9 & 13 & 14 \\ [0.5ex]
 & 8 & \bldr{6} & \bldr{6} & \bldr{4} & \bldr{4} & \bldr{4} & \bldr{4} & 
\bldr{6} & \bldr{8} & \bldr{8} & \bldr{9} & \bldr{7} & 
\bldr{9} & \bldr{9} & \bldr{13} & \bldr{14}
\\ [0.5ex]
 & 9 & \bldr{6} & \bldr{6} & \bldr{4} & \bldr{4} & \bldr{4} & \bldr{4} & 
\bldr{6} & \bldr{9} & \bldr{9} & \bldr{9} & \bldr{7} & 
\bldr{9} & \bldr{9} & \bldr{13} & \bldr{14}
\\ [0.5ex]
 & 10 & 3 & 3 & \bldr{3} & \bldr{3} & \bldr{4} & \bldr{4} & \bldr{5} & 7 &
\bldr{7} & \bldr{7} & 15 & 9 & 9 & 13 & 14 \\ [0.5ex]
 & 11 & 6 & 6 & \bldr{5} & \bldr{5} & \bldr{4} & \bldr{4} & \bldr{6} & 9 &
\bldr{9} & \bldr{9} & 9 & 11 & 12 & 13 & 14 \\ [0.5ex]
 & 12 & 6 & 6 & \bldr{5} & \bldr{5} & \bldr{4} & \bldr{4} & \bldr{6} & 9 &
\bldr{9} & \bldr{9} & 9 & 12 & 12 & 13 & 14 \\ [0.5ex]
 & 13 & 13 & 13 & \bldr{13} & \bldr{13} & \bldr{13} & \bldr{13} & \bldr{13} & 
13 & \bldr{13} & \bldr{13} & 13 & 13 & 13 & 13 & 14 \\ [0.5ex] 
\, \, \turnbox{90}{\qquad \qquad \qquad Secondary Star \, $k_2$} &
14 & 14 & 14 & \bldr{14} & \bldr{14} & \bldr{14} & 
\bldr{14} & \bldr{14} & 14 & \bldr{14} & \bldr{14} & 14 & 14 & 14 & 14 & 14 \\ 
\hline
\end{tabular}
\caption{The stellar type of the new star $k_3$ produced when two stars merge 
as a result of a collision (normal type) or common-envelope evolution 
(bold italic type).}
\label{t:colmtx}
\end{center}
\end{table*}

If $k_1 \in \{ 2,3,4,5,6 \}$ and $k_2 \in \{ 0,1 \}$ then $k_3 = k_1$, the MS 
star is simply absorbed into the giant envelope. 
The initial mass and age of the giant are unaffected by the merger because the  
core mass of the star is not changed. 
Similarly if $k_1 \in \{ 8,9 \}$ and $k_2 = 7$ because the CO 
core of the giant sinks to the centre becoming the core of the new star. 
If $k_1 \in \{ 8,9 \}$ and $k_2 \in \{ 0,1 \}$ then $k_3 = 6$ because the 
MS star mixes with the envelope of the helium giant to form 
a hydrogen-helium envelope around the CO core of the giant. 
An initial mass, $M_{0,3}$, and age, $t_3$, appropriate for the core mass 
and stellar type of the new star must be set. 
We describe our procedure in Section~\ref{s:newstr}.  
This must be done in all merger cases where the new star is of a 
different type to either star involved in the merger, or if the core 
mass of the star has changed as a result of the merger. 

If $k_1 \in \{ 2,3 \}$ and $k_2 \in \{ 2,3,10 \}$ then $k_3 = 3$ because the 
two helium cores merge to form a larger helium core within the existing 
envelope (or mixed envelopes). 
If both merging cores are degenerate ($M_0 < \MHeF$ for the giants) then
enough heat is produced in the merger to ignite the triple-$\alpha$
reaction and the nuclear energy released is greater than the binding energy
of the new core, so $k_3 = 15$.

If $k_1 \in \{ 2,3 \}$ and $k_2 \in \{ 6,11,12 \}$ then $k_3 = 5$ because the 
helium core surrounds the CO (or ONe) core within the 
envelope (or mixed envelopes) to form an EAGB star. 
The same occurs if $k_1 = 6$ and $k_2 = 10$. 

If $k_1 = 6$ and $k_2 \in \{ 6,8,9 \}$ then the two cores mix to form a larger 
CO core, surrounded by the mixed envelopes, so that $k_3 = 6$. 
This is also the result when $k_1 = 6$ and $k_2 \in \{ 11,12 \}$ although 
only one envelope is involved in this case. 
The binding energy of the new envelope must be reduced by
the amount of energy liberated in merging the two degenerate cores.
This removes the envelope in all cases so that in actual fact 
$k_3 \in \{ 11,12 \}$.

If $k_1 \in \{ 8,9 \}$ and $k_2 \in \{ 8,9,11,12 \}$ then $k_3 = 9$ because 
the CO (and possibly neon) cores mix to form a larger core surrounded 
by a helium envelope. 
When $k_1 \in \{ 8,9 \}$ and $k_2 = 10$ we assume that the two stars 
mix to form an evolved HeMS star, $k_3 = 7$, with $M_{0,3} = M_3$ and 
\beq 
t_3 = \left( \frac{ \Mcon }{ \Mcon + M_2 } \right) t_{{\rm\SSS HeMS} 3} \, . 
\eeq 

If $k_1 \in \{ 2,3,4,5,6 \}$ and $k_2 = 7$, $k_3 = 4$ because the giant 
core material mixes with the HeMS star to form a new core that 
contains some unburnt helium. 
All the remaining cases that arise as a result of common-envelope 
evolution, i.e. $k_1 \in \{ 2,3 \}$ and $k_2 \in \{ 4,5,8,9 \}$ as well as 
$k_1 \in \{ 4,5 \}$ and $k_2 \in \{ 4,5,6,8,9,10,11,12 \}$, also have 
$k_3 = 4$ because the merged core is a mixture of processed and 
unprocessed helium. 
The age of the new CHeB star depends on how advanced, in terms of 
central helium burning, the two progenitor stars were.  
How this is taken into account is the subject of Section~\ref{s:newstr}.

\subsubsection{Collision Outcomes}
\label{s:merger}

It now remains to describe the result of all collisions that do not proceed 
via common-envelope evolution. 
We allow the stars to merge without mass-loss, $M_3 = M_1 + M_2$ 
(cf. Bailey 1999). 
The stellar type of the merged product is determined by the nature of the 
colliding stars as given by the collision matrix (Table~\ref{t:colmtx}). 

As in Tout et al.~(1997) the 
collision of two MS stars, $k_1 \in \{ 0,1 \}$ and $k_2  \in \{ 0,1 \}$, 
leads to $k_3 \in \{ 0,1 \}$, with $M_{0,3} = M_3$ and  
$k_3 \in \{ 0,1 \}$, with $M_{0,3} = M_3$ and 
\beq 
t_3 = 0.1 \frac{t_{{\rm MS} 3}}{M_3} \left[ \frac{M_1 t_1}{t_{{\rm MS} 1}} 
 + \frac{M_2 t_2}{t_{{\rm MS} 2}} \right] \, , 
\eeq 
where $t_1$ and $t_2$ are the ages of the colliding stars. 
When two HeMS stars collide we apply the same except that 
$k_1 = k_2 = k_3 = 7$ and the 
$t_{{\rm\SSS MS} i}$ are replaced by $t_{{\rm\SSS HeMS} i}$.  
If $k_1 \in \{ 0,1 \}$ and $k_2 = 7$ then $k_3 = 4$ because the naked helium 
star sinks to the centre forming a CHeB star with $M_{{\rm c} 3} = M_2$.  
A HeWD, $k_1 = 10$, colliding with a MS star, $k_2  \in \{ 0,1 \}$, 
sinks to the centre of the MS star where hydrogen shell burning can begin at 
its surface. 
The result is a GB star, $k_3 = 3$, with core mass $M_{{\rm c} 3} = M_1$. 
Similarly if $k_1 \in \{ 11,12 \}$ and $k_2  \in \{ 0,1 \}$, double shell 
burning begins at the WD surface so that $k_3 = 6$. 
If $k_1 = 10$ and $k_2 = 7$ the naked helium star absorbs the HeWD 
leaving a rejuvenated HeMS star, $k_3 = 7$, with $M_{0,3} = M_3$ and 
\beq 
t_3 = \frac{t_{{\rm\SSS HeMS} 3}}{M_3} \frac{M_2 t_2}
{t_{{\rm\SSS HeMS} 2}} \, . 
\eeq 
However if $k_1 \in \{ 11,12 \}$ and $k_2 = 7$ then $k_3 = 9$ because the WD 
forms the core of an evolved helium giant with $M_{{\rm c} 3} = M_1$. 

When two HeWDs, $k_1 = 10$ and $k_2 = 10$, collide we assume  
that the temperature produced is hot enough to ignite the 
triple-$\alpha$ reaction and that the subsequent nuclear runaway destroys the 
system so that $k_3 =15$ (cf. Section~\ref{s:thmmtd}). 
If $k_1 = 10$ and $k_2 \in \{ 11,12 \}$ then $k_3 = 9$, with 
$M_{{\rm c} 3} = M_2$, because the helium material swells up to form an 
envelope around the CO (or ONe) core, which in 
general is smaller and denser by virtue of a larger mass. 
If $k_1 = 11$ and $k_2 = 11$ then a larger COWD is formed, 
$k_3 = 11$, unless $M_3 \geq \MCh$ in which case the new star is destroyed in 
a type Ia SN and $k_3 = 15$.  
Similarly, if $k_1 = 12$ and $k_2 \in \{ 11,12 \}$ then $k_3 = 12$ unless 
$M_3 \geq \MCh$ causing the ONeWD to undergo an AIC to a NS, $k_3 = 13$ 
(see Section~\ref{s:thmmtd}).  

If $k_1 \in \{ 13 , 14 \}$ and $k_2 \in \{ 0,1,7 \}$ an 
unstable Thorne-\.{Z}ytkow object results, i.e. $k_3 = k_1$ and $M_3 = M_1$.   
Finally, if $k_1 \in \{ 13 , 14 \}$ and $k_2 \in \{ 10,11,12,13,14 \}$ then 
$k_3 = k_1$, unless the new mass of the NS exceeds the maximum 
allowable NS mass of $1.8 \Msun$ in which case we collapse it to a BH.

\subsubsection{The New Star}
\label{s:newstr}

When a new giant or core helium burning star is made ($k_3 \in \left\{ 
3,4,5,6,8,9 \right\}$) then an age, $t_3$, and an initial mass, $M_{0,3}$ must 
be assigned to the star appropriate to its type $k_3$, mass $M_3$
and core mass $M_{{\rm c} 3}$. 

If $k_3 = 3$ then our goal is to find $M_{0,3}$ so as to place the star at 
the base of the GB with its correct core mass. 
The decision as to where the star should start its evolution is quite 
arbitrary, with the base of the GB (BGB) chosen because it simplifies the 
process.  
As discussed in Section~5.2 of PapI the helium core of a GB star is 
degenerate if $M < \MHeF$ (where $\MHeF \simeq 2.0$ is the maximum initial 
mass for which He ignites degenerately in a helium flash) 
and non-degenerate otherwise.  
So the first step is to place $\MHeF$ into eq.~(44) of PapI to find the 
maximum degenerate BGB core mass. 
If $M_{{\rm c} 3}$ is greater than this value then $M_{0,3}$ can be 
fixed by setting $\McBGB = M_{{\rm c} 3}$ and inverting eq.~(44) of PapI. 
Otherwise it is necessary to find the luminosity corresponding to 
$M_{{\rm c} 3}$ using the GB \Mc-$L$ relation given by eq.~(37) of PapI. 
We use a Newton-Raphson procedure to find $M_{0,3}$ such that the 
luminosity at the BGB (given by eq.~(10) of PapI) equals that luminosity, 
taking $M_{0,3} = \MHeF$ as an initial guess. 
Once $M_{0,3}$ has been found we set $t_3$ to the corresponding time taken 
for such a star to reach the BGB as given by eq.~(4) of PapI. 
We note that stars with initial mass $M_0 > \MFGB$ (see Section~5 
of PapI) do not have a GB which means that a GB star cannot be made 
with a core mass greater than the value given by inserting $\MFGB$ 
into eq.~(44) of PapI. 
Such a situation is extremely unlikely, but if 
it does occur then we form an CHeB star. 

We make a CHeB star, $k_3 = 4$, whenever the new core is composed of a 
mixture of burnt and unburnt helium and the surrounding envelope contains 
mostly hydrogen. 
The new core is assigned a fractional age which we determine according to the  
amount of central helium burning that has already occurred in the colliding 
stars. 
Each of these stars is assigned a fraction $y$ where 
\begin{eqnarray*}
y = \left\{ \begin{array}{l@{\qquad}l} 
\DS 0 & k \in \{ 0,1,2,3,10 \} \\ [2ex]
\DS 1 & k \in \{ 6,8,9,11,12 \} \\ [2ex]
\DS \frac{t - t_{\rm\SSS HeI}}{t_{\rm\SSS DU} - 
t_{\rm\SSS HeI}} & 
k \in \{ 4,5 \} \\ [2ex]
\DS \frac{t}{t_{\rm\SSS HeMS}} & k = 7 
\end{array} \right.
\end{eqnarray*}
and then 
\beq 
y_3 = \frac{y_1 \Mcon + y_2 \Mctw}{M_{{\rm c} 3}} \, , 
\eeq 
where $M_{\rm c}$ must be replaced by $M$ for $k = 7$. 
The age of the new star is then 
\beq 
t_3 = t_{{\rm\SSS HeI} 3} + y_3 t_{{\rm\SSS He} 3} 
\eeq 
where $t_{{\rm\SSS HeI} 3}$ and $t_{{\rm\SSS He} 3}$ are dependent 
on $M_{0,3}$. 
The time of central helium ignition, $t_{\rm\SSS HeI}$, the lifetime 
of the CHeB phase, $t_{\rm\SSS He}$, 
the time taken to reach the second dredge-up phase at the start of the TPAGB, 
$t_{\rm\SSS DU}$, and the MS lifetime of a naked helium star, 
$t_{\rm\SSS HeMS}$, are given by eqns.~(43), (57), (70) and (79) of PapI. 
There is a minimum allowed core mass for CHeB stars which corresponds to 
an initial mass of $\MHeF$. 
To find the minimum possible initial mass relevant to $M_{{\rm c} 3}$, 
corresponding to $y_3 = 1$, it is first necessary to find the base of the 
asymptotic giant branch (BAGB) core mass for $M_{0,3} = \MHeF$, 
using eq.~(66) of PapI. 
If $M_{{\rm c} 3}$ is larger than this value we find $M_{\rm min}$ by 
setting $M_{{\rm c} 3} = \McBAGB$ and inverting eq.~(66) of PapI, 
otherwise $M_{\rm min} = \MHeF$. 
We find the maximum possible initial mass corresponding to $y_3 = 0$ 
by setting $M_{{\rm c} 3} = \McHeI$ and solving for $M_{\rm max}$. 
A bisection method can be used to find $M_{0,3}$, where 
$M_{\rm min} \leq M_{0,3} \leq M_{\rm max}$, and  
\beq 
M_{{\rm c} 3} = \McHeI \left( M_{0,3} \right) + y_3 \left( \McBAGB \left( 
M_{0,3} \right) - \McHeI \left( M_{0,3} \right) \right) \, . 
\eeq 

If $k_3 = 5$ we assume that $M_{{\rm c} 3}$ is representative of the helium 
core mass, $\Mche$, of the new EAGB star and find the initial mass 
by setting $M_{{\rm c} 3} = \McBAGB$ and solving eq.~(66) of PapI for 
$M_{0,3}$. 
The age of the star is the time taken to reach the BAGB, 
\beq 
t_3 = t_{{\rm\SSS HeI} 3} + t_{{\rm\SSS He} 3} \, . 
\eeq 

For a new TPAGB star, $k_3 = 6$, $M_{{\rm c} 3}$ is the CO core mass 
and our aim is to put the star at the start of the TPAGB phase so we set 
$M_{{\rm c} 3} = \McDU$.  
To do this it is necessary to find the corresponding helium core mass,  
$\Mche$, just before the transition from the EAGB to the TPAGB. 
Stars with $0.8 < \Mche < 2.25$ undergo second dredge-up during the transition 
so that 
\beq\label{e:mcdurp} 
\McDU = 0.44 \Mche + 0.448 
\eeq
while lower mass stars have 
\beq 
\McDU = \Mche \, .  
\eeq 
Stars with $\Mche \geq 2.25$ on the EAGB would become a NS or BH  
hole before the TPAGB was reached. 
Thus if $M_{{\rm c} 3} \geq 1.438$, given by $\Mche = 2.25$ in 
eq.~(\ref{e:mcdurp}), the star cannot exist with $k_3 = 6$ and we set it up 
so that it immediately becomes a NS or BH by setting $M_{{\rm c} 3} = 
\McSN$ (see eq.~(75) of PapI) in eq.~(\ref{e:mcdurp}).  
Therefore 
\begin{eqnarray*}
\Mche = \left\{ \begin{array}{l@{\qquad}l} 
\DS \left( M_{{\rm c} 3} + 0.35 \right) / 0.773 & M_{{\rm c} 3} \geq 1.438 \\ 
[2ex] 
\DS \left( M_{{\rm c}3} - 0.448 \right) / 0.44 & 0.8 < M_{{\rm c}3} < 1.438 \\ 
[2ex] 
\DS M_{{\rm c} 3} & M_{{\rm c} 3} \leq 0.8 
\end{array} \right.
\end{eqnarray*}
and $\Mche = \McBAGB$ so that $M_{0,3}$ can be found by solving 
eq.~(66) of PapI. 
The age of the star is $t_3 = t_{{\rm\SSS DU} 3}$ by eq.~(70) of PapI. 
At this stage it is possible that $M_{0,3} < 0$ if $M_{{\rm c} 3}$ is less 
than the minimum allowed core mass for the start of the TPAGB. 
This could happen in the rare event of an extremely low mass COWD gaining 
only a very small envelope to make it a TPAGB star. 
If this occurs then we simply add the small amount of mass 
to the WD so that $k_3 = 11$ and $M_{0,3} = M_3$. 

If $k_3 \in \{ 8,9 \}$ we set $M_{0,3}$ and $t_3$ so that the new star 
begins life at the end of the naked helium star MS. 
$M_{0,3}$ must be found by a bisection method because some of 
the parameters involved depend on the chosen value of $M_{0,3}$. 
At each iteration we find the luminosity at the end of the main-sequence, 
$L_{\rm THe}$, by putting $\tau = 1$ in eq.~(80) of PapI, and the HeGB 
luminosity corresponding to $M_{{\rm c} 3}$, calculated with eq.~(84) of PapI, 
for the current guess of $M_{0,3}$. 
We iterate until the two values agree to within a difference of $10^{-4}$. 
%We set the minimum mass for the iteration to $M_{\rm min} = M_{{\rm c} 3}$ 
%and the maximum to $M_{\rm max} = 2 M_3$ so that the corresponding 
%core mass at the base of the HeGB is greater than $M_{{\rm c} 3}$. 
Once $M_{0,3}$ is determined we set the age, $t_3 = t_{{\rm HeMS} 3}$ 
using eq.~(79) of PapI.

\subsection{The Evolution Algorithm}
\label{s:intgrt}

The initial state of a binary system is described by the masses, 
$M_1$ and $M_2$, of the component stars, and the period, $P$, and 
eccentricity, $e$, of the orbit. 
A choice must also be made for the metallicity, $Z$, of the stars. 
In general the system begins with both stars on the ZAMS but it is 
possible to start the binary star evolution (BSE) algorithm with the stars 
in an evolved state. 
The stars are each assumed to have an initial spin frequency on the ZAMS 
independent of the properties of the orbit (given by eq.~(108) of PapI),  
however options exist to begin the stars in co-rotation with the orbit 
or with any given value. 
A number of input parameters and options exist for the BSE code, many of which 
have been introduced in the preceeding sections. 
These are summarized in Table~\ref{t:params} along with their ranges and 
default values. 

\begin{table}
\begin{center}
\begin{minipage}{8cm}
\begin{tabular}{|lccl|}
\hline
 Parameter & Range & Default & Section \\
\hline \hline
${\beta}_{\rm\SSS W}$ & ($0.125 \rightarrow 7.0$) & 0.5 & 
\ref{s:windac} \\
${\alpha}_{\rm\SSS W}$ & ($0.0 \rightarrow 2.0$) & 1.5 & 
\ref{s:windac} \\
${\mu}_{\rm\SSS W}$ & $0.0 \rightarrow 1.0$ & 1.0 & \ref{s:windac} \\
$B_{\rm\SSS W}$ & ($0.0 \rightarrow 10^6$) & 0.0 & \ref{s:windac} \\
$\epsilon$ & $-1.0 \rightarrow 1.0$ & 0.001 & \ref{s:secresp} \\ 
${\alpha}_{\rm\SSS CE}$ & ($0.5 \rightarrow 10.0$) & 3.0 & 
\ref{s:comenv} \\
${\sigma}_{k} \,{\rm km}\,{\rm s}^{-1}$ & $0.0 \rightarrow \infty$\footnote{
A value of zero corresponds to no velocity kick.} & 
$190.0$ & \ref{s:app3b} \\ 
Tidal Evolution & on/off & on & \ref{s:tidal} \\ 
Eddington Limit & on/off & off & \ref{s:secresp} \\ \hline 
Stellar Winds & on/off & on & 7.1 of PapI \\ 
$\eta$ & ($0.0 \rightarrow 2.0$) & 0.5 & 7.1 of PapI \\ 
\hline
\end{tabular}
\end{minipage}
\caption{
Input parameters and options in the BSE algorithm set by the user. 
Column~2 gives the range of values that each parameter may assume 
(suggested range if bracketed) and default values are given in Column~3. 
The relevant section where each parameter or option is discussed is given 
in Column~4 (note that some parameters are introduced in PapI). 
}
\label{t:params} 
\end{center}
\end{table}

Each star has its own stellar evolution time-step, $\delta t_i$, (given by 
eq.~(112) of PapI) which includes restrictions that prevent the star changing 
its mass by more than 1\%, or its radius by more than 10\%.  
For binary evolution it is particularly important to keep the time-steps 
relatively small so that the radius of the primary does not increase 
too much over any one time-step. 
This aids identification of the 
time when the star first fills its Roche-lobe, if indeed it does. 
We also calculate an orbital time-step such that the angular momentum of the 
system changes by 2\%, 
\beq 
\delta t_{\rm b} = 0.02 \frac{J_{\rm orb}}{ | {\dot{J}}_{\rm orb} |} \, , 
\eeq 
and take the actual binary evolution time-step when the system is in a detached 
state to be 
\beq\label{e:btstep}
\delta t = \min \left( \delta t_1 , \delta t_2  , \delta t_{\rm b} \right) \, . 
\eeq
We evolve the system forward by this time, implementing the 
necessary tidal and braking mechanisms as well as evolving each star 
according to the SSE prescription given in PapI. 
If by virtue of the stars growing in size, or the orbital separation 
shrinking, the periastron separation is less 
than the combined radii of the two stars then the stars collide and the 
system is dealt with according to the prescription 
described in Section~\ref{s:comcol}. 

If the primary fills its Roche-lobe then we interpolate within 
the last time-step until the radius of the star only just exceeds its 
Roche-lobe radius ($1 \leq R_1/R_{{\rm L} 1} \leq 1.002$). 
The system enters RLOF and we treat it according to Section~\ref{s:rlof}.  
First we test for dynamical mass transfer and deal with it if 
necessary, by merging or common-envelope evolution, 
otherwise the mass transfer is steady. 
We set the time-step during RLOF to 
\beq 
\delta t_{\rm RL} = k_{\rm m} P 
\eeq 
where initially 
\beq 
k_{\rm m} = \delta t \frac{10^{-3}}{P} \, , 
\eeq 
where $\delta t$ is the previous time-step for the detached system. 
After implementing mass transfer, and any mass loss due to stellar winds, 
and adjusting the separation accordingly, as well as implementing the tidal 
and braking mechanisms, we check whether the primary still fills its 
Roche-lobe. 
If $R_1 < R_{{\rm L} 1}$ then the system leaves RLOF to once more evolve 
as a detached binary.  
If $R_1 \geq R_{{\rm L} 1}$ and $R_2 \geq R_{{\rm L} 2}$, a contact 
system has formed and we treat this as either a collision or 
common-envelope evolution depending on the nature of the stars involved. 
Otherwise we allow the time-step to grow by a factor of two, until 
$R_1/R_{{\rm L} 1}$ is such that $\delta M_1/M_1 = 0.005$, i.e. 
\beq 
k_{\rm m} = \min \left( 2 k_{\rm m} , \frac{0.005 M_1}{|\delta M_1|} 
\right) \, ,  
\eeq 
and the RLOF process is repeated. 

When common-envelope evolution leaves a surviving binary system we 
continue the evolution with the new orbital parameters. 
If at any point the stars merge we follow the evolution of the resulting 
single star.  
If either star explodes as a supernova and the resultant velocity 
kick disrupts the binary we evolve the system forward as two single stars 
that don't interact. 

The CPU time required by the BSE algorithm to evolve $500\,000$ binaries up to 
the age of the Galaxy is approximately $20\,$hr, or $0.144\,$s per 
binary, on a Sun SparcUltra10 workstation (containing a $300\,$MHz processor).

\section{Some Examples and Comparisons}
\label{s:exmple}

To illustrate the BSE algorithm we present some example evolutionary 
scenarios, concentrating in particular on Algol and cataclysmic 
variable evolution. 
These examples also serve to demonstrate the sensitivity of binary 
evolution to the choice of model parameters where, 
unless otherwise specified, the default values given in Table~\ref{t:params} 
are used.  
In addition, we take a selection of evolution scenarios previously 
outlined by other authors and we compare the results of their models 
with those of the BSE algorithm. 

\begin{figure}
\centerline{
\psfig{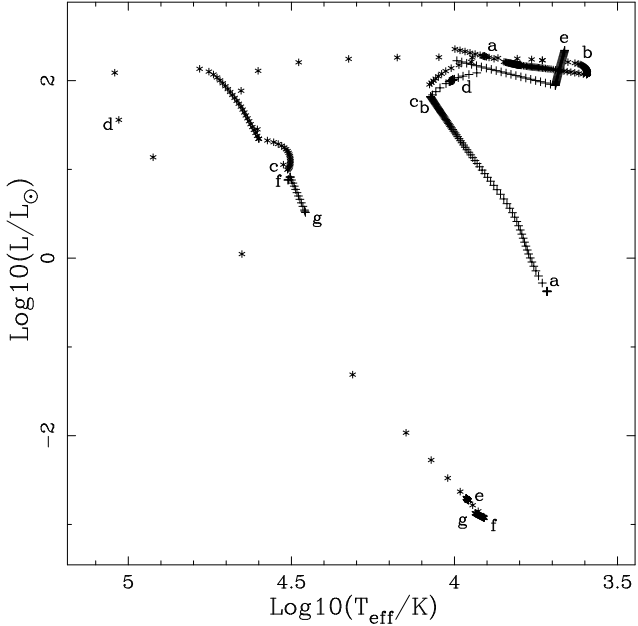}
}
\caption{
Hertzsprung-Russell diagram showing the evolution of a $2.9 \Msun$
star ($*$ points) and a $0.9 \Msun$ star (+ points) in a binary
system with an initial orbit of $P = 8$~d and $e = 0.70$.
The letters represent various times in the evolution of the binary system
(see text for details).
Each point represents an iteration of the BSE algorithm except in regions
of high density, i.e. rapid evolution. 
}
\label{f:alghrd}
\end{figure}

\begin{figure}
\centerline{
\psfig{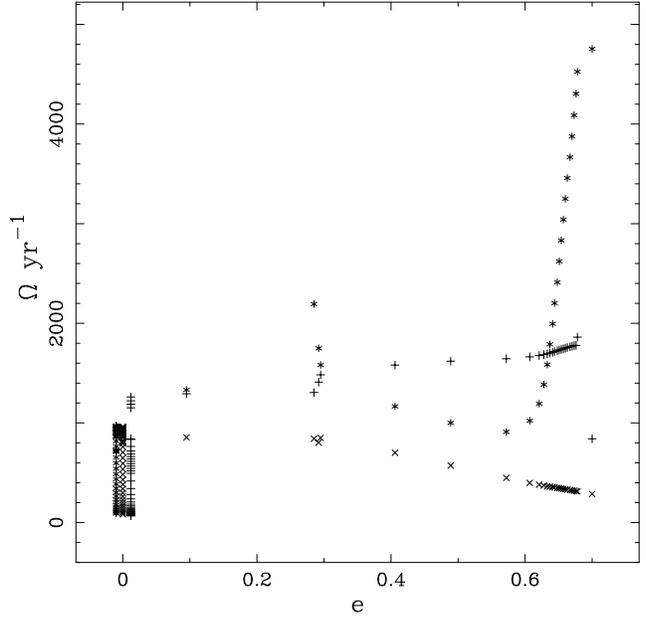}
}
\caption{
The evolution of the spins of the primary ($*$ points) and secondary 
(+ points) stars, and the orbital spin ($\times$ points), as
a function of the eccentricity of the orbit for the example of Algol evolution,
shown until the end of the first phase of mass transfer.
The points at zero eccentricity have been slightly displaced so that
the co-rotation of the stars with the orbit can be clearly seen.
}
\label{f:algspn}
\end{figure}

\subsection{Algol Evolution}

\begin{figure}
\centerline{
\psfig{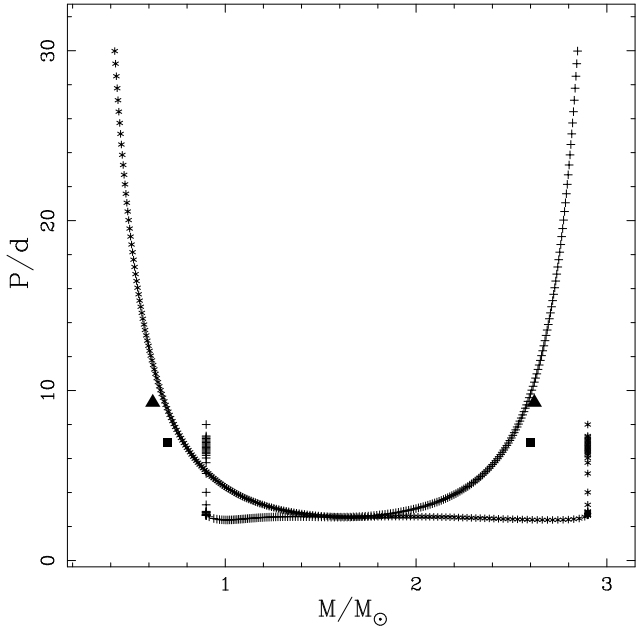}
}
\caption{
The evolution of stellar mass with binary period for the example
of Algol evolution, until the end of the first phase of mass transfer.
The primary mass decreases from $2.9 \Msun$ to $0.42 \Msun$ ($*$ points)
and the secondary mass increases from $0.9 \Msun$ to $2.85 \Msun$ (+ points)
while the period initially decreases owing to circularization of the orbit 
and then increases as a result of the nearly conservative mass transfer.
The observed parameters for the Algol systems RY Gem
(solid triangles) and TT Hya (solid squares) are also shown.
}
\label{f:algper}
\end{figure}

Consider a system in which the initial masses are 
$2.9 \Msun$ and $0.9 \Msun$ in an $8\,$d orbit with $e = 0.7$. 
In this example $Z = 0.02$ and ${\alpha}_{\rm\SSS CE} = 3$. 
Figure~\ref{f:alghrd} shows the evolution of the binary stars in the
Hertzsprung-Russell diagram.  
On the MS tidal friction arising from convective damping of the tide raised 
on the $0.9 \Msun$ secondary acts to circularize the system so that at a 
time $T = 413\,$Myr, when the $2.9 \Msun$ primary reaches the end of its 
MS lifetime, the eccentricity has fallen to 0.28 and $P$ to $3\,$d.
The radius of the primary then increases rapidly as it evolves across the HG  
until $R_1 = 6.2 \Rsun$ when it fills its Roche-lobe. 
By then $P = 2.7\,$d and the orbit has circularized (position~a in 
Figure~\ref{f:alghrd}). 
Mass transfer proceeds on a thermal timescale, with the secondary accreting 
80\% of the transferred mass (limited by its thermal timescale),  
until the primary reaches the 
end of the HG with $M_1 = 0.53 \Msun$, $M_2 = 2.74 \Msun$ and $P = 17\,$d. 
During this phase the equilibrium radius of the primary has exceeded its 
Roche-lobe by as much as a factor of 2. 
Mass transfer continues as the primary begins to ascend the GB, 
but now on a nuclear timescale which allows the secondary to accrete all 
the transferred mass, until the primary's envelope mass becomes so small that 
it shrinks inside its Roche-lobe (position~b). 
Figure~\ref{f:algspn} shows the intrinsic spin of the stars and the 
spin of the orbit as a function of the eccentricity, up to this point. 
We see clearly that co-rotation is achieved as the orbit 
circularizes. 
When RLOF ends $M_1 = 0.42 \Msun$, $M_2 = 2.85 \Msun$ and $P = 30\,$d. 
While the primary was transferring mass on the HG the mass-ratio 
of the system inverted, $q_1 < 1$, so that the more evolved component 
is now the least massive. 
This is the well known Algol Paradox (Hoyle 1955; Crawford 1955). 
Many semi-detached Algol systems have since been observed and their 
parameters determined (e.g. Popper 1980). 
These include RY Gem: $M_1 = 2.6 \Msun$, $M_2 = 0.6 \Msun$ and $P = 9.3\,$d, 
and TT Hya: $M_1 = 2.6 \Msun$, $M_2 = 0.7 \Msun$ and $P = 6.95\,$d; both of 
which could easily be explained by this example.  
Figure~\ref{f:algper} shows the evolution of the mass of the stars with the 
orbital period up to the end of the Algol phase of evolution. 

Shortly after the mass transfer phase has ended the GB primary loses all of 
its envelope leaving a $0.42 \Msun$ HeMS star with a $2.85 \Msun$ MS 
companion which would be a blue straggler in a star cluster (position~c).  
The naked helium star evolves to the HeGB and then to a $0.417 \Msun$ COWD 
after losing its helium envelope (position~d).  
The MS star, now the primary, evolves to the GB and fills its 
Roche-lobe (position~e) when the period has reduced to $18\,$d as a result 
of the transfer of angular momentum from the orbit to the spin of the star. 
As $q_1 = 6.83 > q_{\rm crit}$ the mass transfer is dynamical, a CE forms and 
leaves a binary containing the COWD and the $0.416 \Msun$ He core of the 
primary in an orbit of $0.04\,$d. 
The HeMS star evolves to fill its Roche-lobe (position~f) which has shrunk by  
gravitational radiation, and transfers $0.1 \Msun$ of He-rich material to the 
COWD before the system reaches a contact state at $P = 1\,$min (position~g).  
The two stars merge to a single HeGB star. 

\begin{figure}
\centerline{
\psfig{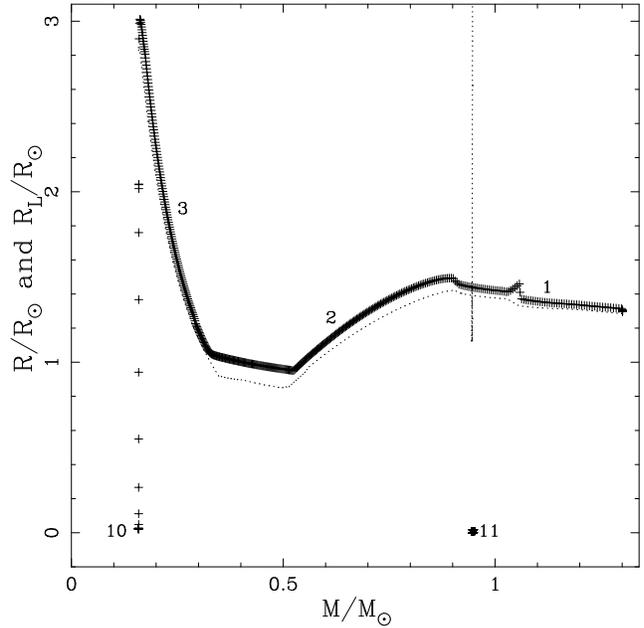}
}
\caption{
The evolution of stellar radius (+ points for the primary and $*$ points for 
the COWD secondary) with mass during a typical phase of CV evolution. 
The numbers indicate the evolution state of each star. 
Dotted lines are the Roche-lobe radii for each star. 
Note that the COWD accretes little material as its Roche-lobe radius shrinks 
and then expands.
}
\label{f:cvsrch}
\end{figure}

\subsection{A Cataclysmic Variable}

Next we use a binary with initially $M_1 = 6.0 \Msun$ and
$M_2 = 1.3 \Msun$, $P = 630\,$d, and a circular orbit, $e = 0.0$, 
to give a quick illustration of CV formation and evolution.
In this example $Z = 0.02$, ${\alpha}_{\rm\SSS CE} = 1$ and 
$\epsilon = 0.001$. 
The relatively large orbital period gives the $6.0 \Msun$ primary enough
room to evolve on to the EAGB before filling its Roche-lobe $78\,$Myr 
after the evolution began. 
At this point the tides raised on the convective envelope of the primary 
have moderately decreased the orbital period, offset slightly by 
orbital changes owing to $0.1 \Msun$ lost from the primary, 
none of which has been accreted by the MS companion. 
Because $q_1 = 4.5 > q_{\rm crit}$ mass transfer is dynamical and a
common-envelope forms.
The He core of the primary and the MS secondary spiral together until their
orbital period is reduced to $0.63\,$d at which point the envelope has been
driven off.

The primary is now a HeHG star of mass $1.5 \Msun$ and very soon
evolves to fill its Roche-lobe again, resulting in another CE system.
Once again the cores do not coalesce before the envelope is removed and a
binary consisting of a $0.94 \Msun$ COWD and a $1.3 \Msun$ MS star in a
circular orbit of $0.45\,$d emerges.
Gravitational radiation acts to reduce the orbital period until 
at $T = 303\,$Myr the MS primary fills its Roche-lobe and a CV is formed. 
Mass transfer is steady and proceeds on a nuclear timescale, driven in part
by gravitational radiation and, when the primary mass is reduced below
$1.25 \Msun$ and it develops a convective envelope, magnetic braking and
tidal friction.
As the primary evolves across the HG it transfers mass on a thermal
timescale, now driven by the evolution. 
Throughout RLOF the mass transfer rate remains below $10^{-7} \Msun
\, {\rm yr}^{-1}$, low enough that nova explosions at the surface of the
white dwarf have allowed only 0.1\% of the transferred material to be
accreted.
RLOF ends with the primary on the GB just before it loses the remainder 
of its envelope becoming a low-mass HeWD of $M_1 = 0.158 \Msun$.
The orbital period is now $4.5\,$d and a gravitational radiation
timescale of $10^{14}\,$yr means that the double-degenerate system cannot
coalesce within the age of the Galaxy.
Figure~\ref{f:cvsrch} shows the radius of each star and the corresponding
Roche-lobe radius, as a function of the stellar mass,
during the CV phase of evolution.

\subsection{Sensitivity to Model Parameters} 

To give an idea of how sensitive the final state of the system is to changes 
in the initial binary parameters and the physical parameters that govern the 
evolution, we first reconsider the Algol example described above. 
If the WD-HeMS system emerged from the CE with a 
separation increased by only a few percent then the COWD would 
accrete $0.15 \Msun$ of He-rich material and explode as an ELD SNIa 
before the system evolved into contact. 
Such an increase in separation could easily be achieved by a less than 
10\% change to ${\alpha}_{\rm\SSS CE}$, well within its uncertainty, 
or small changes in the initial period and/or eccentricity of the system. 

There is no doubt that major changes to the outcome would result if we 
neglected to model tidal evolution for this example but to do so for a 
system that is initially eccentric with a short period would not be 
physically correct. 
However, for the purpose of comparison we can examine the evolution 
of a system that begins with the same component masses and 
semi-latus rectum but in a circular orbit, i.e. $P = 2.9\,$d. 
Apart from the initial circularization phase, this example evolved 
with the tidal evolution activated follows a virtually identical path 
to that of the above example (a result we will discuss further in 
Section~\ref{s:rectum}). 
If tidal evolution is not activated, and therefore any spin-orbit 
coupling is neglected, the orbit experiences a greater degree of 
expansion during RLOF so that $P = 25\,$d at the end of the Algol phase. 
The wider orbit translates to a lower mass-transfer rate during this 
phase which means that more mass is accreted by the companion. 
The subsequent evolution still leads to a CE when the original 
secondary has reached the GB but in this case a lower value of 
${\alpha}_{\rm\SSS CE}$ is required if a COWD-HeMS binary is to 
emerge with $P = 0.04\,$d. 

If the CV example presented above had a common-envelope efficiency of 
${\alpha}_{\rm\SSS CE} = 3$ less energy is required to drive off the CE 
and the WD-MS binary forms with a wider orbital period of $3.0\,$d. 
This allows the MS primary to evolve to the GB before it fills its Roche-lobe 
so that dynamical mass transfer leads to another CE phase in which 
the WD and the giant core coalesce to form an EAGB star. 
A CV is not formed at all. 
If the metallicity of the component stars is taken to be $Z = 0.001$, 
rather than $Z = 0.02$, the system coalesces after the first 
instance of CE evolution and forms an EAGB star. 
This is also the case if the tidal evolution is not followed. 

\subsection{Comparison with Other Authors}

The Algol example described by Tout et al.\ (1997) begins with initial masses 
3.2 and $2.0 \Msun$ and a circular $5\,$d orbit.  
The primary fills its Roche-lobe on the HG and mass transfer proceeds until it 
has lost its envelope on the GB, at which point it is a $0.318 \Msun$ 
naked helium star in a $170\,$d orbit with a $4.65 \Msun$ MS companion. 
The MS star evolves to the AGB, by which time the 
secondary has become a COWD,  before filling its Roche-lobe which results in 
CE evolution and the production of a double-degenerate WD binary 
in a $35\,$min orbit. 
Gravitational radiation decreases the size of the orbit so that steady 
mass transfer from the less massive WD begins and an AM CVn system is 
formed. 
The same binary evolved with the BSE algorithm also forms an Algol system 
but the first phase of RLOF ends when the primary is in the CHeB stage 
and loses its envelope to become a $0.49 \Msun$ HeMS star with a $4.7 \Msun$ 
MS companion. 
Tidal friction within the system has restrained the growth of the separation 
during mass transfer so that the orbital period at this stage is only $110\,$d. 
As a result the new primary fills its Roche-lobe on the GB and the CE 
evolution that follows leads to coalescence and the 
formation of a single CHeB star. 
An AM CVn system is not produced from this set of initial conditions 
when tides are included. 

Tout et al.\ (1997) describe a scenario for the formation of a CV  
with initial masses of $7.0 \Msun$ and $1.0 \Msun$ in a 
circular orbit of $P = 1200\,$d. 
In this example $Z = 0.02$ and ${\alpha}_{\rm\SSS CE} = 1$. 
They find that the $7.0 \Msun$ primary fills its Roche-lobe on the AGB 
when its mass has fallen to $5.36 \Msun$. 
The ensuing CE produces a $1.014 \Msun$ COWD (they did not distinguish ONeWDs) 
in a $35\,$hr orbit with the $1.0 \Msun$ MS star. 
The MS star then fills its Roche-lobe, as magnetic braking reduces orbital  
separation, resulting in steady 
mass transfer with nova explosions on the WD surface.  
Evolution with the BSE algorithm leads to a similar outcome but with  
differences in the states of the stars and the orbit. 
The $7.0 \Msun$ primary fills its Roche-lobe earlier on the AGB when its 
mass has only fallen to $6.8 \Msun$ and the CE produces a $1.183 \Msun$ ONeWD   
in a $18\,$hr orbit with the $1.0 \Msun$ MS star. 
Once again a CV with nova outbursts is formed when the MS primary fills its 
Roche-lobe. 
In this case the differences between Tout et al.\ (1997) and the BSE 
algorithm are primarily due to improvements in the single star evolution 
model. 

As a final example we consider the formation of the double neutron star binary 
described by Portegies-Zwart \& Verbunt (1996). 
In their model stellar evolution is performed using a simple prescription based 
on models with $Z = 0.02$. 
They assume no kick in a supernova explosion, and their treatment of CE 
evolution is similar to that of Iben \& Livio (1993) discussed in 
Section~\ref{s:comenv}. 
Tidal evolution is not included in their model but eccentric orbits are 
instantaneously circularized when the radius of either star exceeds 1/5  
of the separation at periastron. 
The initial binary has masses $13.1$ and $9.8 \Msun$ 
separated by $138 \Rsun$ in a circular orbit. 
Mass transfer, which begins when the primary has evolved to the HG, becomes 
dynamical when the primary reaches the GB leading, via CE evolution, to the  
formation of a binary composed of a $3.7 \Msun$ He star and an 
$18.7 \Msun$ MS companion separated by $411 \Rsun$. 
The He star then undergoes a supernova which leaves a $1.34 \Msun$ NS remnant 
and widens the orbital separation to $463 \Rsun$. 
Another phase of dynamical mass transfer, when the $18.7 \Msun$ star fills its 
Roche-lobe, leads to CE from which the NS emerges with a 
$4.6  \Msun$ He star companion, separated by $1.6 \Rsun$.  
When the new He star explodes as a supernova a double NS binary forms with 
$P = 0.5\,$d. 

This example evolved with the BSE algorithm forms a binary consisting of a 
$3.1 \Msun$ HeMS star separated from an $11.7 \Msun$ MS companion by 
$60 \Rsun$ after the first CE phase. 
The separation in this instance is much smaller because not as much mass 
is transferred and because tides cause the spin of each star to 
synchronize with the binary period so that the orbit does not expand 
during mass transfer.  
Next the HeMS star evolves to the HeGB before a supernova explosion leaves 
a $1.33 \Msun$ NS and widens the separation to $67 \Rsun$. 
The second phase of CE evolution then leads to coalescence of the NS and the 
$2.63 \Msun$ He core which creates an unstable Thorne-\.{Z}ytkow object.  
However, if the initial binary consists of masses $13.6$ and 
$7.7 \Msun$ separated by $100 \Rsun$, the evolution proceeds along a 
similar path to that described by Portegies-Zwart \& Verbunt (1996), forming  
a double NS binary with $P = 1\,$d. 
This system coalesces in $300\,$Myr through gravitational radiation. 
If a velocity kick at supernova is included in the model then the final state 
of the system depends on the magnitude of the kick and the relative 
positions of the two stars when the explosion occurs.

\section{The Effect of Tides on Binary Evolution}
\label{s:tideff}

The examples in the previous section have shown that the evolution of a 
binary is sensitive to the inclusion of tidal evolution in the binary 
algorithm. 
This is true even for initially circular orbits. 
It is instructive to investigate and quantify the systematic effect 
of tides within a population of binaries. 
We do this using the rapid evolution code to generate a series of 
large binary populations from which we calculate the formation rate of 
interesting individual species, such as X-ray binaries, double-degenerates 
and symbiotic stars, and events such as type Ia SNe. 
The results for populations with and without tidal friction
can be compared.
In addition it is useful to investigate the influence of various 
uncertain parameters in the binary algorithm, such as the common-envelope 
efficiency parameter.  
Comparison with observations and work of others 
hopefully constrains the value of some of these parameters. 
The statistics of theoretical binary populations have been presented and 
discussed by many authors in the past. 
Examples include Han (1998): the formation of double-degenerates, type 
Ia SNe and CVs; Portegies Zwart \& Yungelson (1998): 
the formation of binary neutron stars; de Kool (1992) and Politano (1996): 
formation of CVs; Yungelson et al.~(1995): the 
formation of symbiotic stars with a WD secondary; Iben, Tutukov \& Yungelson 
(1995): low-mass X-ray binaries in the Galactic disk; Han et al.~(1995): 
the formation of CVs, Algols, double-degenerates and 
symbiotic stars; and Pols \& Marinus (1994); the production of blue 
stragglers. 
These studies are all flawed by the fact that no attempt was made to 
model the tidal circularization and synchronization of the binary orbit as 
it evolves, and in particular, the interaction of the intrinsic spin of the 
stars with the orbital period.

\subsection{Method}
\label{s:tidmth}
 
Our aim is to evolve a population of binaries according to chosen 
distributions of primary mass, secondary mass and orbital separation, 
in conjunction with a realistic birth rate function and, 
from this population, to calculate birth rates and expected numbers in 
the Galaxy for various individual binary populations and interesting events. 

We first set-up a grid of initial binary parameters $M_1$, $M_2$ and $a$ 
within the limits: 
\begin{eqnarray*}
M_1 & : & 0.8 \: \rightarrow \: 80 \Msun \\ 
M_2 & : & 0.1 \: \rightarrow \: 80 \Msun \\ 
a   & : & 3.0 \: \rightarrow \: 10^4 \Rsun 
\end{eqnarray*}
with the $n_{\SSS {\mathcal{X}}}$ grid points of parameter ${\mathcal{X}}$ 
logarithmically spaced, 
\beq 
\delta \ln {\mathcal{X}} = \frac{1}{n_{\SSS {\mathcal{X}}} - 1} 
\left[ \ln {\mathcal{X}}_{\max} - \ln {\mathcal{X}}_{\min} \right] 
\, . 
\eeq 
For each set of initial parameters we evolve the binary system to an age 
of $15\,$Gyr, or until it is destroyed.  
Each phase of the evolution is followed in detail according to the 
BSE algorithm. 
The lower limit on the primary mass is determined by the lowest mass star that 
will change appreciably in size within $15\,$Gyr and systems with $M_2 > M_1$ 
are not evolved. 
If a particular binary system $j$ evolves through a phase that is to be 
identified with a certain individual binary population $i$, such as 
cataclysmic variables, then the system makes a contribution 
\begin{equation}
\delta r_j = S \Phi \left( \ln M_{1j} \right) 
\varphi \left( \ln M_{2j} \right) 
\Psi \left( \ln a_j \right) \delta \ln M_1 \delta \ln M_2 \delta \ln a 
\end{equation}
to the rate $r_i$ at which that particular population is born. 
This rate depends on the star formation rate $S$, the primary mass 
distribution $\Phi \left( \ln M_1 \right)$, the secondary mass distribution 
$\varphi \left( \ln M_2 \right)$, and the separation distribution 
$\Psi \left( \ln a  \right)$. 
The number of population $i$ in the Galaxy at any time $T$ is then given by 
the sum of $\delta r_j \times \Delta t_{ji}$ for all systems $j$ that 
lived for a time $\Delta t_{ji}$ as a member of that population, where we 
assume that the interval began before $T$ and that $T$ is the 
endpoint of the binary evolution. 

\subsubsection{The Mass Distributions}
 
Our primary mass distribution is the IMF of 
Kroupa, Tout \& Gilmore (1993), derived from 
the stellar distribution towards both Galactic poles as well as the 
distribution of stars within $5.2\,$pc of the Sun, 
\begin{equation}
\xi \left( m \right) = \left\{ \begin{array} {l@{\quad}l} 
0 & m \leq m_0 \\ 
a_1 m^{-1.3} & m_0 < m \leq 0.5 \\ 
a_2 m^{-2.2} & 0.5 < m \leq 1.0 \\ 
a_2 m^{-2.7} & 1.0 < m < \infty
\end{array} \right. 
\end{equation}
where $\xi \left( m \right) dm$ is the probability that a star has a mass, 
expressed in solar units, between $m$ and $m + dm$.  
The distribution is normalized according to 
\beq 
\int^\infty_0 \xi (m) \, dm  = 1 
\eeq 
so that, for $m_0 = 0.1$, $a_1 = 0.29056$ and $a_2 = 0.15571$. 
Then  
\beq 
\Phi \left( \ln M_1 \right) = M_1 \, \xi \left( M_1 \right) . 
\eeq  
The percentage of stars with mass greater than $80 \Msun$ 
is less than 0.005 and hence our upper mass limit for the grid. 

If the component masses are to be chosen independently from the IMF then 
the secondary mass distribution is 
\beq\label{e:secnd1} 
\varphi \left( \ln M_2 \right) = M_2 \, \xi \left( M_2 \right) \, . 
\eeq  
However there is observational evidence (Eggleton, Fitchett \& Tout 1989; 
Mazeh et al. 1992; Goldberg \& Mazeh 1994) to support correlated masses, 
i.e.  
\beq\label{e:secnd2} 
\varphi \left( \ln M_2 \right) = \frac{M_2}{M_1} = q_2 \, , 
\eeq 
which corresponds to a uniform distribution of the mass-ratio $q_2$,  
for $0 < q_2 \leq 1$. 
 
\subsubsection{The Separation Distribution}

We take this to be 
\beq 
\Psi \left( \ln a \right) = k = \mbox{ constant, }  
\eeq 
between the limits 3 and $10^4 \Rsun$.  
Normalization gives $k = 0.12328$. 
This distribution is contrary to the findings of 
Eggleton, Fitchett \& Tout~(1989) but it facilitates comparison 
with other models (e.g. Yungelson, Livio \& Tutukov 1997; Han 1998) 
without upsetting the results. 
We chose the upper limit with a view to including all separations for 
which a binary system is likely to experience some form of mass exchange 
interaction within the lifetime of the Galaxy (Yungelson et al. 1995). 
This was not achieved for all classes of binaries 
(see Section~\ref{s:modres}). 
The choice of lower limit is rather simplistic and a dependence 
on binary mass might be more realistic.

\subsubsection{Star Formation Rate}

We assume that one binary with $M_1 \geq 0.8 \Msun$ is born in the 
Galaxy per year. 
Therefore 
\beq 
S \int^{M_1 = \infty}_{M_1 = 0.8} \Phi \left( m \right) \, dm = 1 , 
\eeq 
which gives $S = 7.6085 \, {\rm{yr}}^{-1}$.  
We fix this rate over the lifetime of the Galaxy. 
It is in rough agreement with the birth rate of WDs in the 
Galaxy, ${\chi}_{\SSS\rm WD} \approx 2 \times 10^{-12} \, {\rm{pc}}^{-3} 
{\rm{yr}}^{-1}$ 
(Phillips 1989), when we note that only stars with $M > 0.8 \Msun$ can 
evolve to white dwarfs in the age of the Galaxy, that all stars are in 
binaries and assume an 
effective Galactic volume of $V_{\rm gal} = 5 \times 10^{11} \, {\rm{pc}}^3$. 
The same assumptions regarding the star formation rate have been made 
previously (e.g. Iben \& Tutukov 1984; Han 1998) facilitating comparison  
with the results of these studies.  

To estimate an effective Galactic volume we use values for the 
Galactic disk of $L_{V, {\rm disk}} = 1.2 \times 10^{10} \Lsun$ 
and $\left( M / L \right)_{\rm disk} = 5$ for the visual luminosity 
and mass-to-light ratio (Binney \& Tremaine 1987) to give 
$M_{\rm disk} = 6 \times 10^{10} \Msun$ for the mass in visible stars. 
We combine this with the local mass density of stars, 
${\rho}_{\odot} = 0.1 \, \Msun \, {\rm{pc}}^{-3}$ (Kuijken \& Gilmore 1989), 
to give $V_{\rm gal} = 6 \times 10^{11} \, {\rm{pc}}^3$.  

\subsubsection{Binary Population Models}

We construct a variety of models, each with slightly different assumptions for 
the initial conditions of the population or the parameters  
that govern the evolution. 
Our standard is Model~A in which the secondary mass is chosen according to 
a uniform distribution of the mass-ratio (see eq.~\ref{e:secnd2})and the 
metallicity of the stars is $Z = 0.02$. 
All binaries are initially circular, with each star having ZAMS rotation 
as given by eq.~(108) of PapI,  
and tidal synchronization is followed in detail.  
The common-envelope efficiency parameter is set at ${\alpha}_{\rm\SSS CE} = 3$ 
so that the treatment of common-envelope evolution in this model is similar 
to that of Iben \& Livio (1993, although the dependence on core and 
envelope mass in the two treatments remains somewhat different). 
Model~B differs from Model~A by not including tidal evolution. 
For Model~C the tides are taken into account but we use 
${\alpha}_{\rm\SSS CE} = 1$. 
Model~D is the same as Model~A except that the secondary mass is chosen 
independently from the same IMF as the primary.  

The assumption that all stars within the population are born with the same 
composition is somewhat naive: nucleosynthesis in successive generations 
of stars enriches the gas from which they form as the Galaxy evolves. 
Currently it is uncertain whether a definite age-metallicity 
relation exists for stars in the Galactic disk but it is clear is that there 
is a large scatter in metallicity at any given age (McWilliam 1997). 
From the data of Edvardsson et al.~(1993) the age-metallicity relation for 
the solar neighbourhood can be represented in terms of the relative iron 
abundance by 
\beq
[{\rm Fe/H}] = 0.35 x - 0.4 \tau - \left( 1.73 + 0.35 x \right) {\tau}^{20}  
\eeq
where $\tau = t/T_{\rm gal}$ is the fractional time since formation of the 
star in terms of the current Galactic age. 
A star born now has $\tau = 0$. 
The scatter in the data is reproduced by a random variable $x$ uniform in  
the range $-1 \leq x \leq 1$. 
The metallicity can be obtained from 
\beq 
[{\rm Fe/H}] \simeq \ln \left( Z / Z_{\odot} \right)  
\eeq 
with $Z_{\odot} = 0.02$. 
This relation shows that solar composition is within the limits of the 
scatter for the most recent 80\% of the Galaxy's life but that for 
older populations the metallicity drops sharply to much lower values. 
While it would be interesting to construct a model with a $Z(t)$ dependence 
this is beyond the scope of this work. 
To represent an older population we simply take the metallicity of the stars 
in Model~E to be $Z= 0.0001$. 

Finally to investigate the changes produced when tidal circularization, as well 
as synchronization, is acting on the population, our Model~F differs from 
Model~A by allowing binaries to form in eccentric orbits. 
We choose the initial eccentricity from a thermal distribution (Heggie 1975) 
\beq 
f(e) = 2e \, , 
\eeq  
between the limits 0 and 1. 
Model~G also allows eccentric orbits but the process of circularization is 
according to the model of Portegies Zwart \& Verbunt (1996, PZV), 
as described in Section~\ref{s:exmple}. 

\begin{table}
\setlength{\tabcolsep}{0.2cm}
\begin{center}
\caption[]{
Model parameters. 
}
\label{t:modpar}
\begin{tabular}{|c|c|c|c|c|c|}
\hline
 & TIDES & ${\alpha}_{\rm\SSS CE}$ & $e$ & $M_2$ & $Z$ \\
\hline
\hline
A & ON & 3.0 & 0.0 & eq.~(\ref{e:secnd2}) & 0.0200 \\ 
B & OFF & 3.0 & 0.0 & eq.~(\ref{e:secnd2}) & 0.0200 \\ 
C & ON & 1.0 & 0.0 & eq.~(\ref{e:secnd2}) & 0.0200 \\ 
D & ON & 3.0 & 0.0 & eq.~(\ref{e:secnd1}) & 0.0200 \\ 
E & ON & 3.0 & 0.0 & eq.~(\ref{e:secnd2}) & 0.0001 \\ 
F & ON & 3.0 & $f(e)$ & eq.~(\ref{e:secnd2}) & 0.0200 \\ 
G & PZV & 3.0 & $f(e)$ & eq.~(\ref{e:secnd2}) & 0.0200 \\ 
\hline
\end{tabular}
\end{center}
\end{table}

The characteristics of each model are summarised in Table~\ref{t:modpar}. 
All models assume the default values given in Table~\ref{t:params} unless 
specified otherwise. 
This means that a velocity kick is imparted to a NS or BH at birth  
(as described in Section~\ref{s:supkck}) and 
the Eddington limit is not imposed during mass transfer in any of these models 
(see Section~\ref{s:secresp}). 
We set the wind accretion parameter to ${\beta}_{\rm\SSS W} = 0.5$ 
in all cases and the 
stars are not assumed to be in co-rotation with the orbit on the ZAMS. 

For each model we use 100 grid points in each dimension so that, recalling   
that only systems with $q_2 \leq 1$ need be evolved, a total of 
$6.54 \times 10^5$ different binaries are evolved in each case. 
We find that a finer grid spacing does not alter the results within the  
tolerance set by the quoted errors. 
For each system we slightly displace the initial parameters from their grid 
positions so that a smooth distribution in mass and separation is achieved.  
This is done uniformly about each grid point, after selection of a random 
variable, with no overlap between parameters attached to adjoining grid points.

\subsection{Results}
\label{s:tidres}

Table~\ref{t:tidcmp} lists the formation rates, as number per year 
in the Galactic disk, of various interesting stars, binaries and events. 
We calculate these assuming an age for the Galactic disk of 
$12\,$Gyr (Eggleton, Fitchett \& Tout 1989; Phelps 1997; Knox, Hawkins 
\& Hambly 1999) and that all stars are born in binaries. 
Also listed is an estimate of the percentage error in the formation rate 
of the individual populations calculated from the standard deviation 
produced by repeated runs of Model~A, each with a different seed for 
the random number generator. 
Errors associated with systems that involve the formation of a NS or BH 
are generally larger owing to the velocity kick distribution. 
The eccentricity distribution used in Models~F and G introduces an 
additional degree of uncertainty in the results, however repeated runs 
of Model~F show that the percentage error only exceeds that quoted in 
Table~\ref{t:tidcmp} for a few cases, and never by more than a factor 
of two. 

\subsubsection{Definitions of Binary Class}

In our models a blue straggler star (BSS) is a MS star that, by 
accreting mass or by merging, appears older than the standard MS 
lifetime applicable to its mass. 
This phase of evolution lasts until the star moves off the MS.  
Cataclysmic variables (CVs) have a WD secondary with a non-degenerate 
Roche-lobe filling companion. 
These we divide into sub-categories, classical CV (CV class) if $k_1 \leq 1$, 
GK Persei systems (GK Per, e.g. Crampton, Cowley \& Fisher 1986) 
if $k_1 = 2$, symbiotic-like binaries (CV Symb) if 
$3 \leq k_1 \leq 6$ and subdwarf B binaries (sdB) if $7 \leq k_1 \leq 9$. 
Binaries in which a MS secondary is accreting from a Roche-lobe 
filling companion are termed {\it Algol}. 
If the mass-ratio of the primary, $q_1 = M_1/M_2$, is greater than unity 
then the system is pre-Algol, if the primary is a MS star and $q_1 < 1$ then 
this is a MS Algol, otherwise the system is a cold Algol, if the 
secondary has $M_2 \leq 1.25 \Msun$, or a hot Algol, if $M_2 > 1.25 \Msun$. 

X-ray binaries have a NS or a BH secondary accreting material from 
either the stellar wind of the companion or by RLOF, 
with an accretion luminosity $L_{\rm X} > \Lsun$. 
The primary star may be of any type. 
If the material is accreted via an accretion disk the 
accretion luminosity is given by 
\beq\label{e:lacc}
L_{\rm X} = \frac{G M_2 {\dot{M}}_2}{2 R_2} .  
\eeq
If the primary is a MS star with $M_1 < 2 \Msun$ the system is a low-mass  
X-ray binary (LMXB), if the primary is a WD then it is a white dwarf  
X-ray binary (WDXB), otherwise it is a massive X-ray binary (MXRB). 
These sources are divided into transient~(t) or persistent~(p) according 
to the criteria of van Paradijs (1996): for a critical luminosity 
\beq\label{e:lxcrit}
\log \left( \frac{L_{\rm X,crit}}{\Lsun} \right) = 
\left\{ \begin{array} {l@{\quad}l}
1.62 + 1.07 \left( P/{\rm hr} \right) & \mbox{ NS secondary} \\
2.22 + 1.07 \left( P/{\rm hr} \right) & \mbox{ BH secondary} 
\end{array} \right. \, , 
\eeq
if $L_{\rm X} > L_{\rm X,crit}$ the X-ray source is persistent, otherwise 
it is a soft X-ray transient (SXT) analogous to a dwarf nova 
(see Section~\ref{s:secresp}). 
SXT outbursts are caused by an instability in the accretion disk, the  
temperature of which is affected by X-ray heating. 
The main effect of this X-ray heating, taken into account in the 
calculation of eq.~(\ref{e:lxcrit}), is to make the flow of material through 
the accretion disk stable to substantially lower mass transfer rates 
than for dwarf novae systems. 

\begin{table*}
\setlength{\tabcolsep}{0.1cm}
\begin{center}
\caption[]{
Formation rates per year in the Galactic disk of various events and systems. 
}
\label{t:tidcmp}
\scriptsize
\begin{minipage}{16cm}
\begin{tabular}{|l|c|c|c|c|c|c|c|c|}
\hline
Model  & A & B & C & D & E & F & G & \%err \\
\hline
\hline
BSS        & $1.295\times 10^{-1}$  & $1.138\times 10^{-1}$ & 
$1.248\times 10^{-1}$ & $5.049\times 10^{-2}$ &
$1.839\times 10^{-1}$ & 
$1.181\times 10^{-1}$ & $1.367\times 10^{-1}$ & 0.31 \\  
\hline
CV class   & $1.987\times 10^{-2}$  & $1.292\times 10^{-2}$ & 
$2.120\times 10^{-2}$ & $1.820\times 10^{-2}$ &
$3.080\times 10^{-2}$ & 
$1.889\times 10^{-2}$ & $3.973\times 10^{-2}$ & 3.65 \\  
GK Per     & $1.398\times 10^{-2}$  & $1.165\times 10^{-2}$ & 
$1.671\times 10^{-2}$ & $1.679\times 10^{-3}$ &
$3.709\times 10^{-2}$ & 
$1.374\times 10^{-2}$ & $3.460\times 10^{-3}$ & 0.93 \\  
CV Symb    & $7.858\times 10^{-4}$  & $1.423\times 10^{-4}$ & 
$2.450\times 10^{-3}$ & $1.898\times 10^{-4}$ &
$1.027\times 10^{-2}$ & 
$8.527\times 10^{-4}$ & $4.440\times 10^{-4}$ & 2.73 \\  
sdB        & $1.135\times 10^{-2}$  & $1.398\times 10^{-2}$ & 
$2.189\times 10^{-3}$ & $1.090\times 10^{-3}$ &
$1.729\times 10^{-2}$ & 
$1.047\times 10^{-2}$ & $1.125\times 10^{-2}$ & 0.49 \\  
\hline
pre Algol  & $1.248\times 10^{-1}$  & $1.040\times 10^{-1}$ & 
$1.244\times 10^{-1}$ & $8.807\times 10^{-2}$ &
$1.861\times 10^{-1}$ & 
$1.112\times 10^{-1}$ & $2.082\times 10^{-1}$ & 0.07 \\ 
MS Algol   & $2.861\times 10^{-2}$  & $2.981\times 10^{-2}$ & 
$2.864\times 10^{-2}$ & $1.558\times 10^{-2}$ &
$2.784\times 10^{-2}$ & 
$2.485\times 10^{-2}$ & $1.077\times 10^{-2}$ & 0.32 \\  
cold Algol & $1.313\times 10^{-2}$  & $1.145\times 10^{-2}$ & 
$1.293\times 10^{-2}$ & $7.007\times 10^{-3}$ &
$3.718\times 10^{-2}$ & 
$1.222\times 10^{-2}$ & $2.255\times 10^{-3}$ & 1.11 \\ 
hot Algol  & $4.910\times 10^{-2}$  & $4.593\times 10^{-2}$ & 
$4.458\times 10^{-2}$ & $1.018\times 10^{-2}$ &
$9.049\times 10^{-2}$ & 
$4.430\times 10^{-2}$ & $2.780\times 10^{-2}$ & 0.18 \\  
\hline
NS LMXBp & $3.487\times 10^{-6}$  & $1.362\times 10^{-7}$ & 
$2.219\times 10^{-7}$ & $5.481\times 10^{-6}$ &
$1.010\times 10^{-5}$ & 
$1.071\times 10^{-6}$ & $4.906\times 10^{-6}$ & 37.56 \\ 
BH LMXBp & $3.601\times 10^{-6}$  & $1.696\times 10^{-6}$ & 
$1.561\times 10^{-6}$ & $3.686\times 10^{-6}$ &
$2.563\times 10^{-5}$ & 
$3.865\times 10^{-6}$ & $5.349\times 10^{-5}$ & 25.36 \\ 
NS MXRBp & $7.481\times 10^{-4}$  & $5.014\times 10^{-4}$ & 
$6.307\times 10^{-4}$ & $1.434\times 10^{-5}$ &
$6.096\times 10^{-4}$ & 
$7.214\times 10^{-4}$ & $5.164\times 10^{-4}$ & 1.37 \\  
BH MXRBp & $1.057\times 10^{-4}$  & $1.612\times 10^{-5}$ & 
$1.092\times 10^{-4}$ & $2.043\times 10^{-6}$ &
$2.425\times 10^{-4}$ & 
$1.085\times 10^{-4}$ & $1.132\times 10^{-4}$ & 3.77 \\  
NS WDXBp & $1.639\times 10^{-3}$  & $2.028\times 10^{-3}$ & 
$1.686\times 10^{-4}$ & $8.393\times 10^{-5}$ &
$3.563\times 10^{-3}$ & 
$1.426\times 10^{-3}$ & $1.347\times 10^{-3}$ & 0.55 \\ 
BH WDXBp & $2.764\times 10^{-4}$  & $3.513\times 10^{-4}$ & 
$2.885\times 10^{-5}$ & $4.987\times 10^{-6}$ &
$4.951\times 10^{-4}$ & 
$2.400\times 10^{-4}$ & $2.085\times 10^{-4}$ & 7.62 \\  
NS LMXBt & $2.359\times 10^{-5}$  & $5.640\times 10^{-6}$ & 
$1.253\times 10^{-6}$ & $2.537\times 10^{-5}$ &
$4.198\times 10^{-5}$ & 
$9.087\times 10^{-6}$ & $3.654\times 10^{-5}$ & 21.27 \\  
BH LMXBt & $9.172\times 10^{-6}$  & $1.691\times 10^{-6}$ & 
$4.854\times 10^{-6}$ & $6.552\times 10^{-6}$ &
$2.323\times 10^{-5}$ & 
$6.542\times 10^{-6}$ & $2.283\times 10^{-5}$ & 20.87 \\  
NS MXRBt & $7.345\times 10^{-4}$  & $8.048\times 10^{-4}$ & 
$8.122\times 10^{-4}$ & $1.823\times 10^{-5}$ &
$1.568\times 10^{-3}$ & 
$7.149\times 10^{-4}$ & $6.515\times 10^{-4}$ & 1.41 \\  
BH MXRBt & $5.447\times 10^{-5}$  & $2.320\times 10^{-5}$ & 
$8.706\times 10^{-5}$ & $2.564\times 10^{-6}$ &
$2.560\times 10^{-4}$ & 
$5.823\times 10^{-5}$ & $4.471\times 10^{-5}$ & 2.81 \\  
NS WDXBt & $8.963\times 10^{-4}$  & $1.088\times 10^{-3}$ & 
$7.332\times 10^{-5}$ & $7.325\times 10^{-5}$ &
$2.296\times 10^{-3}$ & 
$7.753\times 10^{-4}$ & $6.918\times 10^{-4}$ & 0.29 \\  
BH WDXBt & $6.531\times 10^{-4}$  & $8.299\times 10^{-4}$ & 
$8.357\times 10^{-5}$ & $1.017\times 10^{-5}$ &
$1.237\times 10^{-3}$ & 
$5.636\times 10^{-4}$ & $5.784\times 10^{-4}$ & 1.78 \\  
\hline
S-Symb     & $5.353\times 10^{-3}$  & $5.091\times 10^{-3}$ & 
$5.370\times 10^{-3}$ & $2.712\times 10^{-4}$ &
$6.356\times 10^{-3}$ & 
$4.100\times 10^{-3}$ & $4.305\times 10^{-3}$ & 1.79 \\  
D-Symb     & $4.322\times 10^{-2}$  & $4.701\times 10^{-2}$ & 
$4.302\times 10^{-2}$ & $5.748\times 10^{-3}$ &
$3.782\times 10^{-2}$ & 
$3.494\times 10^{-2}$ & $3.763\times 10^{-2}$ & 0.23 \\  
\hline
nHe MSC    & $6.441\times 10^{-3}$  & $1.929\times 10^{-3}$ & 
$2.160\times 10^{-3}$ & $1.657\times 10^{-3}$ &
$1.649\times 10^{-2}$ & 
$5.591\times 10^{-3}$ & $5.059\times 10^{-3}$ & 0.94 \\  
gnt MSC    & $3.366\times 10^{-3}$  & $3.639\times 10^{-3}$ & 
$3.432\times 10^{-3}$ & $1.093\times 10^{-3}$ &
$4.809\times 10^{-3}$ & 
$3.075\times 10^{-3}$ & $8.166\times 10^{-4}$ & 4.75 \\  
\hline
WDWD DD    & $1.131\times 10^{-1}$  & $1.229\times 10^{-1}$ & 
$7.572\times 10^{-2}$ & $1.334\times 10^{-2}$ &
$2.290\times 10^{-1}$ & 
$8.631\times 10^{-2}$ & $7.902\times 10^{-2}$ & 0.16 \\  
WDNS DD    & $8.577\times 10^{-4}$  & $8.982\times 10^{-4}$ & 
$5.265\times 10^{-4}$ & $2.088\times 10^{-5}$ &
$2.842\times 10^{-3}$ & 
$7.986\times 10^{-4}$ & $7.537\times 10^{-4}$ & 1.10 \\  
NSNS DD    & $7.384\times 10^{-5}$  & $8.192\times 10^{-5}$ & 
$2.785\times 10^{-5}$ & $6.972\times 10^{-7}$ &
$1.604\times 10^{-4}$ & 
$7.587\times 10^{-5}$ & $7.295\times 10^{-5}$ & 5.11 \\  
\hline
He DDRch\footnote{DDRch systems are a subset of the corresponding DDs.}  
           & $1.997\times 10^{-2}$  & $2.385\times 10^{-2}$ & 
$6.782\times 10^{-3}$ & $2.344\times 10^{-3}$ & 
$5.361\times 10^{-2}$ & 
$1.927\times 10^{-2}$ & $1.619\times 10^{-2}$ & 1.21 \\ 
CO DDRch   & $2.942\times 10^{-3}$  & $3.598\times 10^{-3}$ & 
$1.549\times 10^{-4}$ & $8.558\times 10^{-5}$ &
$4.959\times 10^{-3}$ & 
$2.538\times 10^{-3}$ & $2.144\times 10^{-3}$ & 0.43 \\  
NS DDRch   & $5.724\times 10^{-5}$  & $6.826\times 10^{-5}$ & 
$2.381\times 10^{-5}$ & $5.096\times 10^{-7}$ &
$1.062\times 10^{-4}$ & 
$5.884\times 10^{-5}$ & $5.240\times 10^{-5}$ & 6.12 \\
\hline
He LMWD    & $1.671\times 10^{-1}$ & $1.709\times 10^{-1}$ & 
$9.715\times 10^{-2}$ & $1.138\times 10^{-1}$ & 
$2.985\times 10^{-1}$ & 
$1.658\times 10^{-1}$ & $1.466\times 10^{-1}$ & 0.42 \\ 
CO LMWD    & $7.019\times 10^{-3}$ & $7.622\times 10^{-3}$ & 
$1.351\times 10^{-3}$ & $1.564\times 10^{-3}$ & 
$1.085\times 10^{-2}$ & 
$6.597\times 10^{-3}$ & $3.002\times 10^{-3}$ & 1.78 \\ 
\hline
He SNIa    & $2.818\times 10^{-3}$  & $8.100\times 10^{-3}$ & 
$3.217\times 10^{-4}$ & $7.214\times 10^{-4}$ &
$2.702\times 10^{-2}$ & 
$2.677\times 10^{-3}$ & $1.242\times 10^{-3}$ & 1.68 \\  
ELD SNIa   & $1.560\times 10^{-2}$  & $1.994\times 10^{-2}$ & 
$5.193\times 10^{-3}$ & $1.665\times 10^{-3}$ &
$3.227\times 10^{-2}$ & 
$1.490\times 10^{-2}$ & $1.102\times 10^{-2}$ & 0.61 \\  
CO SNIa    & $2.567\times 10^{-3}$  & $2.627\times 10^{-3}$ & 
$2.109\times 10^{-4}$ & $1.125\times 10^{-4}$ &
$2.720\times 10^{-3}$ & 
$2.442\times 10^{-3}$ & $1.890\times 10^{-3}$ & 0.59 \\  
AIC        & $3.738\times 10^{-3}$  & $3.975\times 10^{-3}$ & 
$9.771\times 10^{-4}$ & $1.364\times 10^{-4}$ &
$6.286\times 10^{-3}$ & 
$3.442\times 10^{-3}$ & $3.029\times 10^{-3}$ & 0.35 \\  
SNII       & $1.724\times 10^{-2}$  & $1.592\times 10^{-2}$ & 
$1.940\times 10^{-2}$ & $1.142\times 10^{-2}$ &
$2.848\times 10^{-2}$ & 
$1.810\times 10^{-2}$ & $1.921\times 10^{-2}$ & 1.75 \\  
SNIIa      & $3.551\times 10^{-4}$  & $1.453\times 10^{-4}$ & 
$6.189\times 10^{-5}$ & $7.054\times 10^{-6}$ &
$6.840\times 10^{-5}$ & 
$2.851\times 10^{-4}$ & $1.228\times 10^{-4}$ & 1.67 \\  
SNIb/c     & $1.403\times 10^{-2}$  & $1.361\times 10^{-2}$ & 
$1.304\times 10^{-2}$ & $4.043\times 10^{-3}$ &
$2.026\times 10^{-2}$ & 
$1.313\times 10^{-2}$ & $1.510\times 10^{-2}$ & 0.19 \\  
\hline
\end{tabular}
\end{minipage}
\end{center}
\normalsize
\end{table*}

\begin{table*}
\caption[]{
Present number in the Galactic disk of various systems. 
}
\label{t:numcmp}
\footnotesize
\begin{tabular}{|l|c|c|c|c|c|c|c|}
\hline
Model  & A & B & C & D & E & F & G \\
\hline
\hline
BSS        & $6.999\times 10^{7}$  & $4.444\times 10^{7}$ & 
$6.993\times 10^{7}$ & $4.439\times 10^{7}$ & 
$8.823\times 10^{7}$ & 
$8.023\times 10^{7}$ & $6.128\times 10^{7}$ \\
\hline
CV class   & $2.532\times 10^{7}$  & $3.594\times 10^{7}$ & 
$3.554\times 10^{7}$ & $9.908\times 10^{7}$ & 
$3.995\times 10^{7}$ & 
$2.550\times 10^{7}$ & $2.565\times 10^{7}$ \\
GK Per     & $7.242\times 10^{6}$  & $2.775\times 10^{6}$ & 
$8.155\times 10^{6}$ & $1.184\times 10^{6}$ & 
$2.001\times 10^{6}$ & 
$6.953\times 10^{6}$ & $9.842\times 10^{5}$ \\
CV Symb    & $7.070\times 10^{5}$  & $1.255\times 10^{3}$ & 
$2.503\times 10^{6}$ & $1.849\times 10^{5}$ & 
$4.781\times 10^{4}$ & 
$7.982\times 10^{5}$ & $2.486\times 10^{5}$ \\
sdB        & $3.298\times 10^{4}$  & $6.635\times 10^{4}$ & 
$5.548\times 10^{3}$ & $3.335\times 10^{3}$ & 
$5.054\times 10^{4}$ & 
$3.047\times 10^{4}$ & $8.599\times 10^{3}$ \\
\hline
pre Algol  & $6.396\times 10^{6}$  & $1.262\times 10^{7}$ & 
$6.405\times 10^{6}$ & $3.122\times 10^{6}$ & 
$2.411\times 10^{6}$ & 
$5.614\times 10^{6}$ & $7.313\times 10^{5}$ \\
MS Algol   & $6.081\times 10^{6}$  & $1.178\times 10^{7}$ & 
$6.077\times 10^{6}$ & $3.423\times 10^{6}$ & 
$3.806\times 10^{6}$ & 
$5.432\times 10^{6}$ & $2.115\times 10^{5}$ \\
cold Algol & $6.784\times 10^{6}$  & $1.020\times 10^{7}$ & 
$6.795\times 10^{6}$ & $4.658\times 10^{6}$ &
$9.008\times 10^{6}$ & 
$6.007\times 10^{6}$ & $4.516\times 10^{5}$ \\
hot Algol  & $7.132\times 10^{6}$  & $8.216\times 10^{6}$ & 
$7.148\times 10^{6}$ & $3.680\times 10^{6}$ & 
$6.790\times 10^{6}$ & 
$6.320\times 10^{6}$ & $5.034\times 10^{5}$ \\
\hline
NS LMXBp & $4.191\times 10^{0}$  & $1.112\times 10^{1}$ & 
$3.451\times 10^{0}$ & $7.690\times 10^{1}$ & 
$4.409\times 10^{2}$ & 
$1.331\times 10^{0}$ & $4.388\times 10^{1}$ \\
BH LMXBp & $9.140\times 10^{1}$  & $4.106\times 10^{1}$ & 
$5.016\times 10^{1}$ & $3.001\times 10^{1}$ & 
$8.553\times 10^{2}$ & 
$6.998\times 10^{1}$ & $2.196\times 10^{3}$ \\
NS MXRBp & $2.810\times 10^{3}$  & $2.608\times 10^{3}$ & 
$1.600\times 10^{3}$ & $8.178\times 10^{1}$ & 
$3.987\times 10^{3}$ & 
$2.531\times 10^{3}$ & $2.494\times 10^{2}$ \\
BH MXRBp & $2.335\times 10^{2}$  & $2.992\times 10^{1}$ & 
$1.439\times 10^{2}$ & $3.333\times 10^{0}$ & 
$1.876\times 10^{2}$ & 
$2.673\times 10^{2}$ & $1.288\times 10^{1}$ \\
NS WDXBp & $1.233\times 10^{5}$  & $1.494\times 10^{5}$ & 
$1.116\times 10^{4}$ & $7.550\times 10^{3}$ & 
$2.871\times 10^{5}$ & 
$1.055\times 10^{5}$ & $1.023\times 10^{5}$ \\
BH WDXBp & $1.554\times 10^{4}$  & $2.145\times 10^{4}$ & 
$1.841\times 10^{3}$ & $3.066\times 10^{2}$ & 
$2.947\times 10^{4}$ & 
$1.469\times 10^{4}$ & $1.353\times 10^{4}$ \\
NS LMXBt & $2.854\times 10^{4}$  & $1.871\times 10^{4}$ & 
$4.131\times 10^{3}$ & $1.396\times 10^{5}$ & 
$5.545\times 10^{4}$ & 
$2.238\times 10^{4}$ & $1.760\times 10^{5}$ \\
BH LMXBt & $7.240\times 10^{4}$  & $1.085\times 10^{4}$ & 
$4.260\times 10^{4}$ & $5.072\times 10^{4}$ & 
$1.343\times 10^{5}$ & 
$4.405\times 10^{4}$ & $2.456\times 10^{5}$ \\
NS MXRBt & $1.785\times 10^{5}$  & $1.511\times 10^{5}$ & 
$5.432\times 10^{4}$ & $8.452\times 10^{3}$ & 
$7.547\times 10^{4}$ & 
$1.630\times 10^{5}$ & $2.962\times 10^{5}$ \\
BH MXRBt & $4.829\times 10^{4}$  & $7.947\times 10^{3}$ & 
$9.105\times 10^{4}$ & $2.818\times 10^{3}$ & 
$7.353\times 10^{4}$ & 
$6.671\times 10^{4}$ & $1.432\times 10^{5}$ \\
NS WDXBt & $2.802\times 10^{6}$  & $3.406\times 10^{6}$ & 
$2.079\times 10^{5}$ & $2.031\times 10^{5}$ & 
$6.925\times 10^{6}$ & 
$2.423\times 10^{6}$ & $2.177\times 10^{6}$ \\
BH WDXBt & $3.223\times 10^{6}$  & $4.161\times 10^{6}$ & 
$4.313\times 10^{5}$ & $5.088\times 10^{4}$ & 
$6.099\times 10^{6}$ & 
$2.843\times 10^{6}$ & $2.920\times 10^{6}$ \\
\hline
S-Symb     & $2.363\times 10^{2}$  & $1.946\times 10^{2}$ & 
$2.713\times 10^{2}$ & $2.858\times 10^{1}$ & 
$4.470\times 10^{2}$ & 
$1.378\times 10^{2}$ & $1.271\times 10^{2}$ \\
D-Symb     & $3.389\times 10^{3}$  & $3.837\times 10^{3}$ & 
$3.369\times 10^{3}$ & $3.369\times 10^{2}$ & 
$5.044\times 10^{3}$ & 
$2.702\times 10^{3}$ & $2.939\times 10^{3}$ \\
\hline
nHe MSC    & $1.497\times 10^{4}$  & $1.017\times 10^{4}$ & 
$2.116\times 10^{3}$ & $4.578\times 10^{3}$ & 
$4.402\times 10^{4}$ & 
$1.156\times 10^{4}$ & $1.572\times 10^{4}$ \\
gnt MSC    & $4.609\times 10^{4}$  & $7.355\times 10^{4}$ & 
$4.390\times 10^{4}$ & $1.456\times 10^{4}$ & 
$1.292\times 10^{4}$ & 
$3.182\times 10^{4}$ & $5.554\times 10^{2}$ \\
\hline
WDWD DD    & $6.018\times 10^{8}$  & $6.028\times 10^{8}$ & 
$4.919\times 10^{8}$ & $5.759\times 10^{7}$ & 
$1.041\times 10^{9}$ & 
$1.219\times 10^{8}$ & $4.048\times 10^{8}$ \\
WDNS DD    & $2.725\times 10^{5}$  & $1.866\times 10^{5}$ & 
$4.090\times 10^{4}$ & $1.192\times 10^{3}$ & 
$6.959\times 10^{5}$ & 
$1.800\times 10^{5}$ & $2.955\times 10^{5}$ \\
NSNS DD    & $2.427\times 10^{6}$  & $2.160\times 10^{6}$ & 
$1.307\times 10^{6}$ & $7.949\times 10^{4}$ & 
$8.313\times 10^{6}$ & 
$2.367\times 10^{6}$ & $2.249\times 10^{6}$ \\
\hline
He DDRch   & $3.076\times 10^{7}$  & $2.913\times 10^{7}$ & 
$1.003\times 10^{7}$ & $3.959\times 10^{6}$ & 
$1.353\times 10^{8}$ & 
$2.858\times 10^{7}$ & $2.371\times 10^{7}$ \\
CO DDRch   & $2.465\times 10^{7}$  & $3.299\times 10^{7}$ & 
$1.261\times 10^{6}$ & $7.225\times 10^{5}$ & 
$4.125\times 10^{7}$ & 
$2.127\times 10^{7}$ & $1.879\times 10^{7}$ \\
\hline
\end{tabular}
\normalsize
\end{table*}

Symbiotic stars are classed as systems with secondaries of $k_2 \leq 12$ that 
accrete material from the stellar wind of a giant primary at a rate high 
enough to produce an accretion luminosity that exceeds either $10 \Lsun$ 
(Yungelson et al. 1995) or 1\% of the primary luminosity - whichever is the 
smaller. 
Eq.~(\ref{e:lacc}) is used to calculate the accretion luminosity. 
D-type symbiotics (D-Symb) are long-period systems containing a Mira-like 
cool star primary ($k_1 = 6$). 
They are characterized by extensive circumbinary dust shells. 
Shorter period symbiotics with a normal giant primary ($k_1 < 6$) are S-type 
(S-Symb). 
MS stars accreting from a Roche-lobe filling giant could also appear to be 
symbiotic stars (Kenyon 1986) but here we include them in the cold Algol 
population. 
Miscellaneous RLOF systems with a non-degenerate primary ($k_1 < 10$) are 
nHe MSC if the secondary is a helium star ($7 \leq k_2 \leq 9$) and 
gnt MSC if the secondary is giant-like ($2 \leq k_2 \leq 6$). 

Double-degenerate (DD) binaries that consist of two WDs are WDWD DD 
while those containing a WD and either a NS or BH are WDNS DD. 
If the system is composed of a NS-NS, NS-BH or BH-BH pair it is 
a NSNS DD.   
DD systems with a Roche-lobe filling WD (AM~CVn systems) are He DDRch, 
if the the primary is a HeWD, or CO DDRch otherwise. 
NS-NS binaries that evolve into contact and coalesce are NS DDRch. 
As a result of binary interaction it is possible to form WDs of mass 
less than the lowest mass WD that can be formed from single star 
evolution in the lifetime of the Galaxy, $M \la 0.5 \Msun$. 
These low-mass white dwarfs (LMWDs) are recorded as either He LMWD or CO LMWD.  

We also count various types of supernovae. 
HeWDs that explode when their mass exceeds 
$0.7 \Msun$ are recorded as He SNIa, as are HeWD mergers. 
COWDs that explode as a possible 
type Ia SN in an Edge-Lit Detonation, once $0.15 \Msun$ of He-rich material 
has been accreted by the WD, are ELD SNIa, while COWDs that explode 
because their mass exceeds the Chandrasekhar limit are CO SNIa. 
Supernovae that leave no remnant from a primary with $k_1 \leq 9$ are SNIIa. 
AIC represents the accretion induced collapse of a 
Chandrasekhar mass ONeWD to a NS. 
Supernovae that produce a NS or BH from a primary with 
$8 \leq k_1 \leq 9$ are SNIb/c. 
All others are normal type II SNe, or SNII. 

Supernovae are effectively instantaneous {\it events} so there is no record  
of how long they last, whereas the remainder of the populations are 
{\it systems} and must exist for a non-zero amount of time to  
contribute to the birth rate. 
We also count double NS binaries that enter RLOF (NS DDRch) as events because 
they always coalesce immediately as possible $\gamma$-ray bursts. 
Each binary may contribute to the rate of more than one population 
if it evolves through a series of phases identified with the 
individual populations listed in Table~\ref{t:tidcmp}. 
The expected numbers presently in the Galactic disk for all the 
individual populations that exist as systems for a finite time are listed 
in Table~\ref{t:numcmp}. 
These can be converted to space densities by dividing by the effective 
Galactic volume $V_{\rm gal} = 6 \times 10^{11} \, {\rm{pc}}^3$ which 
has been normalized to the local mass density of stars. 

\subsubsection{Comparison of Model Results}
\label{s:modres}

\begin{figure*}
\centerline{
\psfig{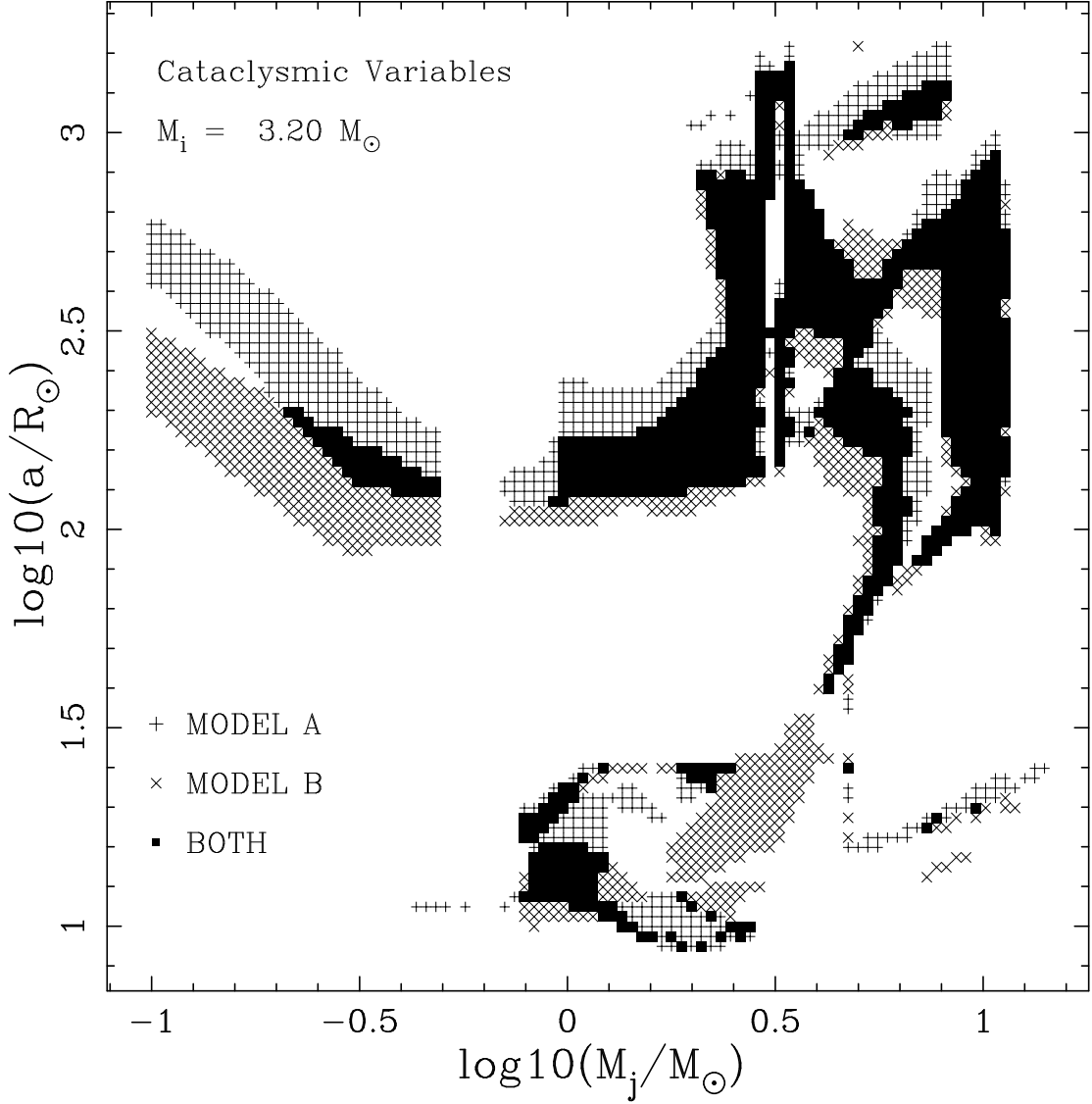}
}
\caption{
The region of parameter space from which cataclysmic variables (CVs) form when
the initial mass of one star is fixed at $3.2 \Msun$.
Systems formed from Model~A are shown as + symbols, from Model~B as $\times$ 
symbols, with any overlap as filled squares. 
The indices $i$ and $j$ can assume the value of either 1 or 2 depending on 
which mass represents the primary star on the ZAMS, 
i.e. if the $3.2 \Msun$ star is the more massive then $i = 1$ and $j = 2$. 
}
\label{f:tides1}
\end{figure*}

Comparison of Models~A and B reveals that most populations show differences in 
their formation rates that are greater than the errors but surprisingly few 
exhibit significant differences. 
Figure~\ref{f:tides1} shows the parameter space for the formation of all CV 
types when the initial mass of one component star is fixed at $3.2 \Msun$. 
The systems that form in Model~A only, Model~B only and in both 
are distinguished. 
There is no special reason for the choice of $3.2 \Msun$ as an 
illustrative example other than that a wide variety of systems are formed 
in this case. 
Systems with low-mass ZAMS companions ($M_2 < 0.5 \Msun$) are CV class while 
GK Per systems are formed with high-mass ZAMS primaries ($M_1 > 7.5 \Msun$). 
The sdB systems form in two main regions of the parameter space: small 
orbital separation ($a < 30 \Rsun$) and large orbital separation 
($a > 500 \Rsun$). 
The remainder, those occupying the middle region of the 
diagram, are either CV class, GK Per or CV Symb, with a large number of the 
systems evolving through a combination of these phases. 
It is evident from Figure~\ref{f:tides1} that the CV population 
behaves as expected when tidal evolution is allowed in the model. 
Wider systems are brought into an interaction distance by the action of 
tidal synchronization while closer systems, that formed CVs without tides, 
now get too close and coalesce during the common-envelope (CE)  
phase that is an integral part of CV formation. 

\begin{figure*}
\centerline{
\psfig{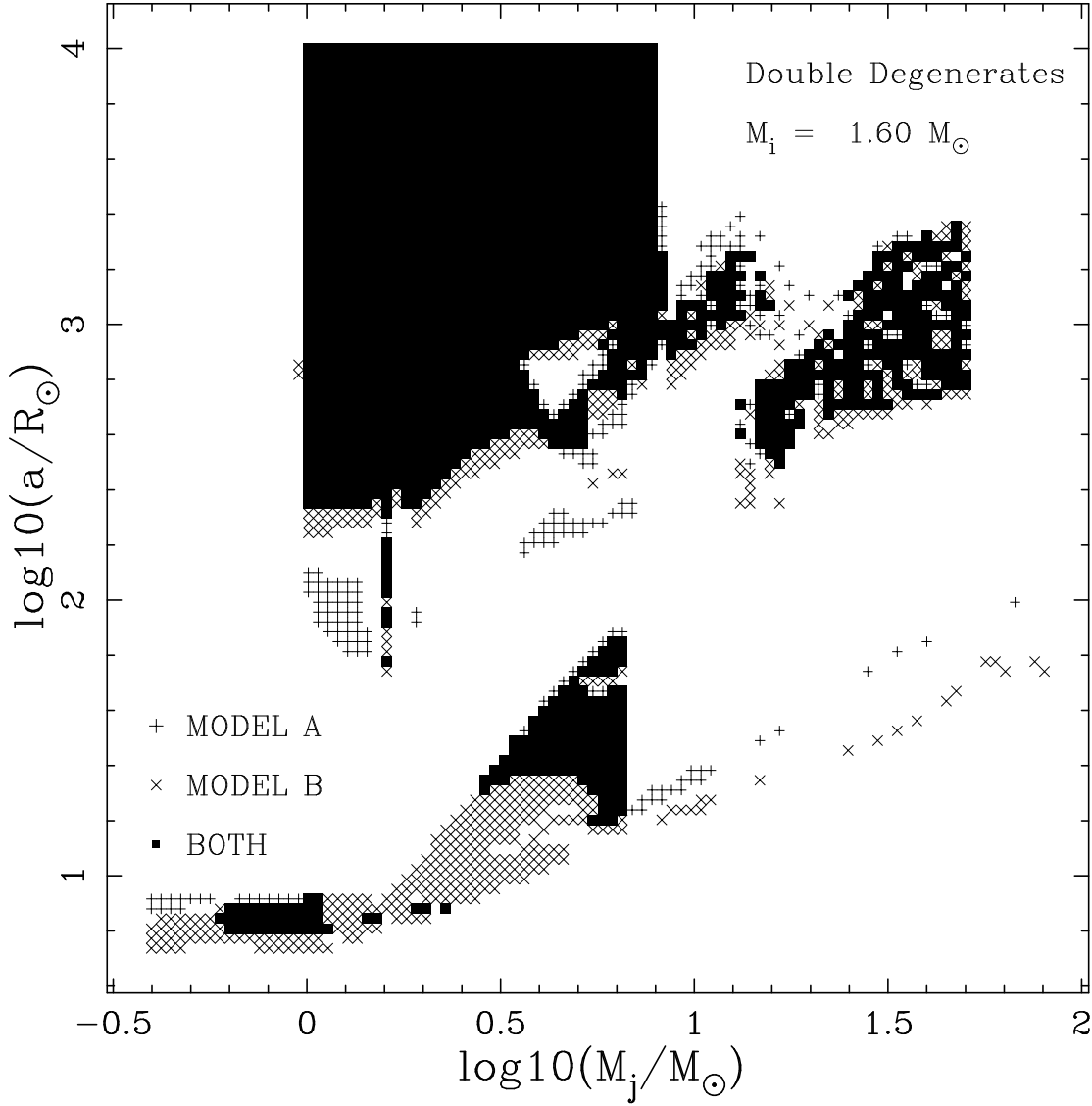}
}
\caption{
The region of parameter space from which double-degenerate (DD) binaries form
when the initial mass of one star is fixed at $1.6 \Msun$.
Systems formed from Model~A are shown as + symbols, from Model~B as $\times$ 
symbols, with any overlap as filled squares. 
The unstable behaviour of some near equal-mass systems is due to the
relatively short lifetime of the GB stage.
In this case there is a range of separations where the first phase of RLOF
begins with both stars on the GB and CE gives a HeWD-HeWD pair.
If the mass of either star changes slightly then one star is still on the MS
when RLOF begins and the subsequent evolution is altered.
}
\label{f:tides2}
\end{figure*}

As a typical example of classical CV formation in Model~A let us consider 
stars of mass $2.0$ and $0.2 \Msun$ in an orbit with initial separation 
$100 \Rsun$. 
The more massive star evolves to fill its Roche-lobe on the EAGB when the 
separation has fallen to $70 \Rsun$ (owing to tidal interaction), 
a CE forms and a HeHG star 
separated from the $0.2 \Msun$ MS star by $1.1 \Rsun$ emerges.  
The HeHG star evolves to a COWD and shortly afterwards the 
MS star fills its Roche-lobe creating a classical CV. 
The same system evolved with Model~B has expanded to $a = 105 \Rsun$ 
(owing to stellar wind mass-loss) when the first phase of RLOF begins. 
As a result the post-CE binary is too wide to interact within $T_{\rm gal}$.  
However, if the same masses are evolved from a closer initial 
separation of $50 \Rsun$, in Model~A the stars coalesce during CE 
while Model~B produces a classical CV. 

A typical CV Symb system can form from a $3.2$ and a $1.0 \Msun$ star  
initially separated by $180 \Rsun$. 
RLOF first occurs when the primary has evolved to the EAGB and tides have 
reduced the separation to $160 \Rsun$. 
Mass transfer is dynamical so that a CE forms from which a close binary 
consisting of a HeHG and a MS star emerges. 
The helium star evolves to a COWD and a second phase of RLOF begins 
when the companion reaches the HG. 
This time the mass transfer proceeds on a thermal timescale so that when the 
new primary reaches the GB its mass has dropped to $0.3 \Msun$ and a 
second CE phase is avoided. 
Stable mass transfer then continues on the GB until the primary loses its 
envelope and a DD system is born.  
This system has evolved through both a GK Per and a CV Symb phase. 
If it were to be evolved without tidal interaction the binary formed after the 
CE phase would be wider. 
The new primary would still fill its Roche-lobe on the HG, but only when 
closer to the GB, so that upon reaching the GB a second 
CE occurs in which the system coalesces. 
Thus a GK Per phase still exists but the CV Symb phase does not. 

Unlike the other types of CV, the sdB population increases in  
formation rate in Model~B.  
As already mentioned, there are two main regions of the parameter space 
shown in Figure~\ref{f:tides1} from which sdB systems form. 
First consider the population formed from close orbits which can be 
characterized by a binary of initial masses $3.2$ and $1.0 \Msun$ 
separated by $20 \Rsun$. 
The primary fills its Roche-lobe on the HG and its mass has fallen to 
$0.6 \Msun$ by the time it reaches the GB so that CE evolution is 
avoided and stable mass transfer continues. 
Eventually it loses its entire envelope to become a HeMS star. 
The companion evolves to the GB and fills its Roche-lobe, by which time 
the HeMS star has become a COWD and the resulting CE evolution leaves a 
COWD/HeMS pair in a close orbit. 
When the new HeMS evolves to fill its Roche-lobe a sdB system is formed. 
The population formed from wider orbits involves a slightly more complicated 
path represented by a binary 
with $M_2 = 3.2 \Msun$, $M_1 = 10.0 \Msun$ and $a = 500 \Rsun$.  
A first phase of RLOF begins when the more massive star is on the HG with the 
mass transfer becoming dynamical when it reaches the GB. 
The ensuing CE leaves a MS/HeMS pair separated by $10 \Rsun$. 
After the helium star evolves to the helium GB it fills its Roche-lobe and 
mass transfer proceeds until all the envelope is lost and the star becomes 
an ONeWD. 
The MS star now begins a third phase of RLOF when it reaches the HG. 
During this phase the ONeWD accretes enough material to swell-up and become a 
TPAGB star so that shortly afterwards the system evolves into 
another CE from which a close HeMS/ONeWD pair emerges. 
Yet another phase of RLOF begins when the new HeMS star fills its Roche-lobe 
and the sdB system is born. 
Mass transfer continues until an AIC changes the ONeWD to a NS. 
The increase in Model~B birth rate for the sdB population is explained by the 
cluster of additional systems about $M_2 = 2.6 \Msun$ and $a = 20 \Rsun$. 
These evolve via a stable RLOF phase followed by two instances of CE 
evolution which produce a close HeMS-COWD pair. 
When the helium star expands to fill its Roche-lobe the sdB phase begins. 
If tidal evolution is included the system formed after the 
first CE is much closer so that coalescence occurs during the second CE. 

A decrease in the common-envelope efficiency parameter 
(Model~C, ${\alpha}_{\rm\SSS CE} = 1$) 
increases the CV birth rate because wider systems are brought within an 
interaction distance after the CE phase. 
Thus systems that don't interact in Model~A can in Model~C. 
This is countered to some degree by closer systems  
that would interact in Model~A now coalescing. 
The net effect is an increase in production. 
It could be argued that a combination of Models~B and C would produce CV birth 
rates similar to those of Model~A. 

Comparing the birth rates of Algols in Models~A, B and C reveals no 
significant variation. 
This is expected because RLOF generally begins while the primary is still 
on the MS when tides have had little effect on the orbit and also 
because a CE phase is not involved. 
Of the gnt MSC systems that form 55\% of the population consist of a GB 
primary and a HG secondary while the remaining 45\% are GB primary and 
GB secondary. 
In most cases the GB-GB systems have previously been GB-HG and before that 
were Algols. 
The general evolution path for these systems begins with RLOF while the 
primary is on the HG and the secondary on the MS, the HG and GB lifetimes of 
the primary increase as it loses mass on the HG, mass transfer has reduced 
$q_1$ below $q_{\rm crit}$ when the primary reaches the GB, and the secondary 
evolves to the HG and then the GB while the Roche-lobe filling 
primary is still on the GB. 
Of the miscellaneous RLOF systems with helium star secondaries 51\% have a 
MS primary and a HeMS secondary, and 42\% a HeMS primary and secondary. 

\begin{figure*}
\centerline{
\psfig{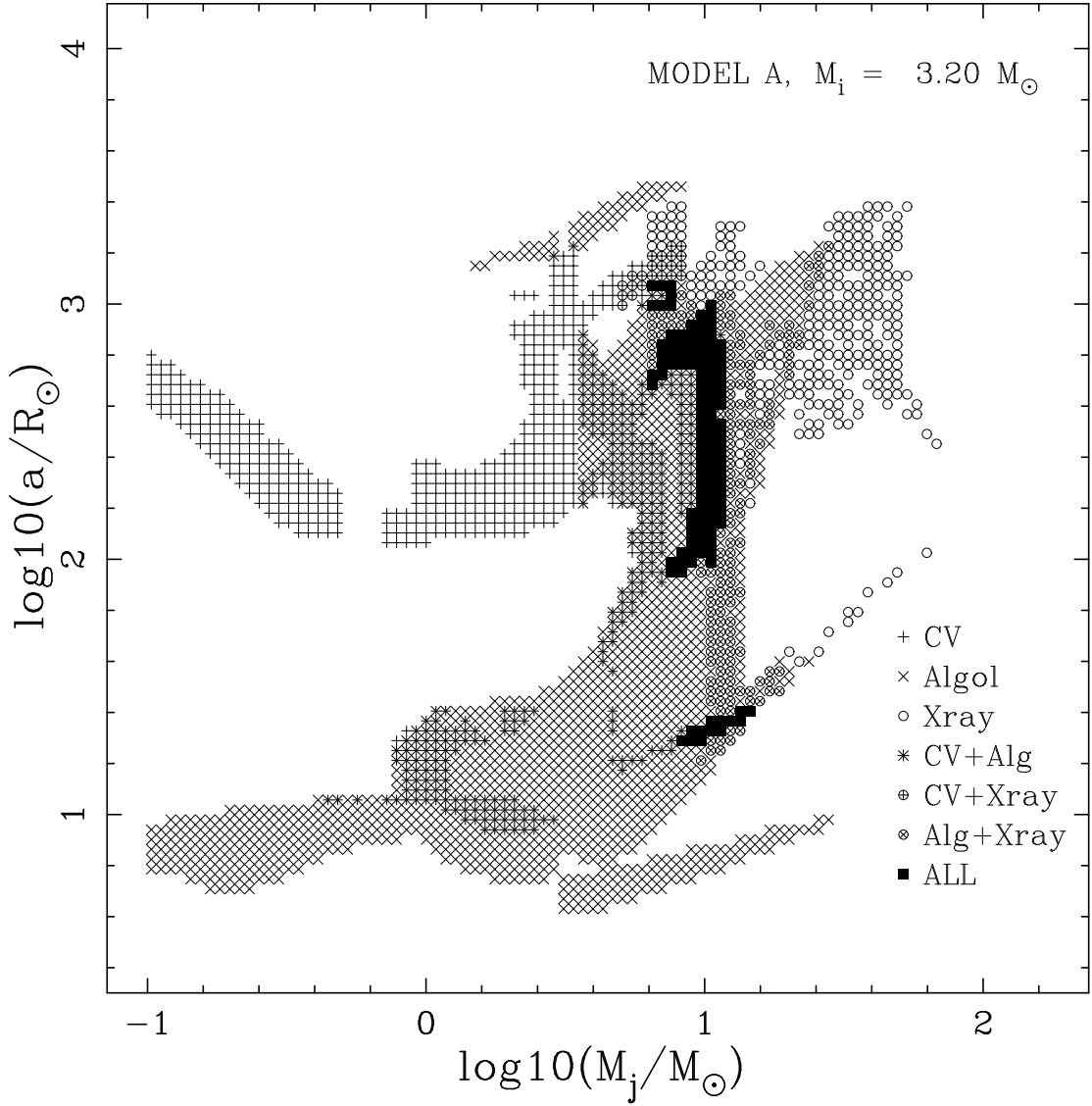}
}
\caption{
The region of parameter space from which CV, Algol and X-ray binaries form when
the initial mass of one star is fixed at $3.2 \Msun$ in Model~A.
}
\label{f:tides3}
\end{figure*}

\begin{figure*}
\centerline{
\psfig{figure=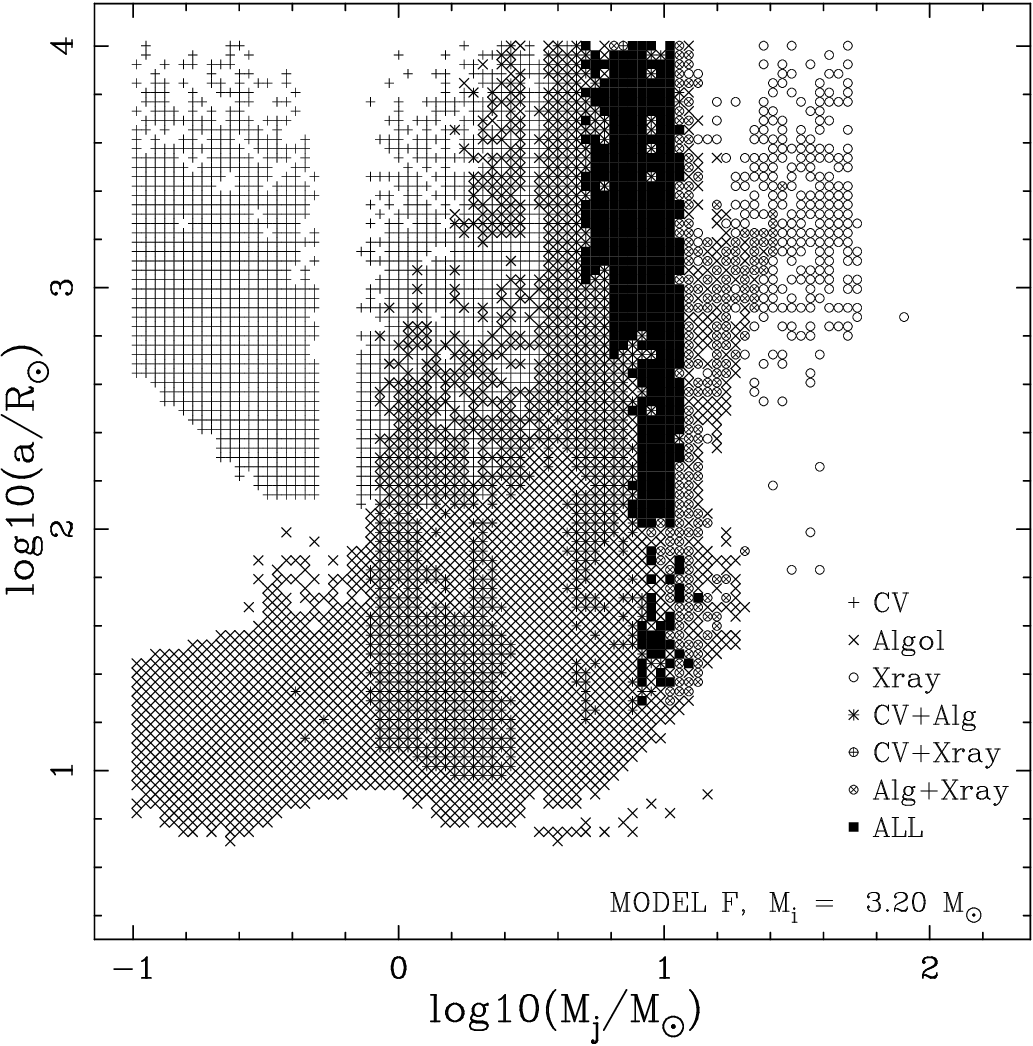,height=15cm}
}
\caption{
The region of parameter space from which CV, Algol and X-ray binaries form when
the initial mass of one star is fixed at $3.2 \Msun$ in Model~F.
}
\label{f:tides4}
\end{figure*}

The birth rate of LMXBs shows an order of magnitude decrease when either 
Model~B or Model~C is compared to Model~A. 
With tides a typical LMXB formation scenario begins with a primary mass 
somewhere in the range $10$ to $12 \Msun$, a low-mass companion with 
$M_2 \simeq 1.0 \Msun$ and $P \simeq 1\,000\,$d.  
The massive primary evolves to the EAGB before filling its Roche-lobe and 
forming a CE system from which a fairly close HeHG-MS pair emerges. 
The helium star explodes in a supernova that leaves a NS. 
This may increase or decrease the separation depending on the size of the 
velocity kick. 
After some time the MS star evolves to fill its Roche-lobe, by which time any 
eccentricity induced in the orbit by the supernova has been removed by the  
tides and a persistent NS~LMXB is formed. 
If tides are not included then the orbit of the HeHG-MS pair is much wider 
so that the second RLOF phase occurs when the primary is on the GB, a CE 
forms and a DD NS-HeWD binary is produced. 
On the other hand if ${\alpha}_{\rm\SSS CE} = 1$ then the orbit of the HeHG-MS 
pair is much closer after the initial CE so that the HeHG star fills its 
Roche-lobe resulting in another CE in which the two stars coalesce. 
There is an increase in the LMXB birth rate for Model~D, for which the 
secondary mass is chosen from the same IMF as the primary, 
because massive stars are more likely to have low-mass companions, 
increasing the number of systems that can become LMXBs. 
This is compensated by a corresponding decrease in the MXRB birth rate. 
All the remaining binary systems show a drop in birth rate for Model~D 
because more systems form with low~$q_2$ decreasing the likelihood of 
interaction. 

Interestingly the formation rate of DD systems increases when the tides are 
not used. 
Han (1998) finds that a phase of stable RLOF followed by CE evolution 
(RLOF+CE) and a combination of two CE phases (CE+CE) are the two main channels 
for forming DDs with the former channel more likely to occur for close orbits. 
Adding tides to the model removes many systems from the closer RLOF+CE channel 
because the separation decreases while the secondary evolves on the HG and GB, 
after the first RLOF phase, so that the CE produces a much closer pair 
of HeMS stars. 
This pair then evolve into contact before a DD can form. 
Figure~\ref{f:tides2} shows the parameter space in the $M_j$-$a$ plane for  
DD formation in Models~A and B when the initial mass of one star is 
fixed at $1.6 \Msun$. 
The systems with $M_1 > 8.0 \Msun$ are WD-NS binaries that survive the 
supernova kick while the range of close systems with $a \la 100 
\Rsun$ are RLOF+CE WD-WD binaries, of which there are noticeably more formed 
in Model~B. 
A typical RLOF+CE WDWD DD formation scenario in Model~B begins with $M_1 = 2.2  
\Msun$, $M_2 = 1.6 \Msun$ and $P = 2\,$d. 
The primary fills its Roche-lobe on the HG when $P = 1.9\,$d. 
RLOF ends on the GB just before the primary becomes a CHeB star, loses what 
little is left of its envelope, and moves to the HeMS. 
Now $M_1 = 0.35 \Msun$, $M_2 = 3.45 \Msun$ and $P = 60\,$d.  
The orbit has widened as a result of the conservative mass transfer. 
The more massive star (now assuming the role of the primary) 
evolves to the GB and overflows its 
Roche-lobe on a dynamical timescale so that a CE forms from which a 
HeMS-HeMS pair emerges with $P = 0.1\,$d. 
Both of these evolve to WDs in a close COWD-COWD DD system. 
If the same initial parameters are evolved with tides the system 
has $P = 30\,$d just before the CE so that a closer HeMS-HeMS pair forms. 
This evolves into contact before either star can become a WD. 
If the same masses are evolved with a longer initial period, $P \simeq 240\,$d, 
then a DD forms via the CE+CE channel in Model~B. 
In this case the primary fills its Roche-lobe on the EAGB, CE evolution 
ensues and a HeHG star with $M_1 = 0.6 \Msun$ in an orbit of $P = 6\,$d with 
the MS star emerges. 
The HeHG star evolves to a COWD. 
The MS star later fills its Roche-lobe on the GB, leading to another  
CE and the formation of a HeWD-COWD DD system with $P = 0.03\,$d. 
Evolving this case with Model~A leaves a closer binary after the first CE so 
that the second CE ends in coalescence of the two stars. 
Systems with wider initial orbits can survive the second CE phase to produce 
a DD binary when either Model~A or B is used. 
Even wider systems form a WDWD DD with no interaction but these have 
much longer periods than those formed via CE and are not easily detected. 

\begin{figure*}
\centerline{
\psfig{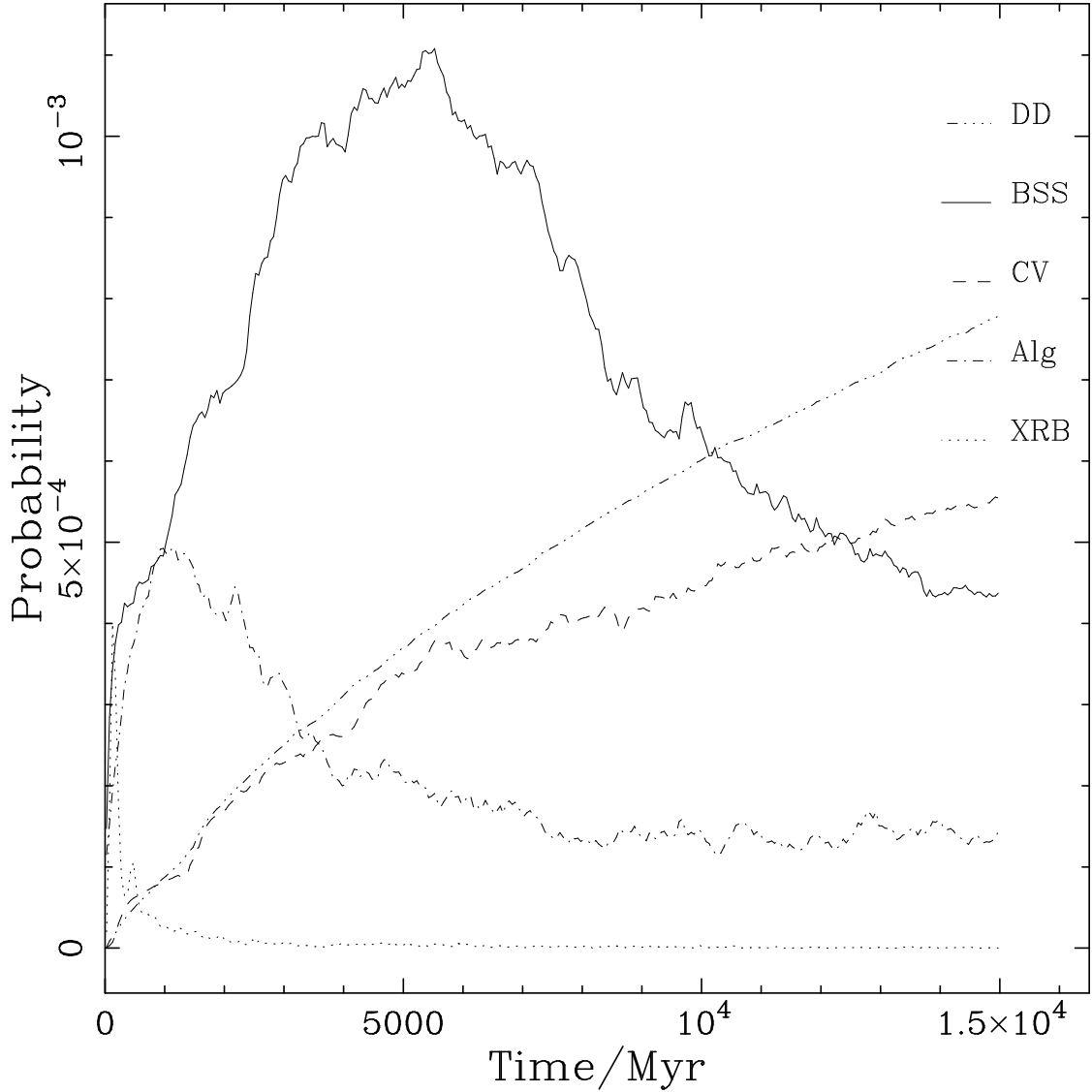}
}
\caption{
The probability, as a function of time, of a binary being found as a member of
certain individual populations of Model~A.
We assume that all stars are born at the same time.
The individual populations shown are: blue stragglers (full line), cataclysmic
variables (dashed line), Algols (dash-dot line), persistent X-ray 
binaries (dotted line) and double-degenerates (dash-dot-dot-dot line). 
Each population includes contributions from each of its sub-populations
listed in Table~\ref{t:tidcmp}.
The numbers for double-degenerates have been reduced by a factor of 10 and
the numbers for persistent X-ray binaries have been increased by a factor
of 10.
}
\label{f:clust1}
\end{figure*}

Inspection of Figure~\ref{f:tides2} reveals a large empty region 
in parameter space between the close group of systems that form DDs via 
RLOF+CE and the wider group of systems that form DDs via CE+CE. 
In this gap the initial period is short enough that the 
second CE leads to coalescence in both Models~A and B. 
However pockets of DDs do exist in Model~A because stable RLOF can occur
on the HG. 
Mass transfer reduces the primary mass and actually causes a net increase in 
the separation. 
This phase of RLOF, coming after the first CE, is non-conservative so that 
under the influence of tidal synchronization the separation 
initially decreases. 
As the primary mass is reduced further and the strength of the tides 
weakens the orbit starts to expand. 
When the primary reaches the GB a second CE occurs but the orbit is wide 
enough, and the size of the primary small enough, that coalescence is 
avoided and a DD formed. 
This is however a very unstable region of phase space. 
If the initial separation is reduced slightly the RLOF phase after the 
first CE occurs on the MS, the primary mass falls until it develops a 
convective envelope when dynamical mass transfer leads to coalescence. 
A slightly larger initial separation means that the RLOF phase starts on the 
GB before the star has lost any mass, so it is much larger and the CE ends 
in coalescence. 
If $M_1$ is increased slightly the RLOF phase after the first CE still starts 
on the HG but, owing to a larger $q_1$, leads to dynamical mass transfer and 
coalescence. 
A reduction in $M_1$ means the primary takes longer to fill its 
Roche-lobe before the first CE so that the tides give a greater reduction in 
the separation, the system formed after the CE is closer, and the 
next RLOF phase starts on the MS, once more ending in coalescence.  
With Model~B these regions do not occur at all because the orbit is always 
closer when the second CE forms so that the binary never survives. 

The lower metallicity population, Model~E, has increased 
birth rates in almost all cases compared with Model~A. 
This behaviour can be primarily attributed to the shorter nuclear lifetimes 
of lower metallicity stars with $M \la 9 \Msun$ (see Figure~5 of PapI).  
For example, a solar mass star with $Z = 0.0001$ evolves to a WD cooling  
cooling track in half the time it takes a solar mass star with $Z = 0.02$. 
This gives it more chance of interacting with its companion when in a 
relatively wide orbit. 
Model~E represents a very young population that would have formed soon 
after a galaxy condensed.  

It should be noted that the choice of separation distribution 
$\Psi \left( \ln a  \right)$ and its limits, particularly the upper cut-off, 
affects the calculated birth rates and numbers. 
Only the BSS and D-Symb populations generate any members from initial 
separations greater than $10^4 \Rsun$ but these are rare and so do not 
affect the absolute numbers very much. 

\subsubsection{The Role of Semi-Latera Recta}
\label{s:rectum} 

Figures~\ref{f:tides3} and \ref{f:tides4} show the parameter space of CV, 
Algol and X-ray binaries that are formed for Models~A and F respectively 
when the initial mass of one star is fixed at $3.2 \Msun$. 
In Model~F we form binaries with eccentric orbits and, as expected, this 
means that systems with wider orbits can interact at periastron and 
contribute to the rates. 
This does not automatically lead to an increase in formation rates because 
closer systems are more likely to be destroyed by a collision or coalescence.  
If we plot semi-latus rectum on the vertical axis of 
Figure~\ref{f:tides4} then it appears virtually 
identical to Figure~\ref{f:tides3}. 
Therefore, if a distribution of orbital angular momentum or $l$ is used to 
determine the initial state of each binary, rather than one of semi-major 
axis or period, the results do not depend on the form of any chosen 
eccentricity distribution. 
In fact eccentricity need not be a free parameter. 
This becomes evident when considering that tidal interaction conserves 
angular momentum and that almost all systems circularize before RLOF. 
Therefore (see eq.~\ref{e:ltoh}) systems with the same initial $l$ end up 
in a circular orbit with the same separation when RLOF starts, and their 
subsequent evolution will be identical.  
The only instances where the initial eccentricity distribution does matter 
are binary populations in which tidal interaction is weak, such as wide 
binaries that may avoid RLOF (e.g. the Barium stars, see 
Karakas, Tout \& Lattanzio 2000)

The circularization method of Model~G usually has increased formation rates 
because it acts to bring systems closer making them more likely to interact. 
Unlike Model~F none of the angular momentum is absorbed by the stars 
so that systems do not get so close and as a result less are destroyed. 
 
The relative fractions of blue stragglers, cataclysmic variables, Algols,
persistent low-mass X-ray binaries and double-degenerates as a function
of time for Model~A are shown in Figure~\ref{f:clust1},
assuming that all binaries were born in a single burst of star formation.
This represents the probability of a
binary observed in a star cluster being of a particular type.
Each population is a combination of its sub-populations as listed in
Table~\ref{t:tidcmp}.
The peak in the blue straggler distribution at approximately $5\,000\,$Myr 
is primarily due to a decrease in progenitor systems and its timing 
depends on the ratio of the mean binary separation, which remains constant, 
to the radius of the largest MS star, which decreases in time 
(Hurley et al. 2001). 
Both the CV and DD probabilities show a steady increase in time, reflecting 
both the long-lived nature typical of these systems and the constant 
production of WDs from the evolving mass distribution.  
As the existence of LMXBs is directly linked to the production 
of NSs or BHs it is no surprise that their probablity distribution peaks 
at an early age. 
The probability of observing an Algol also shows a peak that is linked to 
the evolution of massive stars as these are more likely to fill their 
Roche-lobes on the MS or HG, thus avoiding a common-envelope phase.

\subsection{Comparison with Observations}
\label{s:brfobs}

To determine whether the parameters of Model~A are a good choice we can 
compare with observations.  
For most of the individual binary populations observational birth 
rates or numbers in the Galaxy are very uncertain because of  
selection effects involved when undertaking surveys. 
However enough data exist overall to enable a meaningful comparison with the 
results from which a decision on the best model parameters can be made. 
We should add that caution must be exercised when comparing observationally 
defined classes of binary with the model classes defined in this work. 
Some overlap between classes necessarily exists: owing in part to the 
restrictive nature of the theoretical model and to remaining uncertainty 
about the details of some of these objects in the first place. 

\subsubsection{Cataclysmic Variables}

Ritter \& Burkert (1986) estimate the CV birthrate in our Galaxy to be 
about $10^{-14} \: {\rm pc}^{-3} \: {\rm yr}^{-1}$ or 
$6 \times 10^{-3} \: {\rm yr}^{-1}$ if $V_{\rm gal} = 6 \times 
10^{11} {\rm pc}^{3}$. 
In calculating this rate they corrected the observed local space density 
for the selection effect of only a small fraction of CVs being bright 
enough to be seen even in a distance limited sample. 
They assumed that the mass spectrum of WDs in CVs is not too different 
from that of single WDs. 
The birth rate of classical CVs in Model~A is $2.0 \times 10^{-2} \: 
{\rm yr}^{-1}$ which is only a factor of 3 greater than this so that the  
two are consistent, considering that a binary fraction less than unity 
would reduce the derived rate.  
The intrinsic local space density of CVs quoted by Ritter \& Burkert (1986) 
is $1-2 \times 10^{-4} \: {\rm pc}^{-3}$ 
(6 to $12 \times 10^7$ in the Galaxy), which they note seems rather high, 
and can only be matched by the results of Model~D. 
Yungelson, Livio \& Tutukov (1997) find the present local space density of 
classical CVs in their model to be about $3 \times 10^{-5} \: {\rm pc}^{-3}$ 
which is in agreement with the Hertz et al.~(1990) observed number and 
consistent with the results of all the models except Model~D. 

\begin{figure}
\centerline{
\psfig{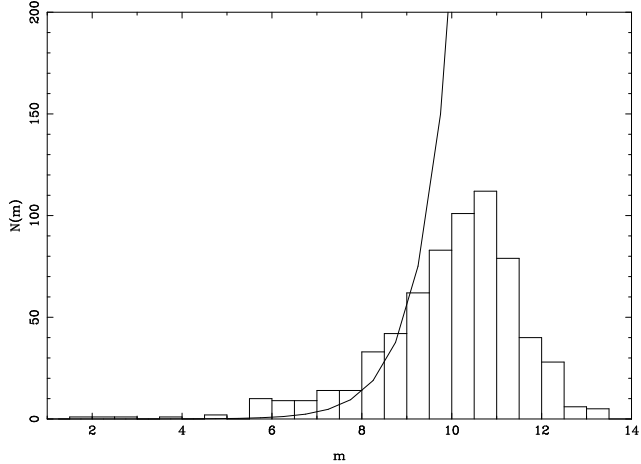}
}
\caption{
The apparent magnitude distribution of Algol systems identified by  
Brancewicz \& Dworak (1980). 
The magnitudes are taken from Wood et al.~(1980). 
A best fit to the data for $m \leq 9.5$ is shown by the full line. 
}
\label{f:algcat}
\end{figure}

\subsubsection{Algols}

Observational data on Algol systems suffers from the lack of 
a comprehensive survey aimed at establishing the local space density. 
The catalogue of eclipsing binaries published by Brancewicz \& Dworak (1980) 
contains parallaxes for 653 systems classified as Algol. 
Apparent magnitudes for these systems can be obtained from the catalogue 
of interacting binaries published by Wood et al.~(1980) although no 
mention is made as to the completeness limit of this catalogue. 
Figure~\ref{f:algcat} shows the distribution of the Algol systems that 
have a primary mass greater than $1 \Msun$ as a function of their apparent 
magnitude.
The apparent brightness of a star is linked to its distance and bolometric 
luminosity by 
\beq 
l = \frac{L}{4 \pi d^2}  
\eeq 
which defines a volume 
\beq 
V = \frac{4 \pi}{3} \left( \frac{L}{4 \pi l} \right)^{3/2} \, .  
\eeq 
Assuming that the stars are uniformly distributed within this volume the 
number of stars out to apparent brightness $l$ is 
\beq\label{e:appbgt}
N_l \left( m \right) = A \, l^{-3/2} n \left( m \right) 
\eeq
where $A$ is some constant. 
The function $n$($m$) depends only on the stellar mass and 
incorporates such quantities as the mass function, the dependence on 
luminosity and how long a star of a certain mass remains visible. 
Integrated over all masses this is constant in eq.~(\ref{e:appbgt}) 
so it acts only to normalize the relation. 

To determine the completeness limit of the Algol sample we made a series of 
least-squares fits of eq.~(\ref{e:appbgt}) to the data  
starting from the lowest magnitude bin, each successive fit includes data 
from the next bin in line. 
The number of Algol systems in the Brancewicz \& Dworak (1980) sample 
that are brighter than 10th magnitude is 282.  
A least-squares fit to this subset of the data is shown in 
Figure~\ref{f:algcat} and gives 300 systems brighter than 10th magnitude 
which is just consistent with the data within the Poisson error. 
If we include systems above 10th magnitude in the fit this leads to a large 
increase in the RMS error between the fitted function and the data as well  
as making the two inconsistent within Poisson error. 
Therefore it seems safe to assume that the Algol sample is complete to 
10th magnitude. 
This corresponds to a solar mass primary at the end of the MS being visible 
to a distance of roughly $180\,$pc. 
Returning to the Brancewicz \& Dworak (1980) sample and counting only systems 
within this distance with primary mass greater than $1 \Msun$ we find 50 
Algol systems corresponding to a local number density of 
$n \simeq 2 \times 10^{-6} \: {\rm pc}^{-3}$ and the number of classical 
Algols with $M_1 \geq \Msun$ currently in the Galaxy is then $1.2 \times 10^6$. 
However, almost half of the systems classified as Algol by 
Brancewicz \& Dworak (1980) are assessed to have an extremely unlikely 
probability of RLOF by Budding (1984) who analysed the 
semi-detached nature of each system. 
Therefore it is likely that this number is an over-estimate. 
On the other hand, only about 1/3 of all classical Algols are likely to 
be observed as eclipsing binaries (see Appendix~\ref{s:app3d} for details). 
The number of Algols with a MS or sub-giant primary more massive than $1 \Msun$ 
is predicted by Model~A to be about $1.4 \times 10^{7}$ 
(combining the MS, cold and hot Algol populations). 
This is an order of magnitude greater than the number predicted by 
observations but is close enough to not be a major worry when 
we consider the uncertainties.  

\subsubsection{Low-Mass X-ray Binaries}

The LMXB catalogue compiled by van Paradijs (1995) lists about 80 bright 
persistent sources in the Galaxy, making no distinction as to whether the 
compact star is a NS or BH. 
This number is in good agreement with the results of Model~A, and in fair 
agreement with Model~B.  
The birth rate of persistent NS LMXBs in Model~A is similar to the rate found 
by Portegies Zwart \& Verbunt (1996) in their standard model that includes 
velocity kicks. 
The relative distribution of NSs and BHs in X-ray binaries depends on 
the choice of single star mass above which BHs are formed rather than NSs 
(see Section~\ref{s:secresp}).  
This mass is not well constrained.  
It also depends on the assumption that a NS collapses to a BH when 
it accretes enough material to take its mass above $1.8 \Msun$ 
(Bombaci 1996; see also PapI). 
If the Eddington limit for mass transfer is applied then the birth rate of 
persistent NS LMXBs in Model~A increases to $3.612 \times 10^{-6}$ while 
the BH LMXB rate drops to $1.446 \times 10^{-6}$. 
The actual number of persistent LMXBs drops in this case, making the model 
inconsistent with the observations. 

\subsubsection{Symbiotic Stars}

About 125 symbiotic stars are observed, with WD secondaries 
about 10 times more prevalent than MS secondaries (Kenyon 1986). 
Estimates of the total number in the Galaxy range from about $1\,000$ 
(Boyarchuk 1975) up to $20\,000$ (Luthardt 1992). 
Of the known symbiotic stars roughly 20\% are D-Symb but this is probably 
biased because these are harder to identify due to dust enshrouding the system. 
Kenyon (1986) predicts from theory that the total number in the Galaxy 
should be about $3\,000$ with S-types and D-types contributing equally, 
although the relative number depends on the period distribution. 
Yungelson et al.~(1995) consider only symbiotics with WD secondaries 
and derive a birth rate of $0.073 \: {\rm yr}^{-1}$ with $3\,370$ present in 
the Galaxy using their standard model. 
These numbers are in good agreement with the predicted numbers from all of 
the models except Model~D which has only about $350$. 
At least 98\% of the symbiotics found in each of the models have 
WD secondaries, the remainder have HeMS secondaries, 
and the birth rates don't appear to fluctuate significantly 
between models, except Model~D.  

As already mentioned, the systems with MS secondaries are part of the 
cold Algol population, the number of which is typically $7 \times 10^5$ 
when only GB or AGB primaries are considered. 
This is much larger than the expected number so perhaps they are only 
observed as symbiotic for some of the RLOF phase or possibly 
only a small fraction are observed as symbiotic at all. 
The birth rate is typically of the order $10^{-2} \: {\rm yr}^{-1}$ so an 
average symbiotic lifetime of $10^4$~yr for these MS+giant systems 
would be consistent with the predicted numbers. 
The large CV Symb numbers are not consistent with these systems being observed 
as symbiotic. 
It could be that they are simply not found, because nova explosions don't 
occur on the WD surface, and the giant primary most likely dominates any 
light from the accretion disk. 

\subsubsection{Double-Degenerates}

Maxted \& Marsh (1999) have conducted a radial velocity survey to measure the 
fraction of DDs among hydrogen-line (DA) WDs. 
They claim that their survey is the most sensitive so far undertaken, in 
terms of the detection of DDs, yet they only found two systems in a 
sample of 46 DA WDs, and these two were known to be DD prior to the 
observations. 
Their detection efficiency is 80\% or higher for periods less 
than $10\,$d and quickly drops to zero for wider orbits. 
Thus it seems that the fraction of DDs with $P < 10\,$d among DA WDs 
has a lower limit of ${\nu}_{\SSS\rm WD} \approx 0.04$. 
This is consistent with the survey of Bragaglia et al.~(1991) who find 
between one and three DDs in a sample of 49 DA WDs. 
Combining their results with a theoretical model that predicts the 
period, mass and mass-ratio distributions of DDs, Maxted \& Marsh (1999) 
find a 95\% probability that the DD fraction lies in the range 0.017 to 0.19, 
independent of the details of the model used. 
Saffer, Livio \& Yungelson (1998) conducted a radial velocity survey of a 
sample of 107 DA WDs and 46 sdB stars. 
The sdB, or helium, stars were included because they are believed to evolve to 
the WD cooling track on a relatively short time, following the 
exhaustion of central helium. 
From this sample they found a possible 23 binaries of which 14 are WD+WD, 
7 sdB+WD and 2 WD+MS.  
Their detection efficiency is only 70\% for periods between 2 and $20\,$d. 
Of the 14 possible WD+WD binaries, seven have now been confirmed as DDs 
(Marsh, private communication) from the sample of 107 DA WDs so this 
suggests ${\nu}_{\SSS\rm WD} > 0.07$. 

Consideration of all these results together requires a reasonable 
theoretical model to predict at least ${\nu}_{\SSS\rm WD} \approx 0.05$ 
for DDs with periods less than $10\,$d, and 
${\nu}_{\SSS\rm WD} \approx 0.1$ when all DDs are taken into account. 
The frequency of DDs among DA WDs can be converted to a DD birth rate if it is 
assumed that one DA WD is born per year (Fleming, Liebert \& Green 1986; 
Phillips 1989): equivalent to our one star of $M > 0.8 \Msun$ per year. 
In terms of the WDWD DD birth rates both Models~A and B satisfy the 
requirements of the observations: they predict $0.113$ and 
$0.123 \: {\rm yr}^{-1}$ respectively when all periods are included;  
if only DDs with $P < 10\,$d are considered the rates drop to $0.053$ 
and $0.059 \: {\rm yr}^{-1}$. 
Model~C is in substantial disagreement with the 
observations: it predicts $0.076 \: {\rm yr}^{-1}$ for all periods and 
only $0.013 \: {\rm yr}^{-1}$ for $P < 10\,$d; Model~D is even worse. 
The predicted rate when eccentric orbits are allowed is also low 
but the small period rate in this case is still $0.051  \: {\rm yr}^{-1}$,  
in good enough agreement with both Model~A and the observations. 
Han (1998) includes periods up to at least $100\,$d in his calculation 
of $0.03 \: {\rm yr}^{-1}$ for the DD birth rate but his model uses 
${\alpha}_{\rm\SSS CE} = 1$ (although Han accounts for the ionization energy 
in calculating the envelope binding energy so this corresponds to a higher and 
variable ${\alpha}_{\rm\SSS CE}$ in this work). 

\begin{figure}
\centerline{
\psfig{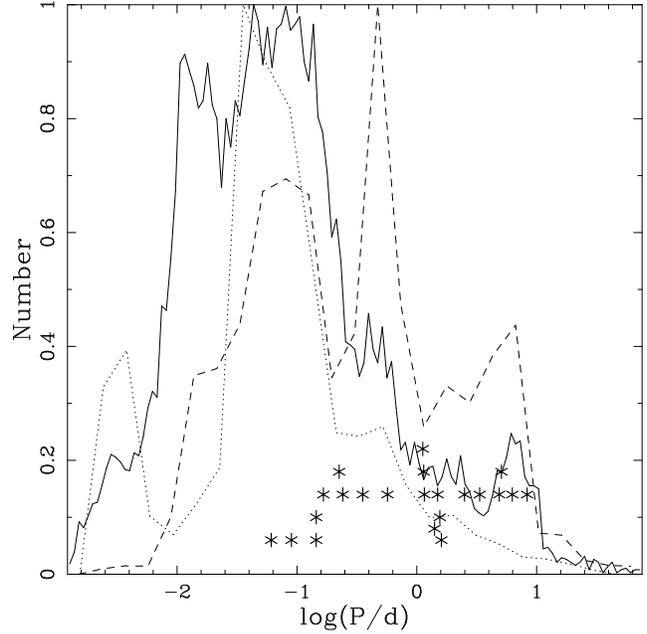}
}
\caption{
The distribution of periods of WDWD DD systems at formation in Model~A
(full line).
The same distribution weighted by the gravitational radiation
timescale of each system is shown for 
Model~A (dashed line) and Model~B (dotted line).
Also shown are the periods of 22 observed DD systems (asterisks) 
provided by Marsh (private communication)
Note that the vertical positioning of these points bears no relation to
any physical parameters.
}
\label{f:npdist}
\end{figure}

The period distribution at formation for DDs in Model~A is shown in 
Figure~\ref{f:npdist}.  
It has a peak at about $0.05\,$d. 
Included are periods for 22 suspected DD binaries provided by Marsh 
(private communication). 
To account for the short lives of short-period systems we also show the period 
distribution weighted by the corresponding gravitational radiation timescale 
for each system.  
The weighted distribution for Model~B shows that the 
observed distribution is much better represented when tidal evolution is 
included in the model. 

If the merging of two WDs is the progenitor of all type Ia SNe  
then it is necessary for an observed population of DDs 
to contain a certain proportion of very close systems with the 
combined mass of the components near or above the Chandrasekhar limit. 
The absence of these systems has proven a major obstacle for the 
proponents of the WD merger progenitor model 
(e.g. Saffer, Livio \& Yungelson 1998).  
However, the merger timescale decreases with shorter periods  
and larger binary masses so, as Figure~\ref{f:npdist} shows, such systems  
are less likely to be observed. 
The statistics of DD searches must be improved before the 
WD merger progenitor model can be ruled out on this basis. 
In Model~A the percentage of DDs with $P < 10\,$d present in the Galaxy now 
that contain two COWDs, have $M_1 + M_2 > \MCh$ and will merge within the 
lifetime of the Galaxy is 0.3. 
The percentage containing a COWD and a HeWD of mass at least $0.15 \Msun$ 
that will merge is 2.3. 
If all periods are included then the percentages drop by a factor of 10. 
Considering that 30 DD systems with measured periods are known  
(Marsh, private communication) we predict that a possible 
type Ia SNe progenitor should soon be discovered (e.g. Maxted, Marsh 
\& North 2000). 

The total number of DDs currently in the Galaxy is predicted by Model~A 
to be $6 \times 10^8$. 
If WDs are detectable for only the first $10^8\,$yr 
after formation then the number drops to $10^7$, more 
than the $3 \times 10^6$ predicted by the standard model of Han (1998). 

Bragaglia, Renzini \& Bergeron (1995) found that, of a sample of 164 
DA WDs, 15 were low-mass WDs with $M < 0.4 \Msun$. 
Low-mass WDs are interesting because they are not predicted by standard 
theory of single star evolution so they are believed to be, or have been,  
members of close binaries. 
They are not however directly representative of the number of DDs because 
they can be produced via other channels of binary interaction, such as in 
WD/NS binaries. 
Our theoretical models include WDs with $M < 0.5 \Msun$ in the LMWD 
populations, so a birthrate of $0.1 \, {\rm yr}^{-1}$ is a lower limit 
for comparison and is in good agreement with all models except Model~C.

\subsubsection{Supernovae}

Cappellaro et al.~(1997) combined the results of five independent SN 
searches to obtain a sample of 110 SNe from which they derived the 
following birth rates for the Galaxy: 
\begin{eqnarray*} 
4 \pm 1 \times 10^{-3} & {\rm yr}^{-1} & \mbox{  SNIa} \\ 
2 \pm 1 \times 10^{-3} & {\rm yr}^{-1} & \mbox{  SNIb/c} \\ 
12 \pm 6 \times 10^{-3} & {\rm yr}^{-1} & \mbox{  SNII} .  
\end{eqnarray*} 
The SNII rate is consistent with all the $Z = 0.02$ models except possibly 
Models~C and G. 
Note that the increase in type II SNe for the lower metallicity population 
(Model~E) is due to supernovae occurring in lower mass single stars 
(see Table~1 in Pols et al.~1998). 
An obvious discrepancy lies in the birth rates of SNIb/c because all the 
theoretical rates are too high, except for Model~D which is in good 
agreement. 
This can be reconciled by assuming a binary fraction smaller than unity.  

The SNII rate is a fairly robust number but the SNIb/c rate is sensitive
to many of the assumptions underlying the model. 
In all models other than Model~D the majority of SNIb/c come from 
naked helium stars formed from relatively low-mass progenitors ($10-20 \Msun$) 
stripped by binary interaction. 
In a model containing only single stars with $Z=0.02$ the SNII rate is the 
same as for Model A but the SNIb/c rate is reduced to $2.7 \times 10^{-3}
\, {\rm yr}^{-1}$ which would be consistent with observations. 
These single SNe Ib/c come from stars initially more massive than $24 \Msun$ 
which become Wolf-Rayet stars prior to explosion. 
However this rate is sensitive to the assumed mass-loss rate for red 
supergiants (see PapI), as well as to our assumption that BH formation 
ignites a supernova explosion, both of which are very uncertain. 
If mass-loss rates are lower or if black holes form without a SN, 
the SNIb/c rate from single stars would be lower.

\subsubsection{$\Gamma$-ray Bursts}

The observed rate of $\gamma$-ray bursts is about $10^{-6} \: {\rm yr}^{-1}$ 
per galaxy (Piran 1996) which is typically an order of magnitude less than 
the merger rate of NS binaries found in all models except Model~D. 
Numerical models of NS mergers suggest that the $\gamma$-rays 
produced are beamed and appear only in certain directions (Davies et al.~1994). 
An optically thick radiation-electron-positron 
plasma, or {\it fireball}, with initial energy larger than its rest mass is 
formed. 
The fireball is highly aspherical and expands along the polar axis to 
form a jet. 
Only when the fireball reaches ultra-relativistic velocities and its 
material becomes optically thin, or reacts with the interstellar medium, do 
the observed $\gamma$-rays emerge. 
If the width of the jet is $\Theta$ then we observe $\gamma$-ray bursts only 
from a fraction $2 {\Theta}^{-2}$ of NSNS mergers (Piran 1996). 
Portegies Zwart \& Yungelson (1998) find the neutron star merger rate in their  
model to be about $2 \times 10^{-5} \: {\rm yr}^{-1}$ which they claim is 
consistent with the observed rate of $\gamma$-ray bursts if the 
$\gamma$-ray emission is beamed into an opening angle of $10^{\circ}$. 
The same reasoning can reconcile the rates for all our models, except 
Model~D, with the observed rate of $\gamma$-ray burst events.

\section{Conclusion}
\label{s:bconc}

The primary purpose of this paper was to describe in detail a rapid evolution 
algorithm for binary stars. 
Recognizing that tidal interaction is an important process in binary 
evolution we have also studied its effect on the results of binary 
population synthesis and made a global comparison with observations. 

The main effects of tides on the evolution of an interacting binary are: 
\begin{itemize}
\item[(a)] circularization of the orbit, generally well before RLOF occurs, 
           and  
\item[(b)] the exchange of angular momentum between the orbit and the spins 
           of the components. 
\end{itemize}
Because as a star expands tidal synchronization transfers 
orbital angular momentum to the stellar rotation, in general the latter
effect tends to bring binary components closer together and cause them to
follow an evolutionary path similar to that of a closer binary if tides
were ignored. 
However, we find that as a result of the sometimes very 
convoluted evolutionary paths of interacting binaries, this general
principle does not always hold and it is difficult to summarize the 
overall effect of tides on binary evolution. 

The primary differences in our population synthesis results when 
tidal evolution within binary systems is ignored are as follows: 
\begin{enumerate} 
\item a 35\% decrease in the birth rate of classical CVs but a 
      23\% increase in sdB CV production; 
\item lower birth rates for LMXBs with the persistent NS LMXB rate 
      decreasing significantly but conversely the number of NS LMXB increases; 
\item a 10\% increase in the birth rate of double-degenerates; 
\item an increase of 190\% in the incidence of exploding HeWDs; 
\item and 55\% more Algols currently in the Galaxy. 
\end{enumerate} 
While these differences may seem substantial they are generally not enough 
to confirm one way or another whether tidal evolution helps to 
explain the observed binary populations. 
This is mainly due to the lack of comprehensive binary searches and 
the selection effects involved that make discrimination on the basis of 
observational evidence a risky business. 
However, tides are present in binary systems whether we like it or not. 
One area where neglecting tidal evolution does seem in clear conflict with 
observations is in the incidence of HeWDs that explode as supernovae. 
This rate is too high by a factor of two to be accounted for by either the 
observed numbers of type Ia or type Ib SNe 
(see Tout et al.~2001 for further discussion). 
 
The common-envelope efficiency parameter ${\alpha}_{\rm\SSS CE}$ is an 
uncertain factor in a phase of evolution that is crucial for the production 
of many types of binary. 
The failure of the model with ${\alpha}_{\rm\SSS CE} = 1$ to produce anywhere 
near enough DDs would seem to favour ${\alpha}_{\rm\SSS CE} = 3$. 
Choosing the secondary mass independently of the primary mass does not 
produce enough DDs, symbiotics or type Ia SN candidates.  
All in all the results favour using the properties of Model~A.  
In fact the standard tidal model, represented by Model~A, is not in 
disagreement with any of the observations, except in the production of  
too many type Ib/c SNe. 
This discrepancy could indicate a lower mass loss rate for helium stars. 

A major conclusion of this work must be that it is extremely difficult to 
set contraints on any of the parameters involved in binary 
evolution from population synthesis of birth rates and galactic numbers 
of the various types of binary. 
This is certainly true while such a large number of parameters remain 
uncertain. 
The task would be simpler if we could find observational tests specific 
to an individual parameter. 
For example, Algols do not require common-envelope evolution 
for formation. 
Therefore, if observational constraints on the number of Algol systems 
currently in the Galaxy improve then this could provide a suitable test 
for models of thermal-timescale mass transfer.  
The effect of varying additional model parameters, such as the binary 
enhanced mass loss, and initial conditions, such as the separation 
distribution, would need to be quantified before such a test could be 
reliably implemented. 
This is beyond the scope of this paper but will be the subject of future 
work. 
In terms of tides it is difficult to isolate direct tests of the tidal 
evolution model with the type of population synthesis performed in this 
paper. 
In fact, to properly constrain the strength of tidal interaction it is 
necessary to study pre-RLOF binaries, for example, 
as has been done previously by Zahn (1975, 1977). 
What we would like to emphasise is that the outcome of binary evolution 
is sensitive to the physical processes of tidal circularization and 
synchronization. 
Therefore, any attempt to constrain uncertain parameters in binary 
evolution by the method of population synthesis must utilise a 
binary evolution algorithm that incorporates a working model of 
tidal evolution, such as that presented here.  

An exciting challenge for the future involves attempting to reproduce  
the individual orbital characteristics of a large number of observed 
binaries with our binary evolution model. 
This will allow multi-parameter fitting and should become a powerful tool 
as the statistics of binary surveys continue to improve. 
Our preliminary work in this area has shown that the observed properties of 
DD binaries are much easier to explain when tidal evolution is included. 
A detailed exploration of the DD parameter space will be the focus 
of another paper. 

Tidal synchronism is important but because orbits generally circularize
before Roche-lobe overflow the outcome of the interactions of systems with
the same semi-latus rectum is almost independent of eccentricity.
Although the inclusion of a distribution of eccentricities seems natural 
it is not necessary in population synthesis of interacting binaries, 
however, the initial separations should be distributed according to the 
observed distribution of semi-latera recta rather than periods or 
semi-major axes. 

A necessity for the near future is a thorough comparison of our BSE 
algorithm with the workings of a detailed evolution code 
(e.g. Nelson \& Eggleton 2001). 
In this way we can improve the algorithm, especially the treatment of the 
Hertzsprung gap phase, and hopefully add more stringent constraints to 
many of the evolution variables. 
Work is already underway to provide more accurate descriptions of the parameters 
$k'_2$ (see Section~\ref{s:beqtid}) and $\lambda$ (see Section~\ref{s:comenv}), 
through investigation of the detailed stellar models provided by 
Pols et al.~(1998). 
The algorithm will also be improved by providing options for how angular 
momentum is lost from the binary system during non-conservative 
mass-transfer (see Section~\ref{s:rlof}), 
reflecting the various possible modes of mass-transfer 
(Soberman, Phinney \& van den Heuvel 1997).  
When this work is completed we will perform additional population  
synthesis calculations in order to quantify how the various improvements 
affect the results presented here.

\section{Availability of the BSE code}
\label{s:bsecde}

The BSE algorithm allows the entire evolution of even the most complicated 
binary systems to be modelled in less than a second of CPU time rather than 
the several hours required when using a full stellar evolution code. 
It is therefore ideal for synthesising large populations of binary stars 
and will prove an extremely useful tool for testing uncertain processes 
in binary evolution. 
To obtain a copy of the BSE package described in this paper send a request to 
the authors who will consider providing the {\sc fortran} subroutines by ftp. 

\section*{ACKNOWLEDGMENTS}

JRH thanks Trinity College and the Cambridge Commonwealth Trust for their
support during this work.
CAT is very grateful to PPARC for support from an Advanced Fellowship.
ORP thanks the Institute of Astronomy, Cambridge for supporting a number of
visits undertaken during this work.
We would like to thank Peter Eggleton, Sverre Aarseth, Philipp Podsiadlowski 
and Gerry Gilmore for many helpful suggestions and comments. 
We thank the anonymous referee for a number of insightful remarks. 
We also thank Tom Marsh for making his data on double-degenerate systems 
available.

\newpage
\appendix
\section{}

\subsection{Orbital Parameters After Supernova}
\label{s:app3b}

\begin{figure}
\centerline{
\psfig{figure=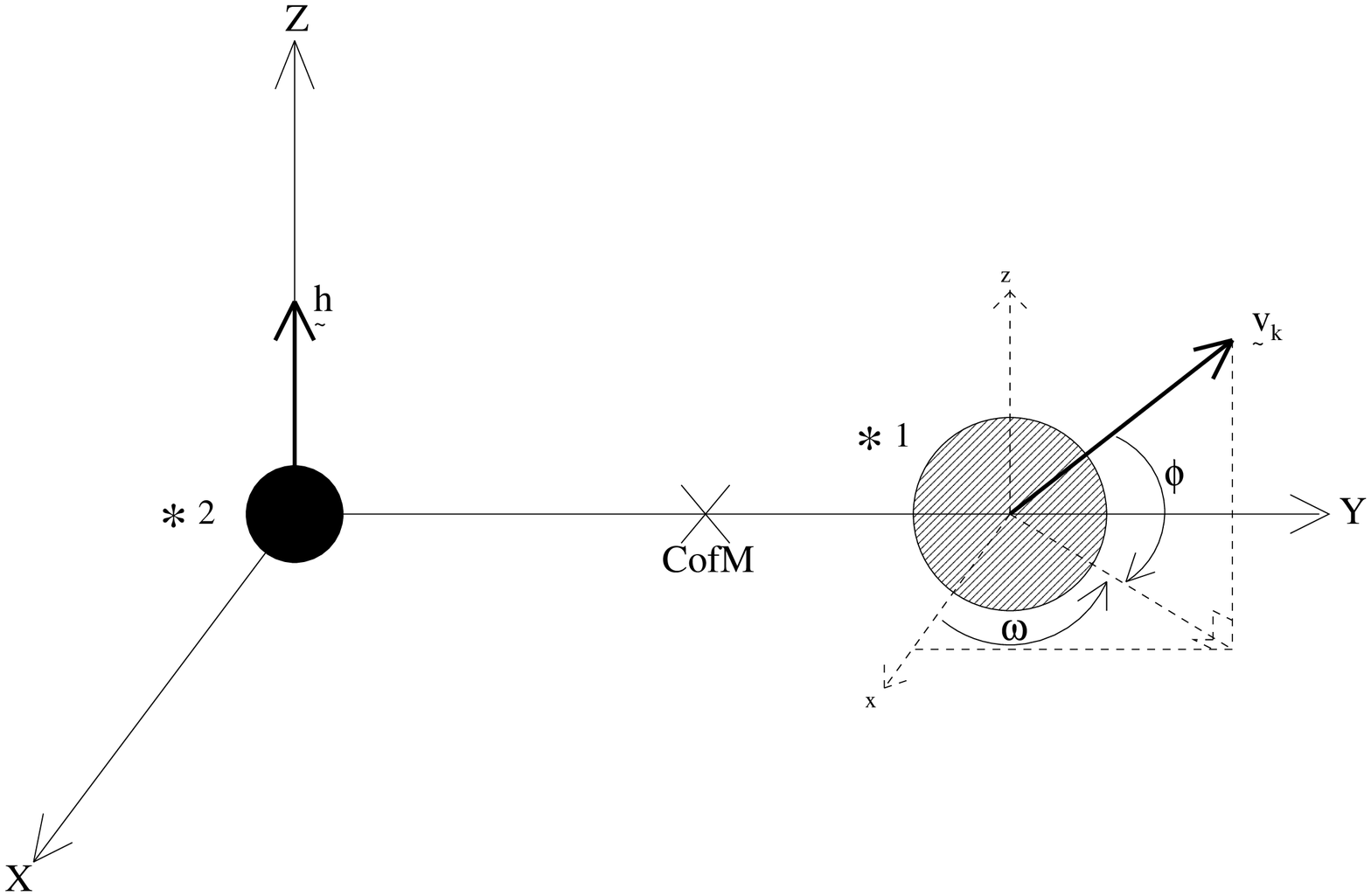,width=8.4cm}
}
\caption{The binary and supernova kick geometry}
\label{f:ffig3x}
\end{figure}

Consider a frame of reference in which the pre-supernova centre-of-mass is at 
rest, the primary star is about to explode and the secondary is at the 
origin. 
The initial orbit is in the XY-plane, so that the initial specific angular 
momentum vector is directed along the positive Z-axis, and the separation 
vector $\vect{r} = \vect{r}_1 - \vect{r}_2$ is directed along the 
positive Y-axis, as shown in Figure~\ref{f:ffig3x}. 
The relative velocity of the stars is 
\beq 
\vect{v} = - v_{\rm orb} \left( \sin \beta \vect{i} + \cos \beta \vect{j} 
\right) 
\eeq 
where $\beta$ is the angle between $\vect{r}$ and $\vect{v}$ and is such that  
\begin{eqnarray}
\sin \beta & = & \left[ \frac{a^2 \left( 1 - e^2 \right)}{r \left( 2 a - r
\right)} \right]^{1/2} \label{e:sinbet} \\
\cos \beta & = & - \frac{e \sin {\rm E}}
{\left( 1 - e^2 {\cos}^2 {\rm E} \right)^{1/2}} \label{e:cosbet}
\end{eqnarray} 
and E is the eccentric anomaly in Kepler's equation
\beq\label{e:kepanom}
{\mathcal{M}} = {\rm E} - e \sin {\rm E}
\eeq
for mean anomaly ${\mathcal{M}}$, which varies uniformly with time between
$0$ and $2 \pi$.
The orbital speed is defined by 
\beq\label{e:relspd} 
v^2_{\rm orb} = {\dot{r}}^2 + r^2 {\dot{\theta}}^2 = 
G M_{\rm b} \left( \frac{2}{r} - \frac{1}{a} \right) \, . 
\eeq 

As well as losing an amount of mass $\Delta M_1$ the primary is subject 
to a kick velocity $\vect{v}_k$ during the supernova explosion so that 
\begin{eqnarray*}
M_1 & \rightarrow & M'_1 = M_1 - \Delta M_1 \\ 
M_{\rm b} & \rightarrow & M'_{\rm b}  = M_{\rm b} - \Delta M_1 \\ 
\vect{v}_1 & \rightarrow & {\vect{v}'}_1 = \vect{v}_1 + \vect{v}_k 
\end{eqnarray*}
where 
\beq 
\vect{v}_k = v_k \left( \cos \omega \cos \phi \vect{i} + \sin \omega 
\cos \phi \vect{j} + \sin \phi \vect{k} \right) \quad , \quad 
v_k = | \vect{v}_k | \, . 
\eeq 
Here $\vect{i}$, $\vect{j}$ and $\vect{k}$ are unit vectors in the X, Y 
and Z directions respectively. 
We assume the separation is constant as the explosion is instantaneous. 
To find the separation at the moment of explosion we randomly choose a mean 
anomaly ${\mathcal{M}}$ and then solve eq.~(\ref{e:kepanom}) 
for the eccentric anomaly E using a Newton-Raphson method. 
Then 
\beq 
r = a \left( 1 - e \cos {\rm E} \right) \quad , \quad r = | \vect{r} | 
\eeq 
in terms of the initial semi-major axis and eccentricity. 
This is necessary because the binary is evolved using values averaged over an 
orbital period so that the exact separation at any one time is not known 
for an eccentric orbit. 

After the supernova the new relative velocity is 
\begin{eqnarray*}
\vect{v}_n & = & \vect{v} + \vect{v}_k \\ 
 & = & \left( v_k \cos \omega \cos \phi - v_{\rm orb} \sin \beta \right) 
\vect{i} + \\ 
& & \left( v_k \sin \omega \cos \phi - v_{\rm orb} \cos \beta \right) 
\vect{j} + v_k \sin \phi \vect{k} \, . 
\end{eqnarray*}
From eq.~(\ref{e:relspd}) it must be true that 
\beq\label{e:newa} 
v_n^2 = G M'_{\rm b} \left( \frac{2}{r} - \frac{1}{a_n} \right) 
\eeq
for the new orbital parameters, where 
\begin{eqnarray}
v_n^2 & = & | \vect{v}_n |^2 \nonumber \\ 
      & = & v_k^2 + v_{\rm orb}^2 - 2 v_{\rm orb} v_k \left( \cos \omega 
\cos \phi \sin \beta + \right. \\ 
      & & \hspace*{3.0cm} \left. \sin \omega \cos \phi \cos \beta 
\right) \, . 
\nonumber 
\end{eqnarray}
This can be solved for the semi-major axis $a_n$ of the new orbit.  
The specific angular momentum of the new system is 
\beq 
{\vect{h}}' = \vect{r} \times \vect{v}_n  
\eeq 
so it follows from eq.~(\ref{e:ltoh}) that 
\beq\label{e:newe}
G M'_{\rm b} a_n \left( 1 - e_n^2 \right) = | \vect{r} \times \vect{v}_n |^2 
\eeq
where 
\beq 
| \vect{r} \times \vect{v}_n |^2 = r^2 \left[ v_k^2 {\sin}^2 \phi + \left( 
v_k \cos \omega \cos \phi - v_{\rm orb} \sin\beta \right)^2 \right] \, . 
\eeq 
This we solve for the new eccentricity $e_n$ of the orbit. 
If either $a_n \leq 0$ or $e_n > 1$ then the binary does not survive the kick. 

The angle $\nu$ between the new and old angular momentum vectors is given by 
\beq 
\cos \nu = \frac{v_{\rm orb} \sin \beta - v_k \cos \omega \cos \phi}
{\left[ v_k^2 {\sin}^2 \phi + \left( v_k \cos \omega \cos \phi - v_{\rm orb} 
\sin \beta \right)^2 \right]^{1/2}} \, . 
\eeq 
An amount of mass $\Delta M_1$ is ejected from the primary, and hence 
from the system, so that the new centre-of-mass has a velocity 
\beq 
\vect{v}_s = \frac{M'_1}{M'_{\rm b}} \vect{v}_k - \frac{\Delta M_1 M_2}
{{M'}_b M_{\rm b}} \vect{v} 
\eeq 
relative to the old centre-of-mass frame. 
This agrees with Brandt \& Podsiadlowski (1995) for initially circular 
orbits. 
Note that the $\phi$ used here is $\pi - \phi$ as defined in their 
coordinate system. 

To determine the kick velocity it is necessary to choose $v_k$, the 
magnitude of $\vect{v}_k$, $\phi$, the angle between $\vect{v}_k$ and the 
orbital plane, and $\omega$, the angle between the projection of 
$\vect{v}_k$ on to the orbital plane and the X-axis, from appropriate 
distribution functions. 
We take the kick speed from a Maxwellian distribution 
\beq 
P \left( v_k \right) = \sqrt{\frac{2}{\pi}} \frac{v_k^2}{{\sigma}_k^3} 
e^{- v_k^2 / 2 {\sigma}_k^2} 
\eeq 
with a dispersion of ${\sigma}_k = 190\, {\rm km} \, {\rm s}^{-1}$  
(Hansen \& Phinney 1997) based on analysis of various pulsar proper motion 
samples.  

For a variable ${\mathcal{X}}$ uniformly distributed between 0 and 1 
we may write 
\beq 
P \left( v_k \right) d v_k = d {\mathcal{X}} 
\eeq 
\beq 
\Rightarrow {\mathcal{X}} = \int_0^{v_k} P \left( v_k \right) d v_k = 
F \left( v_k \right) \, . 
\eeq 
The change of variable 
\beq 
u^2 = \frac{v_k^2}{2 {\sigma}_k^2} 
\eeq 
gives us 
\beq 
F \left( u \right) = {\rm Erf} \left( u \right) - \frac{2}{\sqrt{\pi}} u 
e^{- u^2} 
\eeq 
so that ${\mathcal{X}} = F \left( u \right)$ can be solved for some 
${\mathcal{X}}$, by Newton-Rapshon iteration, to give $v_k$. 

The kick direction is uniform over all solid angles so that  
\beq 
P \left( \phi \right) = \cos \phi \quad , \quad \frac{- \pi}{2} \leq \phi  
\leq \frac{\pi}{2} \, , 
\eeq 
and $\omega$ is uniformly distributed between $0$ and $2 \pi$.

\subsection{Eclipsing Semi-Detached Binaries}  
\label{s:app3d}

\noindent Consider a semi-detached system in which a primary of radius $R$ is 
filling its Roche-lobe and is separated from a companion of radius $r$ by a 
distance $a$. 
The angle that a line which is the inner tangent to the surface 
of both stars makes 
with the line joining the centres of the stars can be given by 
\beq 
\sin \psi = \frac{R + r}{a} \, . 
\eeq 
If we assume that half of the companion must be eclipsed for the 
light-curve to be noticeably changed and that $R \simeq R_{\SS L}$, then 
\beq 
\sin \psi \simeq \frac{R_{\rm\SSS L}}{a} \approx 1/3 \, .   
\eeq 
Define $i$ as the inclination angle between the line-of-sight and the normal 
to the orbital plane of the binary. 
If $i = 0$ then the orbital plane is perpendicular to the line-of-sight and 
the binary will never eclipse; if $i = \pi/2$ then it is edge-on 
and will always eclipse. 
The probability of $i$ between $i$ and $i + di$ is 
\beq 
f (i) \, di = k \, \sin i \, di \, ,
\eeq 
for a normalization constant $k$. 
Only one orientation of the orbit gives $i = 0$ so it is the least 
likely case, whereas a full circle of orientations gives $i = \pi/2$. 
The normalization constraint 
\beq 
\int_0^{\pi/2} f (i) \, di = 1 
\eeq 
determines $k = 1$. 
Now the binary eclipses if 
\beq 
\pi/2 - \psi < i < \pi/2 
\eeq 
where 
\beq 
\psi = \arcsin ( 1/3 ) = \frac{\pi}{2} - \arccos ( 1/3 ) \, .  
\eeq 
Therefore the fraction of systems which are likely to be eclipsing is 
\beq 
P = \int^{\pi/2}_{\arccos ( 1/3 )} \sin i \, di = 1/3 \, . 
\eeq 

\label{lastpage}
\end{document}